\documentclass[fleqn,usenatbib]{mnras}
\usepackage{graphicx}	
\usepackage{amsmath}	
\usepackage[dvipsnames]{xcolor}  
\usepackage{amssymb}	
\usepackage{newtxtext,newtxmath}    
\usepackage{ulem}
\usepackage{macros_vdb} 
\begin{document}
%

\title[Bayesian Inference with Satellite Kinematics]
      {BASILISK: Bayesian Hierarchical Inference of the Galaxy-Halo Connection using Satellite Kinematics--I. Method and Validation}

\author[van den Bosch et al.]{%
   Frank~C.~van den Bosch$^1$\thanks{E-mail: frank.vandenbosch@yale.edu},
   Johannes~U.~Lange$^1$,
   Andrew~R.~Zentner$^2$
\vspace*{8pt}
\\
   $^1$Department of Astronomy, Yale University, PO. Box 208101, New Haven, CT 06520-8101\\
   $^2$Department of Physics and Astronomy \& Pittsburgh Particle Physics, Astrophysics, and Cosmology Center (PITT PACC),\\University of Pittsburgh, Pittsburgh, PA 15260, USA
}


\date{}

\pagerange{\pageref{firstpage}--\pageref{lastpage}}
\pubyear{2013}

\maketitle

\label{firstpage}


\begin{abstract}
   We present a Bayesian hierarchical inference formalism (\Basilisk) to constrain the galaxy-halo connection using satellite kinematics. Unlike traditional methods, \Basilisk does not resort to stacking the kinematics of satellite galaxies in bins of central luminosity, and does not make use of summary statistics, such as satellite velocity dispersion. Rather, \Basilisk leaves the data in its raw form and computes the corresponding likelihood. In addition, \Basilisk can be applied to flux-limited, rather than volume-limited samples, greatly enhancing the quantity and dynamic range of the data. And finally, \Basilisk is the only available method that simultaneously solves for halo mass and orbital anisotropy of the satellite galaxies, while properly accounting for scatter in the galaxy-halo connection. \Basilisk uses the conditional luminosity function to model halo occupation statistics, and assumes that satellite galaxies are a relaxed tracer population of the host halo's potential with kinematics that obey the spherical Jeans equation. We test and validate \Basilisk using mocks of varying complexity, and demonstrate that it yields unbiased constraints on the galaxy-halo connection and at a precision that rivals galaxy-galaxy lensing. In particular, \Basilisk accurately recovers the full PDF of the relation between halo mass and central galaxy luminosity, and simultaneously constrains the orbital anisotropy of the satellite galaxies. \Basilisk's inference is not affected by potential velocity bias of the central galaxies, or by slight errors in the inferred, radial profile of satellite galaxies that arise as a consequence of interlopers and sample impurity. 
\end{abstract} 




\section{Introduction} 
\label{sec:intro}

Accurately constraining the link between galaxies and dark matter haloes, which goes by the catch-all name 'halo-occupation modelling', provides valuable insight regarding the formation and evolution of galaxies in a $\Lambda$CDM cosmology. It describes the link between what we can see (light) and what governs dynamics (mass), and therefore provides a powerful tool to probe (the evolution of) the matter power spectrum. The main techniques that are being utilized to constrain this galaxy-halo connection are galaxy clustering, gravitational lensing, galaxy group catalogues and, to a lesser extent, satellite kinematics.

Since more massive haloes are more strongly clustered \citep[e.g.,][]{Mo.White.96}, the amplitude of galaxy clustering on large, linear scales is often interpreted as indicative of the average mass of the haloes in which the galaxies in question reside. However, this method is severely impeded by the issue of assembly bias; halo bias  depends not only on halo mass, but also on numerous other halo properties, such as halo formation time, halo spin, and halo concentration \citep[e.g.,][]{Gao.etal.04, Wechsler.etal.06, Villarreal.etal.17, Salcedo.etal.18}. Consequently, and contrary to what is assumed in hundreds of studies, large-scale clustering amplitude can not be used as an unbiased estimator of halo mass \citep[see][for a comprehensive review]{Wechsler.Tinker.18}. Rather, one constrains some combination of the various halo properties that correlate with halo bias. Whereas ignoring assembly bias can consequently result in significant, systematic errors \citep[e.g.,][]{Zentner.etal.14}, properly accounting for it \citep[e.g.,][]{Hearin.etal.16} is not only extremely challenging, but also requires additional constraints on halo masses in order to break the various degeneracies.

One of the most powerful alternative methods to constrain the galaxy-halo connection is galaxy-galaxy lensing. As an application of (weak) gravitational lensing, it is one of the most direct probes of halo mass. In practice, though, the signal-to-noise ratio of the tangential shear distortions around {\it individual} galaxies is typically far too small for a reliable estimate of the galaxy's halo mass, and one generally resorts to stacking the data for several thousands of lens-galaxies. Starting with the pioneering work by \citet{Brainerd.etal.96}, this stacking method has been used in numerous studies, and has resulted in accurate measurements of the average relation between stellar mass (or luminosity) and halo mass \citep[][]{Hoekstra.etal.01, Sheldon.etal.04, Mandelbaum.etal.06, Leauthaud.etal.12, Velander.etal.14}. Galaxy-galaxy lensing on small, non-linear scales has also been used in combination with galaxy clustering in attempts to simultaneously constrain the galaxy-dark matter connection and cosmological parameters \citep[][]{Cacciato.etal.09, Cacciato.etal.13, More.etal.15b,  Leauthaud.etal.17, Wibking.etal.19}. However, none of these studies have allowed for assembly bias, and their results therefore have to be taken with a grain of salt (but see also Lange et al. 2019, in prep.).

One can also constrain the galaxy halo connection using galaxy group finders, which try to group together those galaxies that reside in a common dark matter host halo. The mass of the host halo is typically estimated from the line-of-sight velocity dispersion of its member galaxies \citep[e.g.,][]{Eke.etal.04, Robotham.etal.11, Tempel.etal.14}, or from the total luminosity or stellar mass using halo abundance matching \citep[e.g.,][]{Yang.etal.05, Yang.etal.07}. In addition to providing constraints on the galaxy-dark matter connection \citep[e.g.,][]{Berlind.etal.06, Yang.etal.08, Yang.etal.09, Nurmi.etal.13}, galaxy group catalogues have proven particularly powerful for studying the impact of environment of galaxy demographics \citep[e.g.,][]{Weinmann.etal.06, vdBosch.etal.08a, Wetzel.etal.13, Hou.etal.14, WangE.etal.18, Davies.etal.19}. However, it has also become clear that errors in the group finding algorithm and the halo mass assignment can be appreciable \citep[e.g.,][]{Campbell.etal.15, Calderon.Berlind.19}, and thus that the constraints from group catalogs are best combined with additional, independent constraints such as from clustering and/or lensing \citep[][]{Han.etal.15, Sinha.etal.17}.


Satellite kinematics is yet another method to constrain the galaxy-dark matter connection. It uses the notion that satellite galaxies orbiting within the dark matter haloes of their central galaxies are tracers of the gravitational potential, and can therefore be used to probe the relation between halo mass and central galaxy luminosity (or stellar mass). Since individual galaxies typically only have a few (detectable) satellites (with the exception of massive clusters), this method typically relies on the same stacking approach used in galaxy-galaxy lensing. Although it is the oldest technique used to constrain halo masses, starting with the seminal work of \cite{Zwicky.33}, it has been somewhat under-utilized since the advent of clustering and galaxy-galaxy lensing. This is somewhat surprising, as the actual measurements (redshifts) are much easier to obtain than in the case of galaxy-galaxy lensing (tangential shear distortions). The main reason is that the kinematics of dark matter subhaloes, which host satellite galaxies, are believed to be inconsistent with a steady-state tracer population in a spherical, equilibrium potential \citep[e.g.,][]{Wang.etal.17,Wang.etal.18, Adhikari.etal.18}. Consequently, the general notion is that it must be extremely difficult to extract reliable halo masses. In addition, satellite orbits are likely to be anisotropic \citep[][]{Diemand.Moore.Stadel.04, Cuesta.etal.08, Wojtak.Mamon.13}, further complicating the modelling. And finally, most satellite kinematics studies in the past have been extremely conservative in selecting central-satellite pairs, to the extent that the signal-to-noise ratio of the data did not allow for competitive constraints on the galaxy-halo connection \citep[see][for a detailed, historical overview]{Lange.etal.19a}. 

As we demonstrate in this paper, and have demonstrated before \citep{Lange.etal.19a}, these issues are far less severe than has been suggested, and satellite kinematics can be used as a competitive, reliable probe of the galaxy-halo connection. In particular, although individual haloes may not be spherical, and individual satellite galaxies may not obey the spherical Jeans equation, to a good approximation the {\it ensemble} of satellite galaxies can be treated as a steady-state tracer population of the {\it ensemble} of host haloes. And, as we demonstrate in this paper, modeling such ensembles using the spherical Jeans equation yields unbiased estimates of the galaxy-halo connection, as long as one carefully accounts for scatter (i.e., `mass-mixing', see \S\ref{sec:constraining}), sample selection effects (i.e., interlopers, impurity and incompleteness, see \S\ref{sec:fc}), and, to a lesser extent, orbital anisotropy. In addition, \cite{vdBosch.etal.04} demonstrated that by using  iterative, adaptive selection criteria one can boost the number of central-satellite pairs by an order of magnitude, while simultaneously decreasing the fraction of interlopers (galaxies unassociated with the dark matter halo of the central) and increasing the dynamic range of the galaxy-halo connection probed. This sample selection method was subsequently used by \cite{More.etal.09b, More.etal.11} who were able to obtain tight constraints on the galaxy-halo connection. 

Although \cite{More.etal.11} found red centrals to reside in more massive haloes than blue centrals of the same stellar mass, a result that has subsequently been confirmed using galaxy-galaxy lensing \citep[][]{Velander.etal.14, Mandelbaum.etal.16, Zu.Mandelbaum.16}, their inferred stellar mass-to-halo mass ratios are significantly different (by a factor two to three)  than those inferred from  clustering and/or galaxy-galaxy lensing \citep[e.g.,][]{Dutton.etal.10, Leauthaud.etal.12, Mandelbaum.etal.16, Wechsler.Tinker.18}. In \cite{Lange.etal.19a, Lange.etal.19b} we improved on the analysis of \cite{More.etal.09b, More.etal.11} by correcting for sample incompleteness due to fibre collisions in the Sloan Digital Sky Survey \citep[][hereafter SDSS]{York.etal.00} data, by accounting for covariance in the data, and by using forward modeling to correct the model for small, but significant biases. This alleviates the tension with the lensing results mentioned above, demonstrating that satellite kinematics can yield constraints on the galaxy dark matter connection in good agreement with constraints from galaxy-galaxy lensing and/or clustering. 

Here we continue our goal of maturing satellite kinematics into an accurate and precise probe of the galaxy-halo connection. In particular, we develop a Bayesian hierarchical method to analyse satellite kinematics, called \Basilisk\footnote{{\underline{\bf Ba}}ye{\underline{\bf s}}ian h{\underline{\bf i}}erarchica{\underline{\bf l}} {\underline{\bf i}}nference using {\underline {\bf s}}atellite {\underline{\bf k}}inematics}, which is entirely complementary to the forward-modeling-based method that we recently developed, and applied to SDSS-DR7 data, in \cite{Lange.etal.19a, Lange.etal.19b}. \Basilisk has a number of advantages over the standard method for analysing satellite kinematics. First of all, it requires no arbitrary stacking of the data and can be trivially applied to a flux-limited sample, whereas the methodology used by \cite{More.etal.09b, More.etal.11}  and \cite{Lange.etal.19a, Lange.etal.19b} requires volume limited samples. This drastically increases the quantity and dynamic range of the data. In addition, \Basilisk does not make use of any summary statistic (i.e., the satellite velocity dispersion as function of central luminosity), but rather leaves the data in its raw form. This has the advantage that all data is used optimally, thereby allowing to simultaneously constrain halo mass and velocity anisotropy. In addition, as a by-product of the method, \Basilisk yields estimates for the halo mass of each individual, central galaxy.


In this first paper in a series, we introduce \Basilisk and test its performance using mock data.  In \S\ref{sec:standard} we first discuss the standard method of analysing satellite kinematics, in which we highlight some of its shortcomings. \S\ref{sec:method} presents our new, Bayesian hierarchical framework, and our method for correcting for interlopers and fibre collisions. \S\ref{sec:ingredients} discusses the two main model ingredients; the conditional luminosity function that we use to characterize the galaxy-halo connection (\S\ref{sec:CLF}), and our model for the phase-space distribution of satellite galaxies within their host haloes (\S\ref{sec:phase-space}). \S\ref{sec:validation} presents our three-tiered validation process, in which we test the performance of \Basilisk on a series of mock data sets of increasing complexity and realism. In \S\ref{sec:constrain} we examine \Basilisk's ability to constrain the anisotropy of satellite galaxies, and we discuss how central velocity bias and errors in the inferred radial profile of satellite galaxies impacts the inference regarding the galaxy-halo connection. Finally, \S\ref{sec:conc} summarizes our findings and presents a detailed discussion of pros and cons of \Basilisk. 

Throughout this work, we assume a $\Lambda$CDM cosmology with $\Omega_\rmm = 0.3071$, $\Omega_\rmb = 0.0483$, $n_\rms = 0.9611$, $\sigma_8 = 0.8288$ and $h = H_0 / 100 \mathrm{km/s/Mpc} = 0.6777$, the best-fit results from the cosmic microwave background analysis of \cite{Planck.14}. 


\section{Standard Approach}
\label{sec:standard}

Before we outline our new, Bayesian hierarchical approach to satellite kinematics, we first give an overview of what has become the standard method, which basically consists of three steps: (1) selecting a sample of central and satellite galaxies from a galaxy redshift survey, (2) using this data to compute the velocity dispersion of satellite galaxies, with respect to their centrals, as a function of the luminosity or stellar mass of the central, and (3) using these velocity dispersion measurements to constrain the galaxy-dark matter connection. In what follows we describe each of these three steps in detail.

\subsection{Selecting centrals and satellites}
\label{sec:selection}

The standard method to select centrals and satellites, and the one that we will adhere to as well, is to use a cylindrical isolation criterion to identify centrals, and then to assign fainter galaxies within a similar cylindrical volume as corresponding satellites. Due to interlopers and other impurities, discussed below, not every central (satellite) thus selected is indeed a central (satellite). In what follows, we therefore refer to galaxies that are selected as centrals and satellites as primaries and secondaries, respectively.

To be considered a primary, a galaxy must be brighter than any other galaxy in a cylinder defined by radius $\Rh$ and length $2 \dVh$ centred on it. The radius is defined as the physical separation projected onto the sky and the length is measured by the line-of-sight velocity difference (see equation~[\ref{dvdef}] below). We follow \cite{Lange.etal.19a} and apply this criterion in a rank-ordered fashion, starting with the brightest galaxy. Any galaxy located inside the cylinder of a brighter galaxy is removed from the list of potential primaries. All galaxies fainter than the primary and located inside a cylinder defined by $\Rs$ and $\dVs$ centred on the primary
are identified as secondaries. 

The four free parameters that control the selection of primaries and secondaries, $\Rh$, $\dVh$, $\Rs$, and $\dVs$, determine both the completeness and purity of the sample. Increasing the cylinder used to select primaries, i.e., increasing $\Rh$ and/or $\dVh$, boosts the purity among primaries (i.e., it reduces the number of satellites erroneously identified as centrals), but at the cost of a reduced completeness. Similarly, suppressing the number of interlopers, defined as secondaries that are not satellite galaxies within the same halo as the corresponding primary, requires a small secondary-selection cylinder (i.e., small $\Rs$ and/or $\dVs$), which also reduces completeness. A reduced completeness not only complicates the modeling, but also results in data of lower signal-to-noise. As first pointed out in \citet{vdBosch.etal.04}, the fact that brighter primaries typically reside in larger haloes, implies that it is advantageous to scale the cylinder sizes with the luminosity of the primary. We do so using the exact implementation of \cite{Lange.etal.19b}, who adopt $\Rh = 0.5 \,\sigma_{200} \mpch$, $\Rs = 0.15 \, \sigma_{200} \mpch$,  $\dVh = 1000 \,\sigma_{200} \kms$, and $\dVs = 4000 \kms$. Here $\sigma_{200}$ is an estimate for the satellite velocity dispersion in units of $200 \kms$, which scales with the luminosity of the primary as
\begin{equation}\label{sig200}
\log \sigma_{200} = -0.07 + 0.38 \log L_{10} + 0.29 \log^2 L_{10}\,,
\end{equation}
where $L_{10} = L / (10^{10} \Lsunh)$. These criteria were optimized for the SDSS, using an iterative approach, as detailed in \cite{More.etal.09b}. The values of $\Rh$ and $\Rs$ correspond to roughly $1.25$ and $0.375$ times the virial radius, respectively, while the value for $\dVs$ is large enough to include the vast majority of all satellites, even in massive clusters. Note that we do not scale this parameter with $\sigma_{200}$; although this implies an increasing fraction of interlopers with decreasing central luminosity, these interlopers are easily identified as such. In principle one could follow \cite{vdBosch.etal.04} and \cite{More.etal.09b} and apply these selection criteria iteratively, each time updating the $\sigma_{200}(L_\rmc)$ relation based on the inference from the satellite kinematics data selected using the previous $\sigma_{200}(L_\rmc)$. However, as detailed in \S\ref{sec:dataformat}, there is no need for this as \Basilisk's inference is extremely insensitive to moderate changes in equation~(\ref{sig200}). Furthermore, tests with detailed mock data sets have shown that using equation~(\ref{sig200}) yields samples that allow for an accurate recovery of the underlying galaxy-halo connection \citep[see][and \S\ref{sec:validation} below]{Lange.etal.19a}. In \S\ref{sec:tier2} we use mock redshift surveys to assess the completeness, the purity, and the interloper contamination of the above selection criteria when applied to a flux-limited SDSS-like survey. 

\subsection{Characterizing satellite kinematics}
\label{sec:characterizing}

Using a sample of primaries and secondaries, the next step is to quantify the kinematics of these secondaries (assumed to be satellite galaxies) within the host haloes of their associated primaries (assumed to be centrals). Since the typical number of satellite galaxies per central is small, except in nearby clusters, this requires stacking whereby one co-adds all central-satellite pairs for centrals in a given range of luminosity\footnote{One may also stack on stellar mass, or any other (combination) of properties of the central galaxies. For brevity we focus on luminosity throughout.}, $[L_{\rmc,1},L_{\rmc,2}]$. The satellite kinematics are then specified by the line-of-sight velocity distribution (LOSVD), $P(\dV | L_\rmc)$, where $L_\rmc$ is a characteristic luminosity for the luminosity bin in question, and
\begin{equation}\label{dvdef}
\dV = c \, \frac{(z_\rms - z_\rmc)}{1 + z_\rmc}\,,
\end{equation}
with $z_\rms$ and $z_\rmc$ the observed redshifts of the satellite and central, and $c$ the speed of light. The summary statistic that is most often used in the study of satellite kinematics is the satellite velocity dispersion, $\sigma_{\rm sat}(L_\rmc)$,  which characterizes the second moment of this LOSVD. In order to extract $\sigma_{\rm sat}(L_\rmc)$ from  $P(\dv | L_\rmc)$ one has to correct for interlopers, which can be done in a variety of ways, each with its own pros and cons \citep[e.g.,][]{Wojtak.etal.07, Lange.etal.19a}.

There are a number of important shortcomings with this methodology. First, it requires stacking data in some arbitrary luminosity bins in order to measure the corresponding satellite velocity dispersion, $\sigma_{\rm sat}$, as function of luminosity. Larger bins implies fewer independent measurements of $\sigma_{\rm sat}$, but at higher signal-to-noise. However, since luminosity correlates with halo mass, it also implies more `mass-mixing', i.e., combining kinematics from satellites orbiting in haloes of different masses. As discussed in \S\ref{sec:constraining}, properly addressing the impact of mass-mixing is extremely important, non-trivial, yet often ignored. Secondly, because of interlopers it is virtually impossible to extract an unbiased estimate of the velocity dispersion from the LOSVD \citep[see][]{Becker.etal.07, Lange.etal.19a}. Thirdly, by relying solely on the velocity dispersion as a summary statistic, one ignores a wealth of additional information encoded in the detailed shape of the LOSVD and in the correlation between $\dV$ and the projected separation, $\Rp$, of individual central-satellite pairs. This additional information allows one to constrain the density profile of the host halo \citep[e.g.,][]{Prada.etal.03}, and the orbital anisotropy of satellite galaxies \citep[e.g.,][]{Lokas.etal.06, Wojtak.Mamon.13}. The $\dV - \Rp$ correlation also facilitates a more accurate treatment of interlopers \citep[e.g.,][]{Wojtak.etal.07, Lange.etal.19a}.

The method advocated here, and outlined in \S\ref{sec:method}, sidesteps all these shortcomings. It utilizes the full $(\dV,\Rp)$-data, without any binning and without the use of a summary statistic.

\subsection{Constraining the Galaxy-Dark Matter Connection}
\label{sec:constraining}

The final step in using satellite kinematics to constrain the galaxy-halo connection, is to translate the data, $\sigma_{\rm sat}(L_\rmc)$, into corresponding constraints on $P(M | L_\rmc)$, which characterizes the probability that a central of luminosity $L_\rmc$ resides in a halo of mass $M$. Ideally, this is done using forward-modeling based on numerical simulations in which halos are populated with mock galaxies according to a halo occupation model. This has the advantage that model (mock) and data can be treated in the same way, which makes the analysis less susceptible to biases arising from interlopers, sample incompleteness, and sample impurity. However, as discussed in \cite{Lange.etal.19a}, a full-fledged forward modeling approach is computationally unfeasible at the present, and all previous studies have therefore relied on simple halo mass estimators based on the virial theorem \citep[e.g.,][]{Bahcall.Tremaine.81, Zaritsky.White.94, McKay.etal.02, Brainerd.Specian.03}, or on analytical models that use the Jeans equations to predict the satellite kinematics as a function of halo mass \citep[e.g][]{vdBosch.etal.04, Conroy.etal.07, More.etal.09b, More.etal.11, Wojtak.Mamon.13}. In a recent study, \cite{Lange.etal.19b} combined an analytical model based on the Jeans equations with forward modeling, by using the latter to iteratively calibrate and correct the analytical model for small, systematic biases. 

An important issue in trying to infer $P(M | L_\rmc)$ from satellite kinematics is `mass-mixing'. There are good reasons to expect a fair amount of scatter in the galaxy-halo connection, such that central galaxies of a given luminosity occupy haloes of varying masses. Put differently, $P(M | L_\rmc)$ is not a Dirac delta function. Hence, when stacking the satellite kinematics from a number of centrals of similar luminosity, one is combining the kinematics corresponding to a range in halo masses. The extent of mass-mixing is further compounded by the use of luminosity bins of non-zero width. Surprisingly, mass-mixing has been ignored in many previous studies, including \cite{McKay.etal.02}, \cite{Brainerd.Specian.03}, \cite{Prada.etal.03}, \cite{Norberg.etal.08}, and \cite{Wojtak.Mamon.13}. As first pointed out in \cite{vdBosch.etal.04}, and further corroborated in \cite{More.etal.09a}, this can result in a very significant, systematic bias in the inferred halo masses. The origin of this bias is easy to understand: typically, more massive haloes contain more satellite galaxies. Hence, when stacking central-satellite pairs residing in different haloes, the more massive ones receive a larger `weight' in that they contribute more data points. This `satellite-weighting' (i.e., giving each satellite equal weight) results in a systematic overestimate of the average halo mass (see also Appendix~\ref{App:noNsat}). This problem can be avoided, though, by weighting each central-satellite pair by the inverse of the number of satellites around that central. This `host-weighting' gives equal weight to each central, such that the measured velocity dispersion more fairly represents the average halo mass. In fact,  as elucidated in \citet{More.etal.09a}, by simultaneously modeling the satellite-weighted and host-weighted velocity dispersions, one can actually constrain the amount of scatter in halo mass at given primary luminosity.
This idea has been used by \citet{More.etal.09b}, \citet{More.etal.11} and \cite{Lange.etal.19b}, all of whom were able to put tight constraints on the scatter in the galaxy-halo connection.

Finally, using dynamics to infer masses is hampered by the well-known mass-anisotropy degeneracy \citep[][]{Binney.Tremaine.08}. With the exception of \cite{Wojtak.Mamon.13} all previous studies of satellite kinematics have simply assumed isotropic orbits for the population of satellite galaxies. Although a clear oversimplification, and one that is likely to be systematically wrong \citep[e.g.,][]{Diemand.Moore.Stadel.04}, this does not have an important impact on the standard approach outlined here. The reason is simply that the satellite velocity dispersions are averaged over large parts of the host halo. The mass-anisotropy degeneracy predominantly plagues attempt to infer a mass {\it profile} from radially dependent kinematic data. Kinematic tracers that have the same radial distribution but different orbital anisotropies manifest different radial profiles of (projected) velocity dispersion. However, their total velocity dispersion, averaged over the entire system, has little to no dependence on the anisotropy. This is the reason why orbital anisotropy does not enter the virial theorem. As stated above, the cylindrical selection criteria used here to select secondaries only reach out to $\sim 0.375$ times the virial radius, and the resulting satellite velocity dispersion is therefore not averaged over the {\it entire} system. However, as explicitly shown in \cite{vdBosch.etal.04}, even in this case anisotropy only affects the kinematics at the level of a few percent. Although this is advantageous if one is only interested in constraining halo mass, being able to constrain the orbital anisotropy opens up new avenues to test models for galaxy formation and evolution. The new method outlined below allows one to {\it simultaneously} constrain halo mass and orbital anisotropy.

\section{Methodology}
\label{sec:method}

The Bayesian hierarchical method for analysing satellite kinematics presented here differs substantially from the `standard' method outlined above. It is developed with the following goals in mind: (i)  leave the data in its raw form as much as possible, particularly avoiding the use of summary statistics, binning, and/or stacking; (ii) include as much data as possible, by assuring that the method can be applied to flux-limited samples, rather than only to volume limited samples; and (iii) use a sufficiently flexible model that allows 
for proper treatment of mass-mixing and orbital anisotropy.  

After selecting centrals and satellites (or rather, primaries and secondaries) using the same cylindrical isolation criteria as described in \S\ref{sec:selection}, we define a likelihood for the data given the model, and use an affine invariant ensemble sampler, within a Bayesian hierarchical framework, to constrain the posterior of the model parameters. As we detail below, this method achieves all three goals listed above.
\begin{figure*}
\centering
\includegraphics[width=0.8\textwidth]{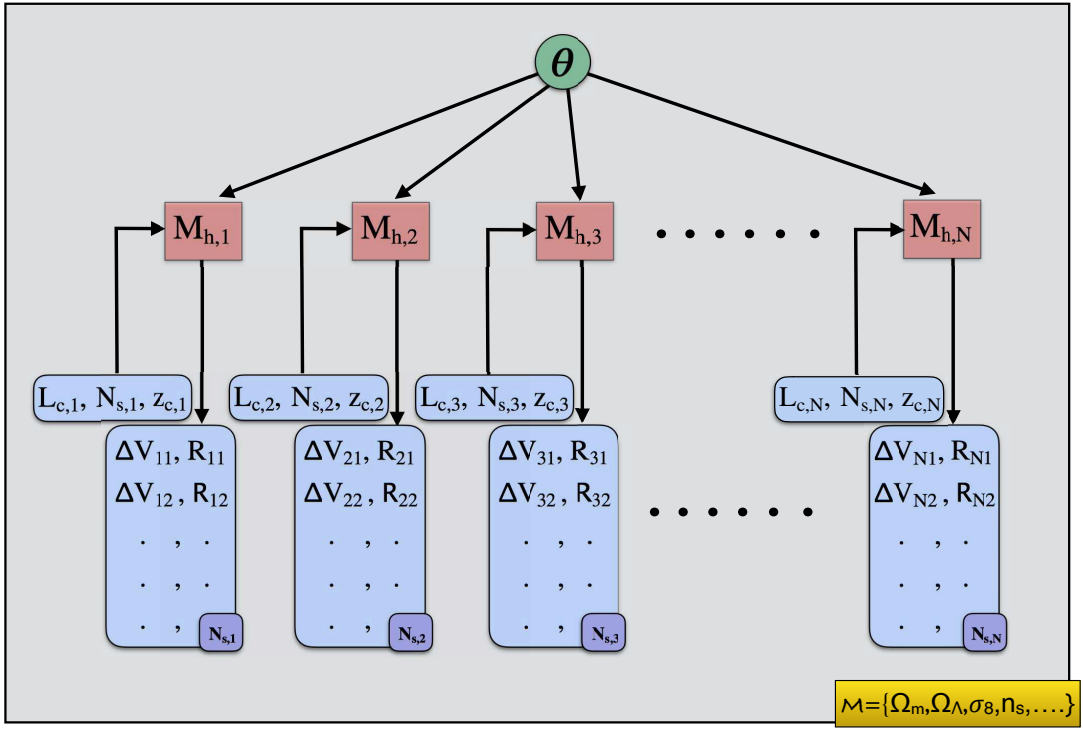}
\caption{Illustration of the hierarchical nature of the problem. Middle row shows the latent variables (halo masses), lower row shows the data, and the upper row depicts the population model. Note how certain aspects of the data ($\Lc$, $\zc$ and $N_\rms$) for each central are used to inform the prior on the latent variable, which is marginalized over when computing the likelihood for the satellite phase-space data $\{\dVij, \Rij|\Lci, \zci, \Nsi\}$. Note that the entire inference is based on a model $\calM$ that depends on cosmology and certain assumptions made (potential of dark matter haloes, radial profile of satellite galaxies, etc).}
\label{fig:diagram}
\end{figure*}

\subsection{Data format}
\label{sec:dataformat}

Running the isolation criteria described in \S\ref{sec:selection} yields a number of central-satellite pairs characterized by the following set of parameters: $(L_\rmc, L_\rms, z_\rmc, \dV, \Rp)$. Here $L_\rmc$ and $L_\rms$ are the luminosities of the central and satellite, respectively, $z_\rmc$ is the redshift of the central, $\dV$ is the line-of-sight velocity difference between central and satellite (equation~[\ref{dvdef}]), and $\Rp$ is the projected separation, which is related to the angular separation, $\vartheta$, according to $\Rp = d_\rmA(z_\rmc) \, \vartheta$, with $d_\rmA$ the angular diameter distance. Finally, associated with each secondary is a weight, $w$, that accounts for spectroscopic incompleteness in the survey, as described in \S\ref{sec:fc}. 

For the purpose of our inference problem, we write our data vector $\bD$ as the union of the $N_\rmc$ data vectors
\begin{equation}\label{dataprim}
\bD_i = \left( \{\dVij, \Rij | j=1,...,\Nsi \} | \Lci, \zci, \Nsi\right)\,.
\end{equation}
Here $\Nsi$ is the number of secondaries associated with primary $i$, and we have made it explicit that we only treat $\Lci$, $\zci$, and $\Nsi$ as {\it conditionals} for the data $\{\dVij, \Rij | j=1,...,\Nsi \}$. In other words, we consider $\Lci$, $\zci$ and $\Nsi$ as `given' and shall not use the distributions of these quantities as constraints on our likelihood. Rather, \Basilisk uses the luminosity function of all galaxies as an additional constraint (see \S\ref{sec:additional}).

The main reason for doing so is to make the method less sensitive to the detailed selection of centrals, which is difficult to model in detail. In particular, this approach makes \Basilisk insensitive to details regarding the $\sigma_{200}(L)$ relation (equation~[\ref{sig200}]) used to define the selection cylinders\footnote{We have explicitly verified this by running \Basilisk on different (mock) samples, extracted from mock redshift surveys such as the ones described in \S\ref{sec:validation} using different values for the coefficients in the $\sigma_{200}(L)$ relation. The resulting posterior distributions are always in excellent, mutual agreement.}. Finally, we emphasize that we ignore the luminosities of secondaries in our inference, which instead relies entirely on their projected phase-space coordinates $\dV$ and $\Rp$.

\subsection{The inference problem}
\label{sec:inference}

Our goal is to use $\bD$ to constrain the galaxy-dark matter connection, which we characterize using the conditional luminosity function (CLF), $\Phi(L|M)$, and a model for the phase-space distribution of secondaries (satellites plus interlopers). The CLF is split in a central component, $\Phi_\rmc(L|M)$, and a satellite component, $\Phi_\rms(L|M)$, as detailed in \S\ref{sec:CLF} below. In particular, we seek to constrain the posterior distribution, $P(\btheta | \bD)$, where $\btheta$ is the vector that describes our model parameters, $\theta_i$ ($i=1,...,N_\rmp)$. From Bayes theorem
\begin{equation}
P(\btheta | \bD) \propto \calL_{\rm SK}(\bD | \btheta) \, P(\btheta)\,,
\end{equation}
where $P(\btheta)$ is the prior probability distribution on the model parameters, and $\calL_{\rm SK}(\bD|\btheta)$ is the likelihood of the satellite kinematics data given the model. Throughout we mainly use uniform, non-informative priors for our model parameters. If we make the reasonable assumption that the data for different primaries is independent, we have that 
\begin{equation}\label{LikelihoodSK}
\begin{split}
\calL_{\rm SK}(\bD|{\btheta}) & = \prod\limits_{i=1}^{N_\rmc} \, \calL(\bD_i|\btheta) \\
& = \prod\limits_{i=1}^{N_\rmc} \, \prod\limits_{j=1}^{\Nsi} P(\dVij, \Rij | \Lci, \zci, \Nsi, \btheta)\,.
\end{split}
\end{equation}

Computing the probability $P(\dV, \Rp | L_\rmc, z_\rmc, N_\rms)$, that a satellite galaxy in a halo at redshift $z_\rmc$, with a central galaxy of luminosity, $L_\rmc$, and with a total of $N_\rms$ detected secondaries has projected phase-space parameters $(\dV, \Rp)$, requires knowledge of the gravitational potential, $\Psi(\bx)$, in which the satellite is moving, as well as some knowledge regarding the phase-space distribution function, $f(\bx,\bv,t)$, of the population of satellites and interlopers. Throughout this work we make the assumption that satellite galaxies are a virialized, steady-state tracer of the gravitational potential well, which implies that $f(\bx,\bv,t) = f(\bx,\bv)$. In addition, we assume that dark matter haloes are spherical NFW profiles with a concentration-mass relation with zero scatter. This implies that $\Psi(\bx)$ is completely specified by a single parameter, which we take to be the halo virial mass\footnote{Throughout this paper, we define virial quantities according to the virial overdensities given by the fitting formula of \cite{Bryan.Norman.98}.}, $\Mh$. Our treatment of the distribution function is described in \S\ref{sec:phase_sat}.

Since we assume that the satellite kinematics are governed solely by host halo mass, the likelihood for data $\bD_i$, given model $\btheta$, can be factored as
\begin{equation}\label{Mmarg}
\begin{split}
\calL(\bD_i|\btheta) & = \int \rmd \Mh \, P(\Mh | \Lci, \zci, \Nsi) \, \times \\
 & \;\;\;\;\;\;\;\;\;\;\;\;\;\;\;\;\;\;\;\;\;\;\;\prod\limits_{j=1}^{\Nsi} P(\dVij, \Rij|\Mh, \Lci, \zci)\,, 
\end{split}
\end{equation}
where, for the sake of brevity, we no longer explicitly write-out that $P(M | \Lci, \zci, \Nsi)$ and $P(\dVij, \Rij | \Mh, \Lci, \zci)$ depend on $\btheta$. This equation describes a marginalization over halo mass, the prior for which is informed by $L_\rmc$, $z_\rmc$, and $N_\rms$ according to the model $\btheta$. In addition, by putting the product operator inside the mass integral, as opposed to outside, we have made it explicit that all secondaries are assumed to belong to the same halo. Note also that the likelihood for $\dV$ and $\Rp$ given a halo mass and redshift is conditional on the luminosity of the central, which arises from the fact that the isolation criteria used to select secondaries depend on the luminosity of the primary (see \S\ref{sec:selection}). As is evident from equation~(\ref{Mmarg}), and illustrated in Figure~\ref{fig:diagram}, the halo masses for the individual primaries serve as latent variables, accentuating the hierarchical nature of our inference procedure.

Using Bayes theorem, we have that
\begin{equation}\label{ProbMass}
P(M|L,z,N_\rms) = \frac{P(N_\rms|M,L,z) \, P(M,L,z)}{\int \rmd M \, P(N_\rms|M,L,z) \, P(M,L,z)}\,.
\end{equation}
which allows us to write the log-likelihood for the satellite kinematics data as
\begin{equation}\label{lnLD}
\ln\calL_{\rm SK}(\bD|\btheta) = \sum\limits_{i=1}^{N_\rmc} \left(\ln G_i - \ln F_i\right)\,,
\end{equation}
with
\begin{equation}\label{Gi}
G_i = \sum_k \wk \, F_{ik} \, \exp[Q_{ik}]\,, \,\,\,\,\,\,\,\,\,\,\,\,\,\,\,\,
F_i = \sum_k \wk \, F_{ik} \,,
\end{equation}
and
\begin{equation}\label{Fi}
F_{ik} = P(\Nsi | \Mk, \Lci, \zci) \, P(\Mk, \Lci, \zci)\,.
\end{equation}
Here $\Mk$ and $\wk$ are the abscissas and weights\footnote{The quadrature weights, $\wk$, which carry one index, should not be confused with the spectroscopic weights, $w_{ij}$, which carry two indices.}  of the Gaussian quadrature used to evaluate the mass-integral, and 
\begin{equation}
Q_{ik} = \sum\limits_{j=1}^{\Nsi} \ln P(\dVij, \Rij|\Mk, \Lci, \zci)\,.
\end{equation}
As detailed in \S\ref{sec:numerics} and Appendix~A, Gaussian quadrature has the advantage that the integrands are always evaluated at the same $\Mk$, which allows for many quantities to be pre-computed, thereby greatly speeding up the Bayesian inference. Throughout we adopt a total of $N_k=21$ quadrature points, which is sufficient to achieve accurate results.

What remains is to specify the probabilities $P(M,L,z)$, $P(N_\rms|M,z,L)$, and $P(\dV, \Rp|M,L,z)$, which we address in the following subsections.

\subsubsection{The probability $P(M,L,z)$.}
\label{sec:PLcen}

Within the CLF formalism that we use to model the halo occupation statistics (see \S\ref{sec:CLF}), the probability that a halo of mass $M$ at redshift $z$ hosts a central of luminosity $L$ is given by $\Phi_\rmc(L|M,z)$. To take account of the fact that not every central is selected as a primary, we have that
\begin{equation}\label{Pmlz}
\begin{split}
P(M,L,z) & = P(L|M,z) \, P(M,z) \\
         & = \calC(M,L,z) \, \Phi_\rmc(L|M,z) \, n(M,z)\,.
\end{split}
\end{equation}
Here $n(M,z)$ is the halo mass function at redshift $z$, and $\calC(M,L,z) = \calC(M|L,z) \, \calC(L,z)$ is a completeness function that expresses the fraction of central galaxies of luminosity $L$ residing in haloes of mass $M$ at redshift $z$ that are selected as primaries by our cylindrical isolation criteria. Since $P(M,L,z)$ appears in both the numerator and the denominator of equation~(\ref{ProbMass}), the factor $\calC(L,z)$ drops out, and the expression for $P(M|L,z,N_\rms)$ only depends on $\calC(M|L,z)$. In \S\ref{sec:tier2} we use detailed mock data to demonstrate that $\calC(M|L,z)$ is virtually independent of halo mass. Hence, we may simply model $P(M,L,z)$ as the product of $\Phi_\rmc(L|M)$ and $n(M,z)$, without having to make any corrections for incompleteness (i.e., we can effectively set $C(M|L,z)=1$). Throughout we compute the halo mass function using the method of \citet{Tinker.etal.08} for our cosmology and halo mass definition, while we assume the CLF to be independent of redshift, at least over the redshift range considered in this study ($0.02 \leq z \leq 0.15$). The functional form that we use to describe $\Phi_\rmc(L|M)$ is presented in \S\ref{sec:CLFcen}.

\subsubsection{The probability $P(N_\rms|M,L,z)$}
\label{sec:PNsat}

The number of secondaries, $N_\rms$, associated with a particular primary consists of both satellites (galaxies that belong to the same dark matter host halo as the primary), and interlopers (those that do not). We now derive the probability $P(N_\rms|M,L,z)$, that a primary of luminosity $L$, residing in a host halo of mass $M$ at redshift $z$, has a number of interlopers, $N_{\rm int}$, and satellites, $N_{\rm sat}$, such that $N_{\rm int} + N_{\rm sat} = N_\rms$.

If we assume that the number of interlopers and the number of satellite galaxies are independent, then
\begin{equation}
P(N_{\rm int} + N_{\rm sat}=N_\rms) = \sum\limits_{n=0}^{N_{\rms}} \,
P(N_{\rm int} = n) \, P(N_{\rm sat} = N_\rms-n)\,.
\end{equation}
If we furthermore assume that both interlopers and satellites obey Poisson statistics, we have that
\begin{equation}
\begin{split}
P(N_\rms|M,L,z) & = \sum\limits_{n=0}^{N_{\rms}} \, 
\frac{\lambda^n_{\rm int} \, \rme^{-\lambda_{\rm int}}}{n!}
\frac{\lambda^{N_\rms-n}_{\rm sat} \, \rme^{-\lambda_{\rm sat}}}{(N_\rms-n)!} \\
& = \frac{\rme^{-\lambda_{\rm tot}}}{N_\rms!}\,
\sum\limits_{n=0}^{N_{\rms}} \, \binom{N_\rms}{n} \, \lambda^n_{\rm int}
\, \lambda^{N_\rms-n}_{\rm sat} = \frac{\lambda^{N_{\rms}}_{\rm tot} \, \rme^{-\lambda_{\rm tot}}}{N_\rms!}\,,
\end{split}
\end{equation}
where $\lambda_{\rm int} = \lambda_{\rm int}(L,z)$ and $\lambda_{\rm sat}(M,L,z)$ are the expectation values for the numbers of interlopers and satellites, respectively, $\lambda_{\rm tot} \equiv \lambda_{\rm int} + \lambda_{\rm sat}$, and we have used the binomial identity in the last step. Hence, we obtain the well known result that the sum of two independent Poisson distributed random variables follows itself a Poisson distribution with a mean that is simply the sum of the means of its components. 

What remains is to specify $\lambda_{\rm int}$ and $\lambda_{\rm sat}$.  The expectation value for the number of satellites brighter than the (redshift dependent) magnitude limit of the survey, in a halo of mass $M$ at redshift $\zc$, that fall within the aperture used to select secondaries around a primary of luminosity $\Lc$, is given by
\begin{equation}\label{lambdasat}
\lambda_{\rm sat} = \lambda_{\rm sat}(M,\Lc,\zc) = f_{\rm ap}(M,\Lc,\zc) \, \int\limits_{L_{\rm min}(\zc)}^{\infty} \Phi_\rms(L|M) \, \rmd L\,.
\end{equation}
Here $\Phi_\rms(L|M)$ is the satellite component of the CLF (see \S\ref{sec:CLFsat}) and $f_{\rm ap}$ is the aperture fraction, defined as the probability for true halo members (satellite galaxies) to fall within the secondary selection cylinder specified by $\Rs$ and $\dVs$ (see \S\ref{sec:selection}). Given that $\dVs$ is much larger than the extent of the halo in redshift space, we have that
\begin{equation}\label{fap}
\begin{split}
f_{\rm ap}(M&, \Lc, \zc) = \\
& 4 \pi \int\limits_0^\infty \bar{n}_{\rm sat}(r|M,\zc) 
\, \Big[\zeta(r,\Rmax) - \zeta(r,\Rmin)\Big] \, r^2 \, \rmd r\,.
\end{split}
\end{equation}
Here $R_{\rm max}\equiv \Rs(\Lc)$ and $R_{\rm min} \equiv R_{\rm cut}(\zc)$ is a cut-off radius used to avoid problems with fibre-collisions, as discussed in \S\ref{sec:fc} below. The function $\bar{n}_{\rm sat}(r|M,z)$ is the average radial profile of satellites around haloes of mass $M$ at redshift $z$, normalized such that
\begin{equation}
4 \pi \int\limits_0^\infty \bar{n}_{\rm sat}(r|M,z) \, r^2 \, \rmd r = 1\,,
\end{equation}
and
\begin{equation}\label{zetafunc}
\zeta(r, R) = \begin{cases}
1 &\quad\text{if } r \leq R \\
1 - \sqrt{1 - R^2 / r^2} &\quad\text{otherwise.} \\ 
\end{cases}
\end{equation}
Note that this neglects the (small) possibility that some primaries are satellites (i.e., what we refer to as impurity). As we demonstrate in \S\ref{sec:tier2}, this impurity is small and does not have a significant impact on any of our results \citep[see also][]{Lange.etal.19a}. The full expression for $\lambda_{\rm sat}(M,\Lc,\zc)$ for our assumed functional forms for the CLF and the radial distribution of satellite galaxies is given in Appendix~\ref{App:comp}. 

For the interlopers, we model the expectation value as the product of an effective `bias', $b_{\rm eff}$, and the expectation value for the number of galaxies with $L_{\rm min}(z_\rmc) < L < L_\rmc$ in a random, cylindrical volume, $V_{\rm cyl}(L_\rmc, z_\rmc)$, equal to that used to select the secondaries around the central of luminosity $L_\rmc$ at redshift $z_\rmc$:
\begin{equation}\label{interlopermodel}
\lambda_{\rm int} = \lambda_{\rm int}(\Lc,\zc) = b_{\rm eff}(\Lc,\zc) \, V_{\rm cyl}(\Lc, \zc) \, \bar{n}_{\rm gal}(\Lc,\zc)\,.
\end{equation}
Here
\begin{equation}\label{phitot}
\bar{n}_{\rm gal}(\Lc, \zc) = \int_{L_{\rm min}(\zc)}^{\Lc} \rmd L \int_0^{\infty} \Phi(L|M) \, n(M,z) \, \rmd M\,,
\end{equation}
is the average number density of galaxies at redshift $\zc$ that are fainter than $\Lc$ but brighter than the survey limit $L_{\rm min}(\zc)$. Since the cylinder used to select secondaries is specified by an opening angle $\theta_{\rm ap} = \Rs(\Lc) / d_\rmA(\zc)$, and accounting for the cut-off radius $\Rmin = R_{\rm cut}(\zc) \equiv \theta_{\rm cut} \, d_\rmA(\zc)$, we have that
\begin{equation}\label{Vcyl}
V_{\rm cyl}(\Lc,\zc) = \omega_{\rm cyl} \, \int_{z_{-}}^{z_{+}} \frac{\rmd^2 V}{\rmd\omega \, \rmd z} \, \rmd z \simeq \omega_{\rm cyl} \, \frac{2 \dVs}{c} \, \frac{\rmd^2 V}{\rmd \omega \, \rmd z}(z_\rmc)\,.
\end{equation}
Here $z_{\pm} = z_\rmc \pm \dVs/c$, the derivative $\rmd^2 V/\rmd \omega \rmd z$ is the comoving volume element at redshift $z$ corresponding to a solid angle $\rmd\omega$ and a depth $\rmd z$, and $\omega_{\rm cyl} = 2 \pi (\cos\theta_\rmc - \cos\theta_{\rm ap})$ is the solid angle of the cylinder centered on the primary. Since $\Rs(\Lc) \ll d_\rmA(z_\rmc)$, we have that both $\theta_\rmc$ and $\theta_{\rm ap}$ are small, which implies that to good approximation
\begin{equation}\label{Vcylred}
V_{\rm cyl}(\Lc,\zc) = \pi \, \left[\Rmax^2 - \Rmin^2\right] \, \frac{2 \dVs}{H(\zc)} \, (1+\zc)^2\,,
\end{equation}
with $H(z)$ the Hubble parameter. What remains is to model the effective bias, describing how the number density of interlopers around primaries is enhanced or suppressed relative to that in a random volume. We simply model this effective bias as having independent power-law dependences on the luminosity and redshift of the central, i.e.,
\begin{equation}\label{beff}
b_{\rm eff}(\Lc,\zc) = \eta_0 \, \left(\frac{\Lc}{10^{10.5}\Lsunh}\right)^{\eta_1} \, \left( 1+z_\rmc \right)^{\eta_2}
\end{equation}
with $\eta_0$, $\eta_1$, and $\eta_2$ three free parameters that fully specify our interloper-model, and whose values are to be determined from the data.

\subsubsection{The probability $P(\dV,\Rp|M,L,z)$}
\label{sec:PdvR}

Since interlopers and satellites have distinct phase-space distributions, we write
\begin{equation}\label{PdVall}
\begin{split}
P(\dV, \Rp|M,L,z) = & f_{\rm int}\, P_{\rm int}(\dV, \Rp|L,z) \, + \\
& [1 - f_{\rm int}] \, P_{\rm sat}(\dV, \Rp|M,L,z) \,
\end{split}
\end{equation}
with 
\begin{equation}\label{fint}
f_{\rm int} = f_{\rm int}(M,L,z) = \frac{\lambda_{\rm int}(L,z)}{\lambda_{\rm tot}(M,L,z)}\,,
\end{equation}
the interloper fraction.  We assume that interlopers have a constant projected number density and a uniform distribution in line-of-sight velocity\footnote{Although a clear oversimplification (see \S\ref{sec:tier2}), this does not significantly impact our inference regarding the galaxy-dark matter connection}. This implies that 
\begin{equation}\label{Pint}
P_{\rm int}(\dV, \Rp |L,z) = \frac{\Rp}{\dVs \, [\Rmax^2 - \Rmin^2]}\,,
\end{equation}
which is properly normalized, i.e.,
\begin{equation}
\int\limits_{-\dVs}^{+\dVs} \rmd \dV \int\limits_{\Rmin}^{\Rmax} \rmd \Rp \, P_{\rm int}(\dV, \Rp|L,z) = 1\,.
\end{equation}

Finally, the probability $P_{\rm sat}(\dV, \Rp |M,L,z)$ is determined by our detailed model for the phase-space distribution of satellite galaxies, which is discussed in detail in \S\ref{sec:phase_sat} below.

\subsection{Correction for Fibre Collisions}
\label{sec:fc}

As demonstrated in \citet{Lange.etal.19a}, it is important to include in the analysis of satellite kinematics a correction for fibre-collision induced incompleteness in the spectroscopic data used. In what follows, we use the SDSS Main Galaxy Sample as a characteristic example. In the SDSS, spectroscopic fibres cannot be placed simultaneously on a single plate for objects separated by less than $55''$ \citep{Blanton.etal.03a}. 
Although some galaxies are observed with multiple plates, yielding spectroscopic redshifts even for close pairs, roughly 65\% of galaxies with a neighbour within $55''$ lack redshifts due to this fibre collision effect. We use this fact to mimic fibre collisions in our mock data sets, as discussed in \S\ref{sec:tier2}.

In order to correct the data for the presence of fibre collisions, we follow \citet{Lange.etal.19a} and start by assigning each fibre-collided galaxy the redshift of its nearest neighbour \citep[see][]{Blanton.etal.05, Zehavi.etal.05}. Although we use these during the identification of primaries\footnote{As shown in \cite{Lange.etal.19a}, ignoring fibre-collided galaxies during the selection of primaries results in a much larger sample impurity.}, during the subsequent analysis only primary-secondary pairs with spectroscopic redshifts for both are used. In addition, each galaxy is assigned a spectroscopic weight, $w_{\rm spec}$, that is computed as follows. For each galaxy we first count the number of galaxies, $n$, brighter than $m_r = 17.6$ within a projected separation less than $55''$. Next, for all galaxies in the survey with $n$ neighbours, we compute the fraction, $f_{\rm spec}$, of those neighbours that have been successfully assigned a redshift. Finally, all galaxies with $n$ neighbours are then assigned a spectroscopic weight equal to $w_{\rm spec} = 1 / f_{\rm spec}$. 

In \citet{Lange.etal.19a} we used these weights to compute fibre-collision-corrected  satellite velocity dispersions, $\sigma_{\rm sat}(L_\rmc)$, and projected surface densities, $\Sigma_{\rm sat}(R_\rmp)$. This works extremely well, except on scales below the fibre-collision scale of $\vartheta_{\rm fc} = 55''$. Therefore, \citet{Lange.etal.19a} decided to exclude all secondaries with a projected separation from their primary less than $60 \kpch$, which is roughly the fibre collision scale at the maximum redshift of their volume limited sample. Using the Tier-2 mocks described below (\S\ref{sec:tier2}), we have tested a number of different fibre-collision-correction schemes for \Basilisk. We find that the following scheme works extremely well; rather than up-weighting the number of secondaries in the data, we down-weight the expectation value, $\lambda_{\rm tot} = \lambda_{\rm sat} + \lambda_{\rm int}$, for the number of secondaries in the model. In particular, we multiply $\lambda_{\rm tot}$ (see \S\ref{sec:PNsat}) with the correction factor
\begin{equation}\label{fcorr}
f_{{\rm corr},i} = \frac{N_{\rms,i}}{\sum_{j=1}^{N_{\rms,i}} w_{ij}}\,,
\end{equation}
where $w_{ij}$ is the spectroscopic weight, $w_{\rm spec}$, for secondary $j$ associated with primary $i$. Since $w_{ij} \geq 1$ we have that $f_{{\rm corr},i} < 1$, thereby correcting the expected number of secondaries for the fact that some are lost as a consequence of fibre collisions. In addition, since correction for fibre collisions is extremely difficult on scales below the fibre-collision scale, we remove all secondaries with $R_\rmp < R_{\rm cut}(\zci) \equiv d_\rmA(\zci) \, \vartheta_{\rm fc}$, with $\zci$ the redshift of primary $i$. Tests with mock data show that his typically removes of order 5 (11) percent of the secondaries when satellite galaxies are assumed to have the phase-space distribution of subhaloes (dark matter particles). Since most primaries only have a single secondary, this cut in $R_\rmp$ also reduces the number of primaries, by roughly the same percentage. Tests with detailed mock data sets indicates that this cut does not significantly affect the constraining power regarding the galaxy-halo connection (see \S\ref{sec:tier2} and \S\ref{sec:tier3}).

\subsection{Additional Observational Constraints}
\label{sec:additional}

\subsubsection{Primaries without secondaries}
\label{sec:Pnull}

The data vector $\bD$ described thus far only contains primaries with at least one secondary. However, running the selection criteria over a spectroscopic redshift survey also yields a complementary data vector, $\bD_0 = (\{L_{\rmc,i},z_{\rmc,i}\} \, | \,i=1,2,...,N_0)$ listing all $N_0$ primaries with zero secondaries. This additional data vector provides additional constraints on the galaxy-halo connection, in particular regarding the satellite component of the CLF, and we therefore include it in our analysis. Since $N_0$ is typically much larger than the number of primaries with at least one primary, $N_{+}$, we bin this data using a $5 \times 5$ uniformly-spaced grid in $(\log L_\rmc, z_\rmc)$ covering the range $[9.5,11.0]$ in $\log L_\rmc$ and $[0.02,0.15]$ in $z_\rmc$. For each bin we compute the probability $P_0 \equiv N_0 / (N_0 + N_+)$ that a primary in that bin has zero secondaries. Since the error distribution of $P_0$ can be very non-Gaussian, the actual constraint that we use in our modeling is $f_0\equiv \log(N_0/N_+) = \log[P_0 / (1-P_0)]$. Assuming that both $N_0$ and $N_{+}$ follow Poisson statistics, we compute the corresponding errors on $f_0$ as $\sigma_{f_0} = (\sqrt{1/N_0 + 1/N_+}) / \ln 10$. 

We define the log likelihood corresponding to this data as 
\begin{equation}\label{calL0}
\ln \calL_{0} \equiv -\frac{1}{2}\sum_i \sum_j \left(\frac{f_{ij} - f_0(L_i,z_j)}{\sigma_{f_0}(L_i,z_j)} \right)^2\,.
\end{equation}
Here $\log L_i$ and $z_j$ are the centres of the bins used to compute $f_0$, and $f_{ij}$ is the corresponding model prediction. The latter is computed using $f_{ij} = \log[P_{ij} / (1 - P_{ij})]$ with
\begin{equation}\label{Pij}
\begin{split}
P_{ij} & = \int \rmd M \, P(N_\rms = 0 |M, L_i, z_j) \, P(M|L_i,z_j) \\
& = \frac{\int \rmd M \, P(L_i | M, z_j) \, n(M,z_j) \, \rme^{-\lambda_{ij}}}{\int \rmd M \, P(L_i | M, z_j) \, n(M,z_j)}\,, 
\end{split}
\end{equation}
the probability that a primary with luminosity $L_i$ at redshift $z_j$ has zero secondaries. Here we have once again used the fact that the completeness, $\calC(\Mh|L,z)$, does not depend significantly on halo mass (see \S\ref{sec:inference}), and we assume that both satellites and interlopers follow Poisson statistics, such that $P(N_\rms=0|M, L_i, z_j) = \rme^{-\lambda_{ij}}$ with $\lambda_{ij}$ the sum of the expectation values for the number of satellites (equation~[\ref{lambdasat}]) and the number of interlopers (equation~[\ref{interlopermodel}]). As always, we evaluate the mass integrals in equation~(\ref{Pij}) using Gaussian quadrature, as detailed in  Appendix~\ref{App:comp}.

\subsubsection{Galaxy Number Densities}
\label{sec:galnumdens}

Since our main goal is to constrain the galaxy-halo connection, it is also advantageous to include constraints from the overall number density of galaxies. In particular, the luminosity function provides important constraints on the CLF \citep[e.g.,][]{Yang.etal.03,vdBosch.etal.03,Cooray.Milosavljevic.05,Cooray.06}, which greatly helps to tighten the posterior in our inference problem.

We follow \cite{Lange.etal.19a, Lange.etal.19b} and use the number density of galaxies in ten, 0.15~dex bins in luminosity, ranging from $10^{9.5}$ to $10^{11} \Lsunh$. For our model, these number densities are computed according to
\begin{equation}\label{numdens}
n_{\rm gal}(L_1,L_2) = \int_{L_1}^{L_2} \rmd L \int_{0}^{\infty} \Phi(L|M) \, n(M,z_{\rm surv}) \, \rmd M\,,
\end{equation}
where $z_{\rm surv}$ is a characteristic redshift for the survey in question. For the mock data samples discussed in \S\ref{sec:validation}, which cover the redshift range $[0.02,0.15]$, we set $z_{\rm surv}$ equal to the redshift of the simulation output used to construct the mock. When analysing SDSS data (van den Bosch et al. 2019, in prep.), we adopt $z_{\rm surv} = 0.1$. We have verified that our results do not depend significantly on this choice; using  $z_{\rm surv} = 0.05$ instead of $0.0$ or $0.1$ yields results that are virtually indistinguishable.

We include the data on $n_{\rm gal}(L_1,L_2)$ in our inference problem by defining the corresponding log-likelihood
\begin{equation}
\ln\calL_{\rm LF}(\bn_{\rm obs}|\btheta) = -\frac{1}{2} \, [\bn(\btheta) - \bn_{\rm obs}]^t \, \boldsymbol{\Psi} \, [\bn(\btheta) - \bn_{\rm obs}]\,.
\end{equation}
Here $\bn_{\rm obs}$ is the data vector for $n_{\rm gal}$ for the ten bins in luminosity, $\bn(\btheta)$ is the corresponding model prediction given by equation~(\ref{numdens}), and $\boldsymbol{\Psi}$ is the precision matrix, which is the inverse of the covariance matrix. The latter is computed using 1000 SDSS-like mocks and the unbiased estimator as described in \cite{Lange.etal.19a}.

\subsection{Numerical Implementation}
\label{sec:numerics}

Probing the posterior $P(\btheta|\bD)$ over our 17-dimensional parameter space requires millions of likelihood evaluations, each of which involves many numerical integrations (see Appendix~\ref{App:comp}). In order to make this problem feasible,  we follow \cite{Lange.etal.19a, Lange.etal.19b} and perform the Bayesian inference under the assumption of a {\it fixed} normalized, radial number density distribution of satellite galaxies, $\bar{n}_{\rm sat}(r|M,z)$, to be defined in \S\ref{sec:phase_sat} below. This has the advantage that $f_{\rm ap}(M,L,z)$ and $P_{\rm int}(\dV, \Rp|L,z)$ are all independent of the model, $\btheta$, while $P_{\rm sat}(\dV, \Rp|M,L,z)$ only depends on a single anisotropy parameter (see \S\ref{sec:phase-space}). Combined with the fact that we perform the mass integration using a Gaussian quadrature with {\it fixed} abscissas, $\Mk$, this implies that we only need to compute (and store) these quantities once for each primary and/or secondary. And the same applies for the halo mass function, $n(M,z)$, which appears in equations~(\ref{Gi}) and (\ref{Fi}). The probabilities $P_{\rm sat}(\dV, \Rp|M,L,z)$ are computed using linear-interpolation over a grid of values that are pre-computed for different anisotropy parameters, as detailed in \S\ref{sec:params}. As a consequence, a single evaluation of the full likelihood 
\begin{equation}\label{Ltot}
\calL_{\rm tot}(\bD + \bD_0 + \bn_{\rm obs}|\btheta) \equiv \calL_{\rm SK}(\bD|\btheta) + \calL_{0}(\bD_0|\btheta) + \calL_{\rm LF}(\bn_{\rm obs}|\btheta)\,, 
\end{equation}
for a mock data set with 5000 satellite galaxies, takes only of order 10 milliseconds using a single, run-of-the-mill CPU. This is sufficiently fast, that it easily allows one to run many different Monte-Carlo Markov Chains for different assumptions regarding  $\bar{n}_{\rm sat}(r|M,z)$, or to find the best-fit radial profile, marginalized over all other model parameters, using a straight-forward $\chi^2$-minimization algorithm.

The method that we use to construct Monte-Carlo Markov Chains is the affine invariant ensemble sampler proposed by \cite{Goodman.Weare.10}. This is the same method that is used by the popular Python code {\tt emcee} developed by \cite{Foreman-Mackey.etal.13} and we refer the interested reader to these two papers for details. Throughout we use 1,000 walkers and the proposal density advocated by \cite{Goodman.Weare.10}. This results in typical acceptance fractions between 0.3 and 0.4. We start the walkers in a small region of parameter space centered on the best-fit model obtained during the `burn-in' stage. Throughout we adopt a Metropolis-Hastings burn-in of 10,000 steps in which we use independent Gaussian proposal distributions for each model parameter. The best-fit model at the end of the burn-in period is always close to the best-fit model subsequently obtained from the entire MCMC. Most of our MCMC chains contain 5 million elements (post burn-in), corresponding to 5,000 steps for each of the 1,000 walkers, and are well converged.

\section{Model Ingredients}
\label{sec:ingredients}

This section describes the model ingredients to be used in combination with the method outlined in the previous section. These include a model for the galaxy-halo connection, and a model for the phase-space distributions of central and satellite galaxies as a function of halo mass. 

\subsection{Galaxy-Halo Connection}
\label{sec:CLF}

We model the galaxy occupation using the conditional luminosity function \citep[CLF;][]{Yang.etal.03, vdBosch.etal.03} approach. The CLF, $\Phi(L|M) \rmd L$, specifies the average number of galaxies with luminosities in the range $L \pm \rmd L/2$ residing in a dark matter halo of virial mass $M$. As already eluded to in \S\ref{sec:inference}, we assume that galaxies can be separated into centrals and satellites, each with their own CLF,
\begin{equation}
\Phi(L | M) = \Phi_\rmc (L | M) + \Phi_\rms (L | M).
\end{equation}
Here, as always, subscripts `c' and `s' refer to central and satellite, respectively.These two populations are described in more detail below. 

\subsubsection{Central Galaxies}
\label{sec:CLFcen}
	
The CLF of centrals is parametrized using a log-normal distribution,
\begin{equation}\label{CLFcen}
\Phi_\rmc (L | M) \rmd L = \frac{\log e}{\sqrt{2\pi \sigma_\rmc^2}} \exp \left[ -\left(\frac{\log L - \log\bar{L}_\rmc}{\sqrt{2} \sigma_\rmc} \right)^2\right] \frac{\rmd L}{L}.
\end{equation}
The mass dependence of the median luminosity, $\bar{L}_\rmc$, is parametrized by a broken power-law:
\begin{equation}\label{averLc}
\bar{L}_\rmc (M) = L_0 \frac{(M / M_1)^{\gamma_1}}{(1 + M / M_1)^{\gamma_1 - \gamma_2}}.
\end{equation}
which is characterized by three free parameters; a normalization, $L_0$, a characteristic halo mass, $M_1$, and two power-law slopes, $\gamma_1$ and $\gamma_2$. 

Motivated by the fact that several hydrodynamical simulations suggest that the scatter, $\sigma_\rmc$, increases with decreasing halo mass \citep[e.g.,][]{Sawala.etal.17, Pillepich.etal.18}, we allow for a mass-dependent scatter using
\begin{equation}\label{scatter}
\sigma_\rmc(M) = \left\{ \begin{array}{ll}
\sigma_{12} & \mbox{if $\log\Mh \leq 12$} \\
\sigma_{12} +  \frac{\log\Mh-12}{2} \, (\sigma_{14} - \sigma_{12}) & \mbox{if $12 < \log\Mh < 14$} \\
\sigma_{14} & \mbox{if $\log\Mh \geq 14$} \\
\end{array} \right.
\end{equation}
Hence, the scatter is characterized by two free parameters, $\sigma_{12}$ and $\sigma_{14}$, that indicate the log-normal scatter in haloes of mass $\Mh = 10^{12}\Msunh$ and $10^{14}\Msunh$, respectively.
\begin{table*}
\begin{tabular}{l | l | l | c | l}
Parameter & Description & Equation & Prior & Default \\
 (1) & (2) & (3) & (4) & (5) \\
\hline\hline
$\log M_1 / (\Msunh)$ & characteristic mass of mass--luminosity relation for centrals & (\ref{averLc}) & U$[8.0, 15.0]$ & $11.20$\\
$\log L_0 / (\Lsunh)$ & normalization of mass--luminosity relation for centrals & (\ref{averLc}) & U$[8.0, 12.0]$ & $9.95$\\
$\gamma_1$ & low-mass slope of mass--luminosity relation for centrals & (\ref{averLc}) &  G$[3.5,0.2]$ & $3.5$\\
$\gamma_2$ & high-mass slope of mass--luminosity relation for centrals & (\ref{averLc}) & U$[0.0, 5.0]$ & $0.25$\\
$\sigma_{12}$ & logarithmic scatter in luminosity at a halo of mass $10^{12} \Msunh$ & (\ref{scatter}) & U$[0.001, 1.0]$ & $0.15$\\
$\sigma_{14}$ & logarithmic scatter in luminosity at a halo of mass $10^{14} \Msunh$ & (\ref{scatter}) & U$[0.001, 1.0]$ & $0.15$\\
$\alpha_{12}$ & the logarithmic slope of the satellite CLF at a halo of mass $10^{12} \Msunh$ & (\ref{alpha}) & U$[-2.0,2.0]$ & $-1.2$\\
$\alpha_{14}$ & the logarithmic slope of the satellite CLF at a halo of mass $10^{14} \Msunh$ & (\ref{alpha}) & U$[-2.0,2.0]$ & $-1.2$\\
$b_0$ & determines the normalization of the satellite CLF & (\ref{satCLFnorm}) & U$[-3.0, 3.0]$ & $-1.2$\\
$b_1$ & determines the normalization of the satellite CLF  & (\ref{satCLFnorm}) & U$[-3.0, 3.0]$ & $1.5$\\
$b_2$ & determines the normalization of the satellite CLF & (\ref{satCLFnorm}) & U$[-3.0, 3.0]$ & $-0.2$\\
$\eta_0$ & normalization of effective bias of interlopers & (\ref{beff}) & U$[0.0, 100.0]$ & $0.0$\\
$\eta_1$ & power-law dependence of effective bias of interlopers on luminosity of primary & (\ref{beff}) & U$[-2.0, 2.0]$ & $0.0$\\
$\eta_2$ & power-law dependence of effective bias of interlopers on redshift of primary  & (\ref{beff}) & U$[-2.0, 2.0]$ & $0.0$ \\
$\calR$ & ratio of scale radius of satellite distribution wrt that of dark matter & (\ref{nsatprof}) & U$[0.1, 10.0]$ & $1.0$\\
$\gamma$ & central slope of radial profile of satellite distribution & (\ref{nsatprof}) & U$[0.0, 1.5]$ & $1.0$\\
$\beta$ & anisotropy parameters (\CA$\;$ models) & (\ref{betadef}) &  U$[-9.0,0.9]$ & $0.0$ \\
$\log[r_\rma/r_\rms]$ & anisotropy radius (\OM$\;$ models) & (\ref{raOM}) & U$[-1.0,1.5]$ & --\\
\hline
\end{tabular}
\caption{Model parameters that quantify the galaxy-halo connection, the interloper fraction, and the phase-space distribution of satellite galaxies within their host haloes. Column (2) gives a description, while column (3) lists the equation in which the parameter is introduced. Column (4) indicates the prior used in our Bayesian analysis, where U$[a,b]$ indicates a uniform prior over the range $[a,b]$ and G$[a,b]$ indicates a Gaussian prior with a mean $a$ and dispersion $b$. Finally, column (5) indicates the default value used when constructing our mock data sets. Note that for $\eta_0$, $\eta_1$, and $\eta_2$, the parameters characterizing the effective bias parameter of the interlopers, these fiducial values only apply to the Tier-1 mock.}
\label{tab:parameters}
\end{table*}

\subsubsection{Satellite Galaxies}
\label{sec:CLFsat}
	
We model the satellite CLF as a modified Schechter function:
\begin{equation}\label{satCLF}
\Phi_\rms (L | M) = \frac{\phi_\rms^*}{L_\rms^*} \left( \frac{L}{L_\rms^*} \right)^{\alpha_\rms} \exp \left[ - \left( \frac{L}{L_\rms^*} \right)^2 \right].
\end{equation}
Thus, the luminosity function of satellites, for a given halo mass, follows a power-law with slope $\alpha_\rms$ with an exponential cut-off above a critical luminosity, $L_\rms^*(M)$, which is related to the characteristic luminosity of central galaxies in haloes of the same mass according to
\begin{equation}
L_\rms^*(M) = 0.562 \, \bar{L}_{\rmc} (M).
\end{equation}
As shown in \citet{Yang.etal.09}, this relation provides a good description of the luminosities of centrals and satellites as inferred from the SDSS galaxy group catalogue of \citet{Yang.etal.07}\footnote{We have tested that treating the ratio $L_\rms^*(M)/\bar{L}_{\rmc}(M)$ as a free parameter in \Basilisk does not significantly impact any of our results.}. Motivated by the results of \cite{Yang.etal.08}, who found evidence for a steeper slope (more negative value of $\alpha_\rms$) in more massive groups, we allow for a mass-dependent power-law slope using
\begin{equation}\label{alpha}
\alpha_\rms(M) = \left\{ \begin{array}{ll}
\alpha_{12} & \mbox{if $\log\Mh \leq 12$} \\
\alpha_{12} +  \frac{\log\Mh-12}{2} \, (\alpha_{14} - \alpha_{12}) & \mbox{if $12 < \log\Mh < 14$} \\
\alpha_{14} & \mbox{if $\log\Mh \geq 14$} \\
\end{array} \right.
\end{equation}
Hence, similar to the scatter, the logarithmic slope is characterized by two free parameters, $\alpha_{12}$ and $\alpha_{14}$, that indicate the slope in haloes of mass $\Mh = 10^{12}\Msunh$ and $10^{14}\Msunh$, respectively. Finally, the normalization $\phi_\rms^*(M)$ is parametrized by
\begin{equation}\label{satCLFnorm}
\log \left[ \phi_\rms^*(M) \right] = b_0 + b_1 \log M_{12} + b_2 (\log M_{12})^2.
\end{equation}
where $M_{12} = M/(10^{12}\msunh)$. 

Note that our particular characterization of the CLF is very similar to that adopted in a number of previous studies \citep[][]{Cacciato.etal.09, Cacciato.etal.13, More.etal.09b, vdBosch.etal.13, Lange.etal.19a, Lange.etal.19b}.

\subsection{Phase-space distributions}
\label{sec:phase-space}
	
The CLF described above specifies the abundance of central and satellite galaxies as function of luminosity and halo mass. We now describe our model for the positions and velocities of these galaxies with respect to their host halo.

\subsubsection{Central Galaxies}
\label{sec:phase_cen}

Throughout this work, we assume that central galaxies are located at the dark matter halo centre and have zero velocity in the rest frame of the dark matter halo. It is known, though, that in reality centrals can have small velocity offsets \citep[][]{vdBosch.etal.05b, Behroozi.etal.13a, Guo.etal.15a, Guo.etal.15b, Guo.etal.16, Ye.etal.17}. But, as previously shown in \cite{Lange.etal.19a}, this does not have a significant impact on the inferences from satellite kinematics. We explicitly demonstrate this assertion in \S\ref{sec:cenbias}.

\subsubsection{Satellite Galaxies}
\label{sec:phase_sat}

In the case of satellite galaxies, the phase-space model determines the probability $P_{\rm sat}(\dV, \Rp | \Mh, z, L_\rmc)$, which characterizes the satellites' projected phase-space distribution and plays the key role in our likelihood evaluation (see \S\ref{sec:PdvR}). In the following, we assume that satellites have a spherically symmetric radial profile $n_{\rm sat}(r|M,z)$. It is known that satellite populations of individual haloes can have varying degrees of asphericity \citep[e.g.,][]{Zentner.etal.05b,Azzaro.etal.07,Wang.etal.08}. However, since we combine the data from a large number of individual haloes with random orientations, this assumption of spherical symmetry will not affect our inferences substantially. We assume that the radial profile as a function of the radial distance $r$ from the halo centre is given by a generalized Navarro--Frenk--White (gNFW) profile,
\begin{equation}\label{nsatprof}
n_{\rm sat}(r|M,z) \propto \left( \frac{r}{\calR \, r_\rms} \right)^{-\gamma} \left( 1 + \frac{r}{\calR \, r_\rms} \right)^{\gamma - 3}\,.
\end{equation}
Here $\calR$ and $\gamma$ are free parameters and $r_\rms$ is the scale radius of the dark matter halo, which is related to the halo virial radius via the concentration parameter $\cvir = \rvir / r_\rms$.  This gNFW profile has sufficient flexibility to adequately describe a wide range of radial profiles, from satellites being unbiased tracers of their dark matter halo ($\gamma = \calR = 1$), to cored profiles that resemble the radial profile of surviving subhaloes in numerical simulations ($\gamma=0$, $\calR \sim 2$). This also brackets the range of observational constraints on the radial distribution of satellite galaxies in groups and clusters \citep[e.g.,][]{Carlberg.etal.97, vdMarel.etal.00, Lin.etal.04, Yang.etal.05a, Chen.08, More.etal.09b, Guo.etal.12a, Cacciato.etal.13, Watson.etal.10, Watson.etal.12, Lange.etal.19b}.

We also assume that the host haloes of satellite galaxies are spherical NFW haloes that are completely specified by their mass, i.e., we adopt the concentration-mass relation of \citet{Maccio.etal.08} without scatter.  In the most general case, under the assumption of spherical symmetry, one then has that
\begin{equation}\label{fundamental}
\begin{split}
P_{\rm sat}(\dV, \Rp &|M,L,z) = \\
& \frac{2 \pi \Rp}{\lambda_{\rm sat}}  \int\rmd {\rm z} \int \int \rmd v_{R} \rmd v_{\phi} \, f(E,J^2)\,.
\end{split}
\end{equation}
Here ${\rm z}$ is the coordinate along the line-of-sight, not to be confused with the redshift $z$, 
$\lambda_{\rm sat} = \lambda_{\rm sat}(M,L,z)$  is the expectation value for the number of satellite galaxies (equation~[\ref{lambdasat}]), and $f(E,J^2)$ is the distribution function (DF), which for a spherically symmetric system is a function of energy, $E$, and angular momentum, $J=|\bJ|$. Typically one of three assumptions is made: (i) the DF is isotropic, such that $f(E,J^2) = f(E)$, (ii) the DF depends on energy and angular momentum only through the quantity $Q = E + J^2/(2 r^2_\rma)$, where $r_\rma$ is a free parameter known as the `anisotropy radius', such that $f(E,J^2) = f(Q)$, or (iii) the distribution function is separable, such that $f(E,J^2) = g(E) \, h(J^2)$. Models that make assumption (ii) are known as Osipkov-Merritt models \citep[][]{Osipkov.79, Merritt.85} and have an anisotropy profile that increases from isotropic in the center ($r \ll r_\rma$) to radially anisotropic at larger radii. Models that make assumption (iii) can have constant anisotropy or a radially varying anisotropy \citep[e.g.,][]{Louis.93, Cuddeford.Louis.95, Wojtak.etal.09}. In each case, the computation of the DF for a given $n_{\rm sat}(r|M,z)$ and halo potential, $\Psi(r|M,z)$, involves at least a 1D integration\footnote{In the case where $f = f(E)$ or $f = f(Q)$ this integral is known as the Eddington formula \citep[][]{Binney.Tremaine.08}. If the DF is separable, $h(J^2)$ needs to be of a special form for the inversion from $n_{\rm sat}$ to DF to reduce to a 1D integration.}. Together with equation~(\ref{fundamental}), which involves a 3D integration, this makes the computation of $P(\dV, \Rp |M,L,z)$ prohibitively expensive. In addition, we lack a good prior on the functional form of $f(E,J^2)$, further disincentivizing the use of equation~(\ref{fundamental}). 
\begin{figure*}
\centering
\includegraphics[width=0.8\textwidth]{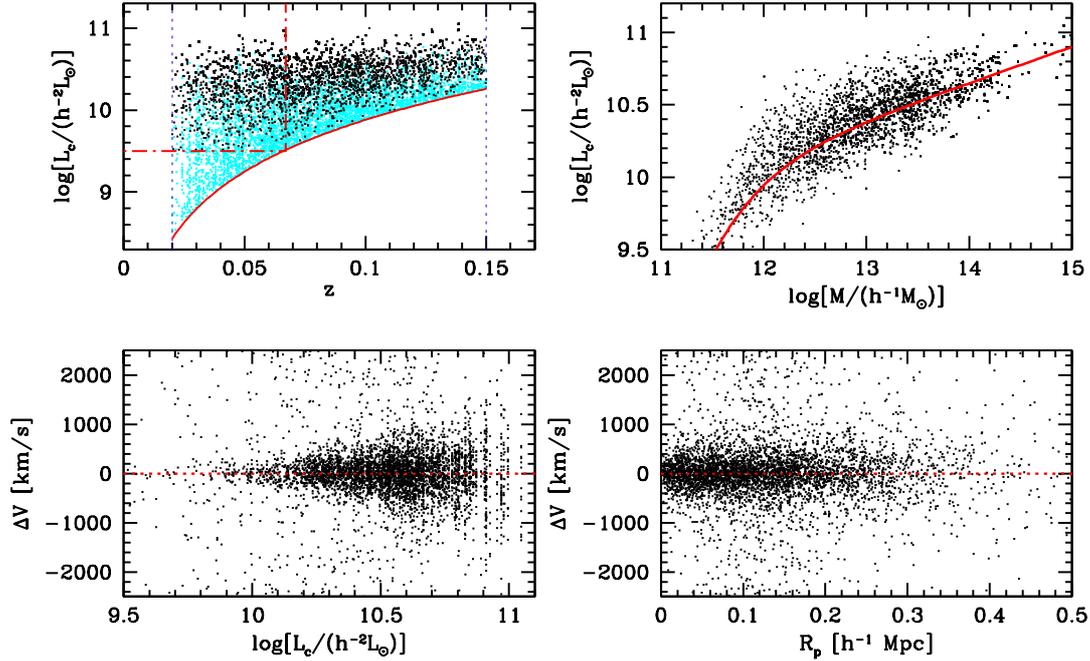}
\caption{Tier-1 mock data. The upper left-hand panel plots luminosity as a function of redshift, with black and cyan dots indicating primaries and secondaries, respectively. The solid red curve indicates the apparent magnitude limit of the (mock) survey, while the red dot-dashed lines mark the volume-limited subsample used in previous SDSS-based analyses of satellite kinematics \citep[in particular][]{More.etal.09b, More.etal.11, Lange.etal.19b}. The latter is shown to underscore that \Basilisk bases its analysis on much more data. The upper right-hand panel plots the luminosity of the primary as a function of its halo mass, with the red solid line marking the expectation value $\langle L_\rmc | M \rangle$ corresponding to the CLF used to construct the mock. Note the large amount of scatter in $M$ at given $L_\rmc$, highlighting the importance of properly taking `mass-mixing' into account.  Finally, the lower panels plot $\Delta V$ as a function of the primary luminosity (left) and the projected separation between primary and secondary (right). Note the obvious increase in the variance of $\Delta V$ with increasing $L_\rmc$, reflecting that more luminous primaries reside in more massive haloes, the contribution of indivual clusters and large groups at the luminous end, and the uniform `background' due to interlopers.}
\label{fig:mock1}
\end{figure*}

We therefore opt for an alternative, approximate method to compute $P(\dV, \Rp|M,L,z)$. Rather than using DFs, we write
\begin{equation}\label{split}
P_{\rm sat}(\dV, \Rp|M,L,z) = P(\Rp|M,L,z) \, P(\dV|\Rp,M,z)\,.
\end{equation}
and make the simplified assumption that $P(\dV|\Rp,M,z)$ is a Gaussian with a projected velocity dispersion, $\sigma_{\rm los} = \sigma_{\rm los}(\Rp|M,L,z)$:
\begin{equation}\label{PdVgauss}
P(\dV|\Rp,M,z) = \frac{1}{\sqrt{2\pi}\,\sigma_{\rm los}} \, \frac{\exp\left[-(\dv)^2/2\sigma^2_{\rm los}\right]}{{\rm erf\left[\dVs/\sqrt{2}\sigma_{\rm los}\right]}}\,.
\end{equation}
The division by the error function is required by the normalization condition, and the fact that our selection criterion only includes satellites for which $|\dV| < \dVs$. Although there is no {\it a priori} reason for the assumption of Gaussianity, the LOSVDs of (non-rotating) dynamical systems often are very close to Gaussian. Indeed, as we demonstrate in this paper, this assumption is adequate for the purpose of constraining the galaxy-halo connection and allows for an extremely efficient computation of $P(\dV, \Rp|M,L,z)$.

The probability $P(\Rp|M,L,z)$ in equation~(\ref{split}) derives from the (normalized) radial number density distribution of satellite galaxies, $\bar{n}_{\rm sat}(r|M,z)$, according to
\begin{equation}\label{PrP}
P(\Rp|M,L,z) = \frac{2\, \pi\, \Rp \, \bar{\Sigma}(\Rp|M,z)}{f_{\rm ap}(M,L,z)}\,,
\end{equation}
where
\begin{equation}\label{ProjNsat}
\bar{\Sigma}(\Rp|M,z) = 2 \int_{\Rp}^{r_{\rm vir}(M,z)} \bar{n}_{\rm sat}(r|M,z) 
\frac{r\,\rmd r}{\sqrt{r^2 - R^2}}\,,
\end{equation}
is the projected, normalized number density distribution of satellite galaxies. The division by the aperture fraction is required by the normalization condition. The projected velocity dispersion is related to the intrinsic, radial velocity dispersion, $\sigma_r^2(r|M,z)$, according to the following Abel integral
\begin{equation}\label{sigmalos}
\begin{split}
\sigma^2_{\rm los}(\Rp |M,z) = \frac{2}{\bar{\Sigma}(\Rp)} & \int_{\Rp}^{\rvir(M,z)} \left[ 1 - \beta(r) \frac{\Rp^2}{r^2} \right] \\
& \bar{n}_{\rm sat}(r|M,z) \, \sigma_r^2(r|M,z) \, \frac{r \, \rmd r}{\sqrt{r^2 - \Rp^2}\,,}
\end{split}
\end{equation}
where
\begin{equation}\label{betadef}
\beta(r) \equiv 1 - \frac{\sigma^2_\rmt(r)}{2\sigma^2_\rmr(r)}\,,
\end{equation}
is the local anisotropy parameter, relating the tangential and radial velocity dispersions.

The radial velocity dispersion follows from the Jeans equation. If we assume a constant orbital anisotropy, such that $\beta(r) = \beta$, then this Jeans equation reduces to 
\begin{equation}\label{sig2beta}
\begin{split}
\sigma^2_r(r|M,z) = & \frac{G}{r^{2\beta}} \, \frac{1}{\bar{n}_{\rm sat}(r|M,z)} \\ &\int_{r}^{r_{\rm vir}(M,z)} r'^{2\beta-2} \, \bar{n}_{\rm sat}(r'|M,z) \, M(r') \, \rmd r'\,,
\end{split}
\end{equation}
where $M(r)$ is the halo mass enclosed by radius $r$. In addition to these constant anisotropy (hereafter \CA) models, we will also consider Osipkov-Merritt (hereafter \OM) models, for which the anisotropy parameter scales with radius as
\begin{equation}\label{raOM}
\beta(r) = \frac{r^2}{r^2 + r^2_\rma}\,.
\end{equation}
Hence, the orbits are close to isotropic ($\beta \simeq 0$) at small radii ($r \ll r_\rma$), and become more and more radially anisotropic at larger radii. This is very reminiscent of the orbital anisotropy of dark matter particles \citep[e.g.,][]{Ascasibar.Gottlober.08,  Wojtak.etal.08, Wojtak.etal.13} and subhaloes \citep[e.g.][]{Diemand.Moore.Stadel.04} in numerical simulations, and is therefore a realistic model to describe the kinematics of satellite galaxies \citep[but see][and Appendix~\ref{App:anisotropy}]{Cuesta.etal.08, Sawala.etal.17}.

For an \OM-model, the Jeans equation for the radial velocity dispersion becomes
\begin{equation}\label{sig2OM}
\begin{split}
\sigma^2_r(r|M,z) = & \frac{G}{r^2 + r^2_\rma} \, \frac{1}{\bar{n}_{\rm sat}(r|M,z)} \\ & \int_{r}^{r_{\rm vir}(M,z)} \frac{r'^2 + r^2_\rma}{r'^2} \, \bar{n}_{\rm sat}(r'|M,z) \, M(r') \, \rmd r'\,,
\end{split}
\end{equation}
\citep[][]{Merritt.85}. Note that an \OM-model with $r_\rma \rightarrow \infty$ is equivalent to a \CA-model with $\beta=0$ (both are isotropic throughout). Detailed expressions for $\sigma^2_{\rm los}(\Rp|M,z)$ for a tracer population with a gNFW profile orbiting within a NFW host halo are given in Appendix~\ref{App:comp}, for both the \CA$\,$ and \OM-model.
\begin{figure*}
\centering
\includegraphics[width=\textwidth]{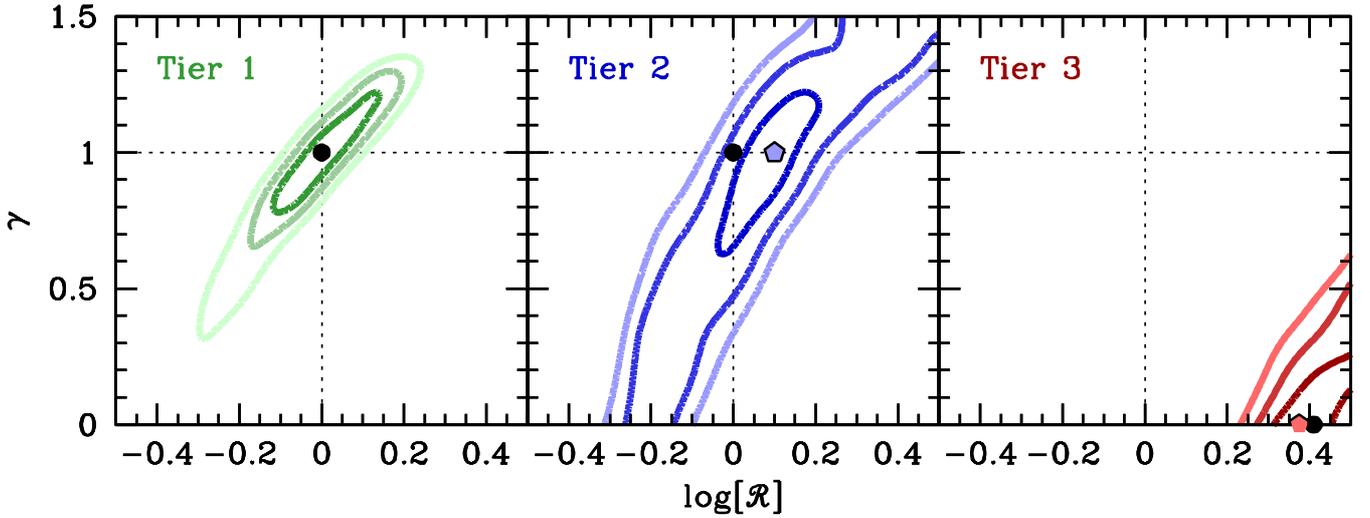}
\caption{Constraints on the two gNFW parameters, $\calR$ and $\gamma$, that characterize the radial distribution of satellite galaxies. Different panels show results for different tier mocks, as indicated, and contours, from dark to light, correspond to the 68, 95 and 99 percent confidence levels obtained from $\Delta\chi^2_{\rm tot}$ as described in the text.  Thick, solid black dots indicate the true input values, while filled pentagons indicate the best-fit model. In the case of the Tier-1 mock, the latter coincides with the former, and is therefore not visible.}
\label{fig:contour}
\end{figure*}

Finally, in order to account for non-zero redshift errors in the data, the line-of-sight velocity dispersion used in equation~(\ref{PdVgauss}) is the quadratic sum of $\sigma_{\rm los}$ given by equation~(\ref{sigmalos}) plus two times the typical redshift error, $\sigma_{\rm err}$. 

\subsubsection{Summary}
\label{sec:params}

Altogether, our model has a total of 15 free parameters: 11 to describe how galaxies populate dark matter haloes, of which 6 describe the CLF of centrals $(M_1,L_0,\gamma_1,\gamma_2,\sigma_{12},\sigma_{14})$ and 5 quantify the satellite component $(\alpha_{12}, \alpha_{14}, b_0, b_1, b_2)$, 3 parameters to specify the effective bias that characterizes the number density of interlopers $(\eta_0, \eta_1, \eta_2)$, and one parameter to characterize the orbital anisotropy (either $\beta$ or $r_\rma$).

As already mentioned in \S\ref{sec:numerics}, the two additional parameters, $\calR$ and $\gamma$, that describe the radial profile of satellite galaxies, are not treated as free parameters in the MCMC analysis. Rather,  we separately constrain the posteriors for different sets of $(\calR, \gamma)$, as this vastly increases the speed.
In the case of the anisotropy parameter, we pre-compute $P_{\rm sat}(\dV, \Rp|M,L,z)$ over a range of anisotropy parameters, and then use linear-interpolation to compute these quantities for given $\beta$ or $r_\rma$. In the case of the \CA-model, we pre-compute the matrix $\calM_{ijkl} \equiv P_{\rm sat}(\dVij, \Rij|\Mk, \Lci, \zci, \beta_l)$ for 10 values of $\beta$ that uniformly sample the parameter
\begin{equation}
\calB \equiv -\log[1 - \beta]\,,
\end{equation}
over the interval $[-1.0,1.0]$, which corresponds to $\beta$ covering the range $[-9.0, 0.9]$. In the case of the \OM-model, we pre-compute $\calM'_{ijkl} \equiv P_{\rm sat}(\dVij, \Rij|\Mk, \Lci, \zci, r_{\rma,l})$ using 10 values of $r_\rma$, that uniformly sample $\log[r_\rma/r_\rms]$ over the range $[-1.0,1.5]$.

Table~\ref{tab:parameters} lists all our model parameters used to quantify the galaxy-halo connection, the number density of interlopers, and the phase-space distribution of satellite galaxies. It also lists their prior ranges used when fitting data and their default values used to create the mock catalogues described in \S\ref{sec:validation}. These default values are similar to the constraints inferred by \cite{More.etal.11} and \cite{Cacciato.etal.09} analysing satellite kinematics, galaxy-galaxy lensing, and galaxy clustering in the SDSS, and therefore give a realistic description of the galaxy-halo connection at low redshift. Note that we adopt non-informative, uniform priors for all parameters except $\gamma_1$, for which instead we use a Gaussian prior with a mean of $3.5$ and a standard deviation of $0.2$. This is motivated by the fact that the slope of the $L_\rmc(M_\rmh)$ relation is poorly constrained at the low mass end, which in turn owes to the fact that we have very few satellites for primaries with $L_\rmc \lta 10^{10} \Lsunh$ (see Fig.~\ref{fig:mock1}). The value of $3.5$ is consistent with constraints from a variety of independent studies \citep[e.g.,][]{Yang.etal.09, Cacciato.etal.13, Lange.etal.19b}, all of which find best-fit values in the range $3.2 \lta \gamma_1 \lta 3.7$. We have verified that this prior has no impact on any of our results, other than restricting the posterior constraint on $\gamma_1$.	
	
\section{Validation}
\label{sec:validation}

We now proceed with a three-tiered validation process of our method. In tier 1 we use highly idealized mocks in which we draw dark matter haloes from an analytical halo mass function, and in which we assume perfect identification of centrals and satellites and ignore survey incompleteness effects (i.e., fibre collisions). In addition, we use the same model to assign satellites and interlopers their phase-space coordinates as used in our analysis. In tier 2 mocks we add complexity by constructing the mocks from dark-matter-only $N$-body simulations, yielding more realistic interlopers, and by including spectroscopic incompleteness due to fibre collisions and other redshift failures. When populating the dark matter halos with mock galaxies, though, we still adopt the same analytical model for their phase-space distributions as in our model. Finally, in the third and final tier, we construct mocks by assigning satellite galaxies the locations and velocities of dark matter subhaloes in the $N$-body simulations. All mocks are constructed to resemble the SDSS Main Galaxy Sample. In particular, we adopt an apparent magnitude limit of $m_{\rm lim} = 17.6$ in the $r$-band and we assume that galaxy redshifts have a velocity error of $\sigma_{\rm err} = 15 \kms$ \citep{Guo.etal.15b}. For the Tier 2 and 3 mocks we also mimick the SDSS footprint on the sky and model the impact of incompleteness due to fibre collisions. We now describe each tier in detail, and highlight some important aspects of our modeling approach that provide valuable insight.
\begin{figure*}
\centering
\includegraphics[width=\textwidth]{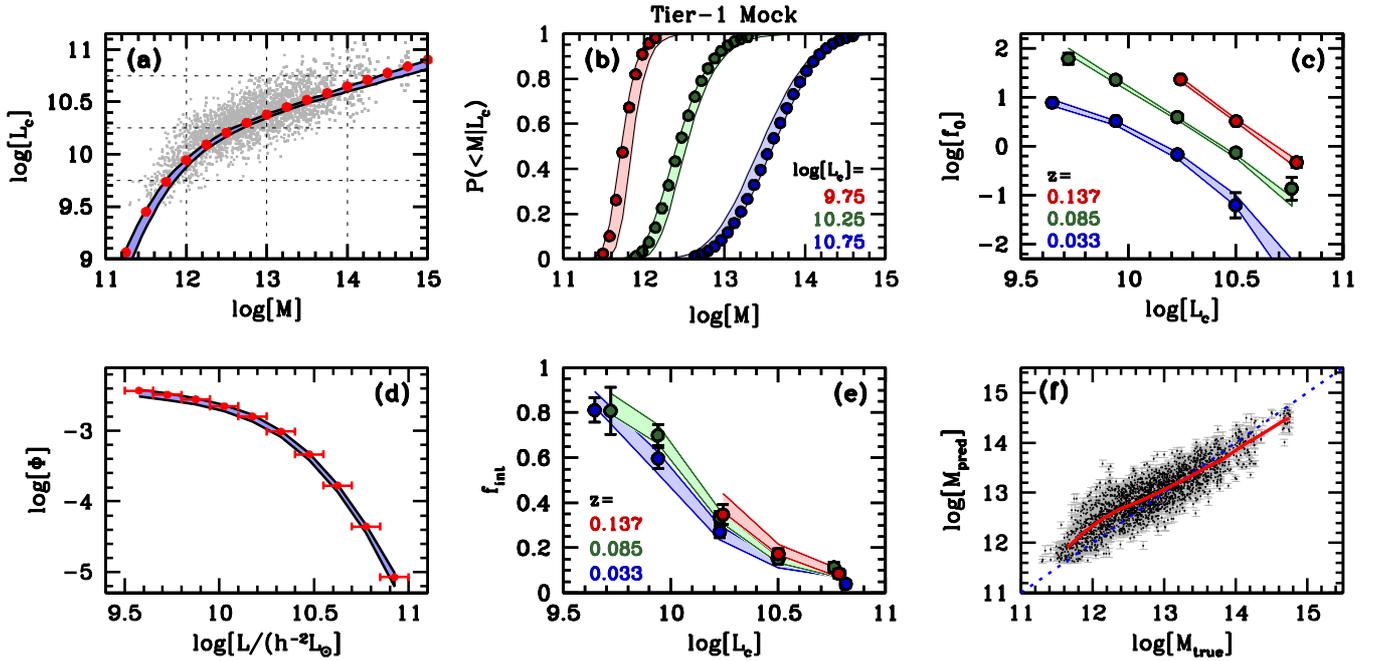}
\caption{Results for the analysis of the Tier-1 Mock data. In panels (a)-(e) solid dots always reflect the true input values of the mock, while shaded regions mark the 95\% confidence interval inferred from the MCMC. Panel (a) plots the luminosity of central galaxies, $L_\rmc$, as a function of halo mass, $M$. Grey dots indicate the actual primaries from the mock, and are shown for comparison and to highlight the amount of scatter in this relation.  Panel (b) plots the cumulative distributions for halo mass, $M$, for three different central luminosities, as indicated. Panel (c) plots $\log[f_0]$, which is the statistic we use to describe the fraction of primaries with zero secondaries (see \S\ref{sec:Pnull}), as function of the luminosity of the primary, for three different redshift bins, the median values of which are indicated.  The luminosity function of all galaxies (centrals plus satellites) is shown in panel (d), while panel (e) plots the interloper fractions as function of luminosity and redshift. Finally, panel (f) plots for each primary the predicted halo mass, $M_{\rm pred}$ (equation~[\ref{Mpred}]) versus the true halo mass, $M_{\rm true}$. Errorbars reflect the 95 percent confidence intervals on $M_{\rm pred}$ as inferred from the MCMC, while the solid, red line indicates the running average. See text for a detailed discussion.}
\label{fig:mock1res}
\end{figure*}

\subsection{Tier-1: idealized mocks}
\label{sec:tier1}

In the first step of our validation process we consider highly idealized mock data sets that are constructed as follows:
\begin{enumerate}

\item Draw a redshift, $z_\rmc$, in the range $[0.02,0.15]$, sampled according to the corresponding comoving volume, i.e.,  $P(z) \propto dV/dz$, and compute the corresponding luminosity limit, $L_{\rm min}(z_\rmc)$, defined as the minimum luminosity for a galaxy at that redshift to have an apparent magnitude brighter than the survey limit $m_{\rm lim}$.

\item At this redshift, draw a halo mass from the halo mass function, $n(M,z_\rmc)$  covering the range $\log[M/\Msunh] \in [10,15]$. These mass limits are purely numerical convenience; the upper limit is large enough that the abundance of more massive haloes is sufficiently small, while the lower limit is low enough that the probability that its central is brighter than the apparent magnitude limit of the survey is negligible.

\item Draw a luminosity for the central galaxy from $\Phi_\rmc(L_\rmc|M)$. If $L_\rmc < L_{\rm min}(z)$ or $L_\rmc < 10^{9.5} \Lsunh$, discard the halo and galaxy and go back to (i). The reason for discarding primaries with $L_\rmc < 10^{9.5} \Lsunh$ is that the number is small and virtually all their secondaries are interlopers. Hence, they add little in terms of constraining power for the model.

\item Compute the aperture radius, $\Rs(L_\rmc)$, and the expectation value for the number of interlopers in the secondary-selection-cylinder, $\lambda_{\rm int} \equiv \langle N_{\rm int}|L_\rmc,z\rangle$. The latter is computed using equation~(\ref{interlopermodel}) with the effective bias set to unity, i.e., ($\eta_0,\eta_1,\eta_2)=(1,0,0)$. Draw the actual number of interlopers from a Poisson distribution with a mean equal to $\lambda_{\rm int}$, and for each of these interlopers draw a  $R_\rmp$ and $\dV$ with respect to the primary from the phase-space probability distribution given by equation~(\ref{Pint}).

\item Use equation~(\ref{lambdasat}) with $f_{\rm ap} = 1$ to compute the expectation value, $\lambda_{\rm sat}$, for the number of satellites in the halo in question, and draw the actual number of satellites from a Poisson distribution with a mean equal to $\lambda_{\rm sat}$. For each satellite draw a position within the halo, assuming a spherically symmetric distribution characterized by $n_{\rm sat}(r|M,z)$. Next, compute the local velocity dispersion, $\sigma^2(r|M,z)$, using the Jeans equation for an isotropic DF (equation~[\ref{sig2beta}] with $\beta=0$), and draw the component of the velocity vector along the line-of-sight, $\dV$, from a Gaussian with a velocity dispersion equal to $\sigma^2(r|M,z)$. Compute the corresponding $R_\rmp$, draw a luminosity for the satellite, $L_\rms$, from $\Phi_\rms(L|M)$, and correct $\dV$ for redshift errors by adding a random velocity drawn from a Gaussian with a dispersion of $\sqrt{2} \sigma_{\rm err}$. If $L_\rms < L_\rmc$ and $R_\rmp < \Rs(L_\rmc)$ and $|\dV| < \dVs$, add this satellite to the list of secondaries for the primary in question.

\item Repeat this procedure until the total number of secondaries exceeds the target number. Note that we keep track of primaries that end up with zero secondaries (zero interlopers and zero satellites), which we use to compute $f_0(L_\rmc, z)$ for our mock data set.

\end{enumerate}
\begin{figure*}
\centering
\includegraphics[width=0.95\textwidth]{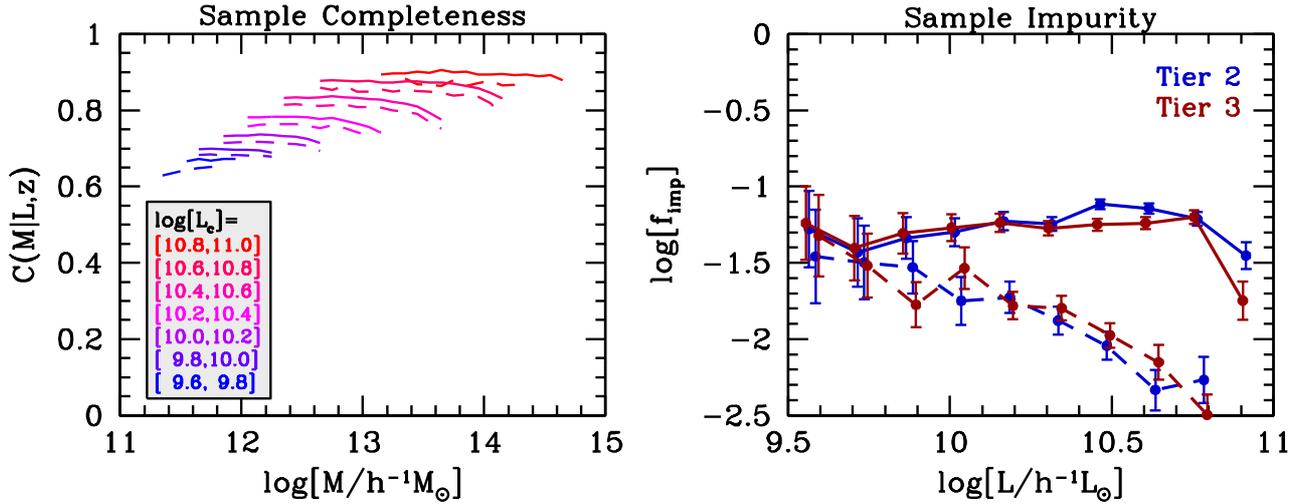}
\caption{{\it Left panel:} The completeness, $C(M|L,z)$, in our Tier-2 mock for different bins of central luminosity (different colors, as indicated), and different redshifts bins; $z=[0.02,0.09]$ (dashed lines) and $z=[0.09,0.15]$ (solid lines).  $C(M|L,z)$ is defined as the fraction of centrals of luminosity $L$ at redshift $z$ in haloes of mass $M$ that are selected as primaries by our cylindrical isolation criterion (cf. equation~[\ref{Pmlz}]). Note that this completeness has only a very weak dependence on halo mass, which implies that it can be ignored in the modelling (see \S\ref{sec:PLcen}). {\it Right panel:} Impurity fraction, $f_{\rm imp}$, defined as the fraction of primaries that are not centrals, plotted as function of the luminosity of the primary. Results are shown for our fiducial Tier-2 and Tier-3 mocks (solid blue and red curves, respectively), and for corresponding mocks in which we remove satellites that are brighter than their centrals, i.e., in which, by construction, each central is its brightest halo galaxy (dashed lines). Errorbars are computed assuming Poisson statistics. Note that the impurity fraction for the Tier-1 mock is zero by construction.}
\label{fig:impurity}
\end{figure*}

We use this method to construct a mock that has 5,000 secondaries around 2,379 primaries covering the redshift range $[0.02,0.15]$. The mock also contains 15,361 primaries with zero secondaries. Satellites are assumed to be unbiased, isotropic tracers of the mass distribution of their host halos (i.e., $\gamma = \calR = 1.0$ and $\beta=0$), and the halo occupation statistics are given by a CLF with the fiducial parameters listed in Table~\ref{tab:parameters}. Hence, this mock data set is generated using exactly the same model as used to compute the likelihood, and the results of the likelihood analysis discussed below therefore merely serves as a sanity check of \Basilisk's inference procedure.

Fig.~\ref{fig:mock1} shows some properties of this mock data set. The upper-left panel plots the luminosity as a function of redshift. The two vertical lines mark the redshift limits, while the red, solid curve corresponds to an apparent $r$-band magnitude of $17.6$. Black and cyan dots correspond to primaries and secondaries, respectively. The red dot-dashed lines demarcate the volume-limited survey that was used in the satellite kinematics analyses of \cite{More.etal.09b}, \cite{More.etal.11}, and \cite{Lange.etal.19b}. The method developed here can be applied to a flux-limited sample, thereby greatly increasing the amount of data that can be used. Note that, when evaluating $\ln\calL_0$ using equation~(\ref{calL0}), we only sum over those $\log L_\rmc$ and $z_\rmc$ bins that lie entirely above the flux limit of the survey, i.e., for which $\log L_i - \Delta\log L/2 > \log L_{\rm min}(z_j+\Delta z/2)$ with $\Delta \log L = 0.15$ and $\Delta z = 0.026$ the bin widths. This is the case for 16 out of the total of 25 bins. 
The upper-right panel of Fig.~\ref{fig:mock1} plots the luminosity of primaries in the mock as a function of their halo mass, with the red, solid line indicating the expectation value, $\log\langle L_\rmc | M \rangle$, computed from the central CLF, $\Phi_\rmc(L|M)$, used to construct the mock. Although the mock only assumes a scatter in $\log L_\rmc$ at fixed halo mass of 0.15 dex (see Table~\ref{tab:parameters}), it is clear that at fixed $L_\rmc$, the primaries cover a huge range in halo mass \citep[see][for a detailed discussion]{More.etal.09a}. Properly accounting for this `mass-mixing' is one of the main challenges in satellite kinematic. Finally, the lower two panels of Fig.~\ref{fig:mock1} plot $\dV$ of the primary-secondary pairs in the mock as functions of $\log L_\rmc$ (lower-left) and $\Rp$ (lower-right). In addition to an obvious increase in the dispersion of $\dV$ with increasing $L_\rmc$, which is the signal of interest, a roughly uniform contribution from interlopers is apparent. 

We analyse this mock data using the method outlined in \S\ref{sec:method}. The first step is to determine the best-fit radial profile of the satellite galaxies, $n_{\rm sat}(r|M)$, properly marginalized over all other model parameters.  Using a $10 \times 9$ grid in $(\gamma,\calR)$-space, we use the downhill simplex method \citep[][]{Nelder.Mead.65} to find the best-fit \CA-model for each $(\gamma,\calR)$-model (results for the \OM-model are very similar). Since this Tier-1 mock has no fibre collisions, or other form of incompleteness, we set the weights of all secondaries, $w_{ij}$, to unity, and the cut-off radius, $R_\rmc$, to zero. The left-hand panel of Fig.~\ref{fig:contour} shows the 68, 95 and 99 percent confidence levels thus obtained. Confidence levels are computed assuming that $\Delta \chi^2_{\rm tot}$ obeys a Chi-square distribution with two degrees of freedom, where $\chi^2_{\rm tot} = -2 \ln\calL_{\rm tot}$. The contours trace out a narrow region centered on the input model ($\gamma = \calR = 1$), indicated by a solid, black dot. Hence, \Basilisk yields an unbiased, and well-constrained estimate of the radial profile of the satellites, at least for this highly-idealized mock. 

The next step is to quantify the full posterior distribution of our model parameters using the best-fit radial profile (i.e., $\calR = \gamma = 1.0$).  We do so constructing a MCMC of 5 million elements (5,000 steps for 1,000 walkers). The resulting medians and 95 percent confidence intervals for all parameters are listed in Table~\ref{tab:results}, while Fig.~\ref{fig:mock1res} shows a number of posterior predictions. In particular, the solid dots in panels (a)-(e) indicate the true values, while the shaded regions always mark the 95 percent confidence intervals inferred from our MCMC. Panel (a) shows the constraints on the luminosity-halo mass relation for central galaxies. For completeness, the grey dots show the actual mock data. Clearly, the model is successful in recovering the expectation value for the central luminosity given the mass of its halo. Panel (b) shows the cumulative distributions $P(<M|L_\rmc)$ for three different central luminosities, as indicated. Again, the posterior predictions are in excellent agreement with the true distributions, indicating that \Basilisk not only recovers the {\it average} relation between light and mass, but the full distribution. This is also apparent from Table~\ref{tab:results}, which shows that both $\sigma_{12}$ and $\sigma_{14}$ are tightly constrained, and in excellent agreement with the true values. An important reason for this success is the fact that we include the number of secondaries, $N_\rms$, as a constraint. As we demonstrate in Appendix~\ref{App:noNsat}, ignoring this observable, as has been done in several previous studies, results in a strongly biased inference on the galaxy-halo connection.
\begin{table*}
\begin{tabular}{l r | r r r | r r r | r r r }
          &       & \multicolumn{3}{c}{Tier 1} & \multicolumn{3}{c}{Tier 2} & \multicolumn{3}{c}{Tier 3} \\
Parameter & Input & $p_{2.5}$ & $p_{50}$ & $p_{97.5}$ & $p_{2.5}$ & $p_{50}$ & $p_{97.5}$ & $p_{2.5}$ & $p_{50}$ & $p_{97.5}$ \\
(1) & (2) & (3) & (4) & (5) & (6) & (7) & (8) & (9) & (10) & (11)\\
\hline\hline
$\log M_1 / (\Msunh)$ &  11.20 &  11.15 & 11.27 & 11.40 & 11.13 & 11.24 & 11.35 & 11.07 & 11.18 & 11.30  \\
$\log L_0 / (\Lsunh)$ &   9.95 &   9.91 &  9.99 & 10.07 &  9.87 &  9.94 & 10.02 &  9.89 &  9.96 & 10.03 \\
$\gamma_1$    &   3.50 &   3.02 &  3.45 &  3.87 &  3.08 &  3.49 &  3.91 &  3.04 &  3.45 &  3.87 \\
$\gamma_2$    &   0.25 &   0.20 &  0.23 &  0.26 &  0.23 &  0.25 &  0.28 &  0.23 &  0.25 &  0.27 \\
$\sigma_{12}$ &   0.15 &   0.13 &  0.15 &  0.18 &  0.14 &  0.17 &  0.19 &  0.12 &  0.15 &  0.17 \\
$\sigma_{14}$ &   0.15 &   0.14 &  0.17 &  0.19 &  0.13 &  0.14 &  0.16 &  0.13 &  0.15 &  0.17 \\
$\alpha_{12}$ &  -1.20 &  -1.61 & -1.28 & -0.94 & -1.21 & -0.74 & -0.32 & -1.01 & -0.69 & -0.37 \\
$\alpha_{14}$ &  -1.20 &  -1.37 & -1.16 & -0.90 & -1.37 & -1.16 & -0.84 & -1.35 & -1.11 & -0.82 \\
$b_0$         &  -1.20 &  -1.47 & -1.28 & -1.12 & -1.26 & -1.06 & -0.90 & -1.06 & -0.93 & -0.82 \\
$b_1$         &   1.50 &   1.41 &  1.62 &  1.86 &  1.39 &  1.60 &  1.85 &  1.24 &  1.41 &  1.59 \\
$b_2$         &  -0.20 &  -0.30 & -0.23 & -0.17 & -0.33 & -0.27 & -0.20 & -0.26 & -0.20 & -0.14 \\
$\eta_0$      &   1.00 &   0.81 &  0.96 &  1.18 &  0.49 &  0.58 &  0.71 &  0.40 &  0.48 &  0.58 \\
$\eta_1$      &   0.00 &  -0.24 & -0.13 &  0.01 & -0.20 & -0.06 &  0.07 & -0.30 & -0.16 &  0.02 \\
$\eta_2$      &   0.00 &  -0.98 &  0.96 &  1.93 & -1.73 &  1.23 &  1.95 & -0.52 &  1.33 &  1.96 \\
$\beta$       &   0.00 &  -0.13 &  0.05 &  0.22 & -0.13 &  0.18 &  0.44 &  0.07 &  0.27 &  0.46 \\
$\log[r_\rma/r_\rms]$ &--& 0.64 &  0.92 &  1.40 &  0.89  &  1.25  &  1.48  &  1.12 &  1.38 &  1.49 \\ 
\hline
\end{tabular}
\caption{Confidence intervals for posteriors of model parameters inferred from \Basilisk for the three tier mocks described in the text. Column (1) lists the model parameter, column (2) the input value used to construct the mock, and the remaining columns list the 2.5, 50 and 97.5 percentiles of the corresponding posterior distributions. Results for the Tier-1 mock are in columns (3)-(5), for the Tier-2 mock in columns (6)-(8), and for the Tier-3 mock in columns (9)-(11). Note that these percentiles all correspond to the MCMCs obtained using the \CA-model. Results for the \OM-model are extremely similar, and therefore not shown. However, the last row lists the 2.5, 50 and 97.5 percentiles for $\log[r_\rma/r_\rms]$ obtained from a separate MCMC that uses the \OM-model. Note that the input values for the effective bias parameters of the interlopers, $\eta_0$, $\eta_1$ and $\eta_2$, are only valid for the Tier-1 mock, while the input value for $\beta$ only applies to the Tier-1 and Tier-2 mocks.}
\label{tab:results}
\end{table*}

Panels (c) and (d) of Fig.~\ref{fig:mock1res} plot the additional data used to constrain the model: the former plots $f_0 = \log[N_0/N_{+}]$ (see \S\ref{sec:Pnull}) and the latter plots the luminosity function of all galaxies (centrals plus satellites). Note that $f_0$ is plotted as a function of the luminosity of the primary and for three  different redshift bins, as indicated. The same holds for the interloper fractions plotted in panel (e). The posterior predictions for $f_0(L_\rmc,z_\rmc)$,  $\Phi(L)$, and $f_{\rm int}(L_\rmc,z_\rmc)$ are all in excellent agreement with their true values. In particular, as is evident from Table~\ref{tab:results}, the posterior constraints on the parameters $\eta_0$, $\eta_1$, and $\eta_2$ that model the interlopers are in excellent agreement with their input values\footnote{The parameter $\eta_2$ which characterizes the redshift dependence of the effective bias of interlopers (see equation~\ref{beff}) is extremely poorly constrained. We find this to be true in all cases, and for all mocks. Although this implies that we may thus ignore a potential redshift dependence of the effective bias, we will continue to treat $\eta_2$ as a free parameter in what follows.}, further elucidating the success of \Basilisk. Finally, panel (f) plots the predicted halo mass, $M_{\rm pred}$, versus the true halo mass, $M_{\rm true}$, 
for each primary. The former is computed using
\begin{equation}\label{Mpred}
M_{\rm pred} = \int_{0}^{\infty} P(\Mh|L_\rmc,z_\rmc,N_\rms) \, \Mh \, \rmd\Mh\,,
\end{equation}
with $P(\Mh|L_\rmc,z_\rmc,N_\rms)$ given by equation~(\ref{ProbMass}), and serves as a latent variable in our hierarchical Bayesian framework. Errorbars reflect the 95 percent confidence intervals on $M_{\rm pred}$ as inferred from the MCMC, while the solid, red line indicates the running average. Although there is a small systematic bias, in that $M_{\rm pred}$ is too high (low) when $M_{\rm true}$ is small (large), the bias is small compared to the primary-to-primary variance; averaged over all 2373 primaries, we obtain $\langle \log(M_{\rm pred}/M_{\rm true}) \rangle = 0.12$, while the halo-to-halo scatter is $0.31$. This demonstrates that our inferences regarding the probability distribution $P(\Mh|L_\rmc,z_\rmc,N_\rms)$, which we marginalize over in our evaluation of the likelihood for the satellite kinematics data (cf. equation~[\ref{Mmarg}]), is not significantly biased.

\subsection{Tier 2: simulation-based mocks}
\label{sec:tier2}

The idealized Tier-1 mock discussed above is based on a number of unrealistic oversimplifications. First of all, it is assumed that all primaries are centrals, and that each central is selected as a primary (i.e., it is effectively assumed that purity = completeness = 100\%). In reality, though, the selection criteria are imperfect and some selected primaries will be satellites, giving rise to impurities. In addition, not all centrals will pass our selection criteria giving rise to sample incompleteness. Realistic redshift surveys also suffer from fibre collisions, which results in an additional, spectroscopic incompleteness that is correlated with local (projected) density. Another shortfall of the Tier-1 mocks is that they assume that interlopers are distributed randomly and uniformly in space, which ignores clustering and redshift space distortions. Finally, when constructing the Tier-1 mocks, we assume the same zero-scatter concentration-mass relation for dark matter haloes as used in the modeling; in reality, haloes have a fair amount of (roughly log-normal) scatter in concentrations \citep[e.g.,][]{Bullock.etal.01, Maccio.etal.07}.

To allow for all these complications, we construct our Tier-2 mocks using a high-resolution $N$-body simulation from which we construct a mock redshift survey that is similar to the SDSS DR7. Our Tier-2 mocks are based on the SMDPL simulation \citep {Klypin.etal.16}, which uses $3840^3$ particles to trace structure formation in a cubic volume of $(400 \mpch)^3$, adopting cosmological parameters that are compatible with the CMB constraints from \cite{Planck.14}. We use \texttt{halotools} \citep{Hearin.etal.16c} to populate dark matter haloes at $z=0.0$, identified with \texttt{ROCKSTAR}, according to our fiducial CLF model (see Table~\ref{tab:parameters}).

We populate each host halo with $M_{\rm vir} \ge 3 \times 10^{10} \Msunh$ with mock galaxies using the same method as outlined above for the Tier-1 mocks; i.e., we draw centrals and satellites from the CLF, all above a luminosity limit of $10^{8.5} \Lsunh$, and assign them phase-space coordinates within the host halo. In particular, each central galaxy is given the position and velocity of its halo core, defined as the region that encloses the innermost $10$ percent of the halo virial mass. These positions and velocities are calculated by \texttt{ROCKSTAR} as detailed in \citet{Behroozi.etal.13a}. For the satellites, we draw positions from a spherical distribution with radial profile $n_{\rm sat}(r|M)$, and one-dimensional velocities from a Gaussian distribution with dispersion $\sigma(r|M)$, given by equation~(\ref{sig2beta}) with $\beta=0$ (i.e., we assume isotropy). Both the positions and the velocities are with respect to the core of the host halo, and we use the measured concentration of each halo to determine individual halo profiles and satellite kinematics.
\begin{figure*}
\centering
\includegraphics[width=0.87\textwidth]{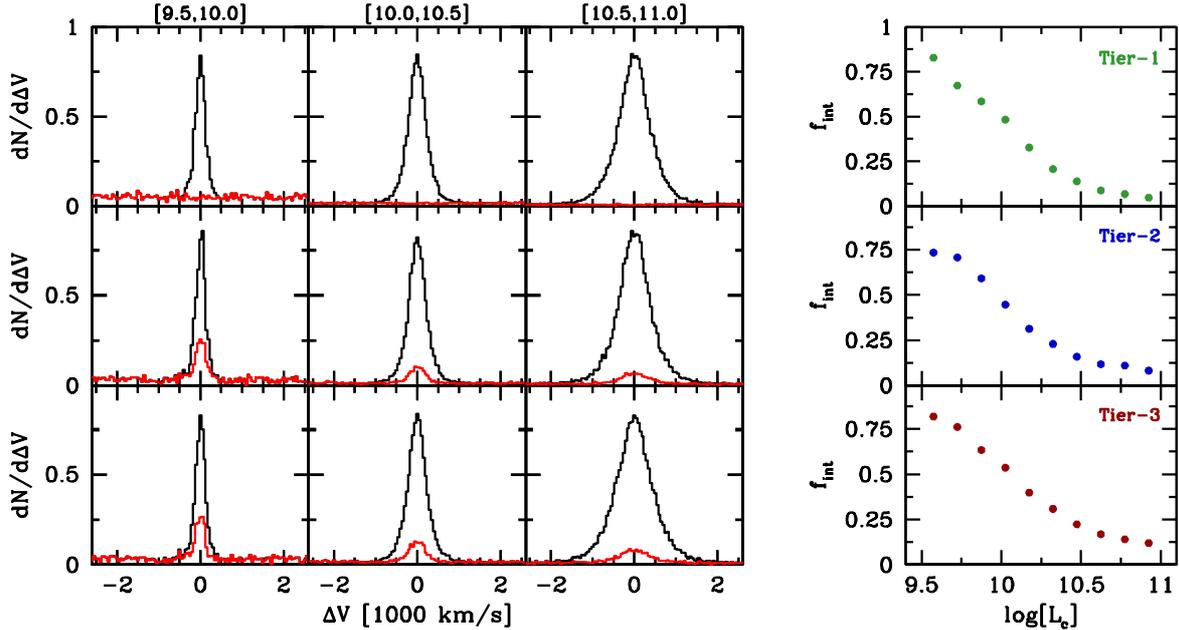}
\caption{Interlopers in the Tier-1 (upper panels), Tier-2 (middle row of panels) and Tier-3 (lower panels) mocks. Panels on the left show the velocity distributions of secondaries around primaries with $\log[L_\rmc]$ in the range indicated at the top of each columns. The contribution due to interlopers is marked in red. Right-hand panels plot the interloper fraction as function of the luminosity of the primary. }
\label{fig:fint}
\end{figure*}

In the next step, we simulate the SDSS observations, following the procedure outlined in \cite{Lange.etal.19a}. First, we place a virtual observer with a random position and orientation into the simulation volume. We use this virtual observer to convert the $(x, y, z)$ coordinates of each galaxy into sky coordinates plus a cosmological redshift. If necessary, the simulation box is repeated periodically until the entire cosmological volume out to $z = 0.17$ is filled. Next, we only keep galaxies with $m_r \leq 17.6$ that lie within the SDSS DR7 survey mask. Redshift-space distortions are simulated by adding $(1 + z) v_{\rm los} / c$ to each galaxy with cosmological redshift $z$ and line-of-sight peculiar velocity $v_{\rm los}$. A random redshift error from a Gaussian with scatter $\sigma_{\rm err} = 15\kms$ is added in order to simulate spectroscopic redshift errors in the SDSS \citep{Guo.etal.15b}. Finally, we simulate the effect of spectroscopic incompleteness. As discussed in \S\ref{sec:fc}, the SDSS suffers from fibre collisions whereby galaxies with a neighbour within $55''$ have a 65\% chance of not having a spectroscopic redshift. We first construct a decollided set of target galaxies \citep[][]{Blanton.etal.03a}, defined as galaxies without neighbouring targets within $55''$. We randomly assign 65\% of all galaxies that are not part of this decollided set a redshift, with the remaining 35\% making up our `fibre-collided' set (galaxies that lack a redshift due to fibre collisions). Finally, we randomly remove an additional $1\%$ of all redshift to simulate other redshift failures. As demonstrated in \citet{Lange.etal.19a} this approach captures all the salient features of spectroscopic incompleteness in the SDSS DR7. Once the mock is completed, we select primaries and secondaries as described in \S\ref{sec:selection}, and assign spectroscopic weights to all secondaries using the method described in \S\ref{sec:fc}.

\begin{figure*}
\centering
\includegraphics[width=\textwidth]{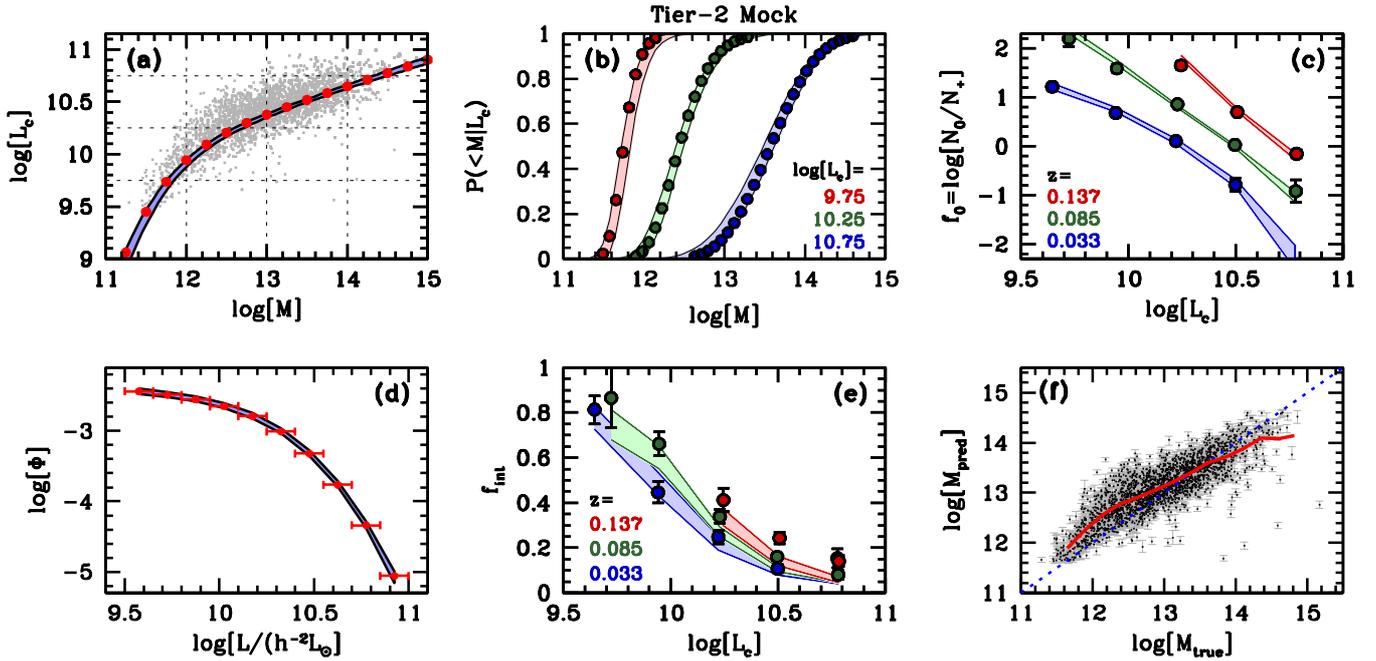}
\caption{Same as Fig.~\ref{fig:mock1res}, but for the Tier-2 Mock. Note that we have excluded secondaries with $\Rp < R_{\rm cut}(z_\rmc)$ from the analysis in order to avoid problems due to fibre-collisions; see text for details.}
\label{fig:mock2res}
\end{figure*}

We analyse the Tier-2 mocks in exactly the same fashion as the Tier-1 mocks described above. Note, though, that the spectroscopic weights, $w_{ij}$, used for the fibre collision correction are no longer unity, and that we use a non-zero cut-off radius, $R_\rmc(z_\rmc)$, equal to $55''$. Before showing the results from such an analysis, we first discuss some statistics of the mocks. A full Tier-2 mock, which mimicks the SDSS-DR7, contains $\sim 37,000$ secondaries around $\sim 19,000$ primaries (with at least one secondary). From such a mock, we construct eight subsamples of roughly equal size, whereby each primary plus its corresponding secondaries are randomly assigned to one of the subsamples. Since the CPU-cost for the MCMC analysis is proportional to the number of secondaries, analysing smaller subsamples is faster, and therefore ideal for testing. Furthermore, as is evident from Figs.~\ref{fig:mock1res} and~\ref{fig:mock2res}, samples with of order 5,000 secondaries already yield extremely tight constraints. This implies that we can easily subdivide the actual SDSS-DR7 data in of order 10 subsamples, and still achieve exquisite constraints on the galaxy-dark matter connection for each. These subsamples can be random, in which case we can test for consistency among them, or based on secondary properties of the primaries (i.e., split by color, size, bulge-to-disk ratio, etc.). The latter will allow for a much richer characterization of the galaxy-halo connection, and give valuable insight regarding galaxy assembly bias \citep[][]{Zentner.etal.14, Hearin.etal.16, Zentner.etal.19}. In what follows, we focus on one of the subsamples of our Tier-2 mock, consisting of $4,567$ secondaries around $2,373$ primaries. In addition, the subsample contains $23,234$ primaries with zero secondaries. Note that this is almost twice as many as for the Tier-1 mock with $5,000$ secondaries. As we will see below, this is mainly because the (more realistic) number density of interlopers in the Tier-2 mock is roughly half of that in the Tier-1 mock.

The left-hand panel of Fig.~\ref{fig:impurity} plots the completeness $C(M|L,z)$ in our Tier-2 mock, defined as the fraction of centrals of luminosity $L$ residing in haloes of mass $M$ at redshift $z$ that are selected as primaries by our cylindrical isolation criterion. The completeness is plotted as function of halo mass for 7 different luminosity bins (different colors, as indicated) and two different redshift bins (dashed and solid lines correspond to $z=[0.02,0.09]$ and $[0.09,0.15]$, respectively). In each case we plot  $C(M|L,z)$ over the 5 to 95 percentile range of the corresponding halo mass. Note that the completeness of centrals is fairly high, increasing from $\sim 65$ percent at $L_\rmc = 10^{9.6} \Lsunh$ to $\sim 90$ percent at  $L_\rmc = 10^{11} \Lsunh$. Most importantly, though, the completeness at given $L$ and $z$ has virtually no halo mass dependence. As discussed in \S\ref{sec:PLcen} this implies that we can simply ignore $C(M|L,z)$ altogether, or, equivalently, assume that it is equal to unity throughout.

The typical impurity in our Tier-2 mocks, defined as the fraction of primaries that are not centrals, is $\sim 5$ percent. The solid, blue curve in the right-hand panel of Fig.~\ref{fig:impurity} plots the impurity fractions, $f_{\rm imp}$, in our Tier-2 mock as a function of the central luminosity. As is evident, $f_{\rm imp}$ has little dependence on $L_\rmc$. For comparison, the blue, dashed curve shows the results for a similar Tier-2 mock in which we do not allow for satellites to be brighter than their central. In particular, if the luminosity of a satellite galaxy is drawn to be brighter than that of its central, the satellite is discarded \citep[see][for details]{Lange.etal.19a}. At the bright end, this results in a dramatic reduction of the impurity fraction, bringing the total impurity fraction of the mock sample to $0.5\%$. This indicates that most of the impurity arises from the fact that not all centrals are brightest halo galaxies. At the low-luminosity end, impurity mainly arises from the selection criterion used, which occasionally incorrectly identifies a satellite (typically in the outskirts of a more massive halo) as a primary. Analysing both Tier-2 mocks yields constraints on the galaxy-halo connection that are virtually indistinguishable, indicating that impurity at these levels does not significantly impact our results. This is in agreement with the conclusions reached by \cite{Lange.etal.19a,Lange.etal.19b},
and with the results presented below. 

Finally, it is illustrative to compare to properties of interlopers in our Tier-1 and Tier-2 mocks. In the former, these were `put in by hand' by drawing line-of-sight velocities and projected radii assuming that interlopers have a uniform phase-space distribution. In the latter, these arise from true projection effects in redshift space. The middle row of panels in Fig.~\ref{fig:fint} shows that the interlopers in the Tier-2 mocks have velocity distributions that differ notably from a purely uniform distribution. In particular, their $\dv$-distributions reveal a pronounced peak near $\dv=0$. This aspect of the phase-space distribution of interlopers has been pointed out in several previous studies \citep[e.g.,][]{vdBosch.etal.04, Wojtak.etal.07, Mamon.etal.10}. These have shown that these `peaks' are mainly due to satellite galaxies between one and two halo virial radii that are {\it bound} to the halo of the primary. Hence, their kinematics reflect the underlying gravitational potential of the host halo, and failing to identify them as `interlopers' is not expected to cause a significant bias in the inferred halo masses. This was also pointed out in \citet{vdBosch.etal.04} and will be confirmed below. The right-hand panels of Fig.~\ref{fig:fint} shows that the interloper fraction as function of central luminosity in the Tier-2 mock is very similar to that in the Tier-1 mock. 
\begin{figure*}
\centering
\includegraphics[width=0.95\textwidth]{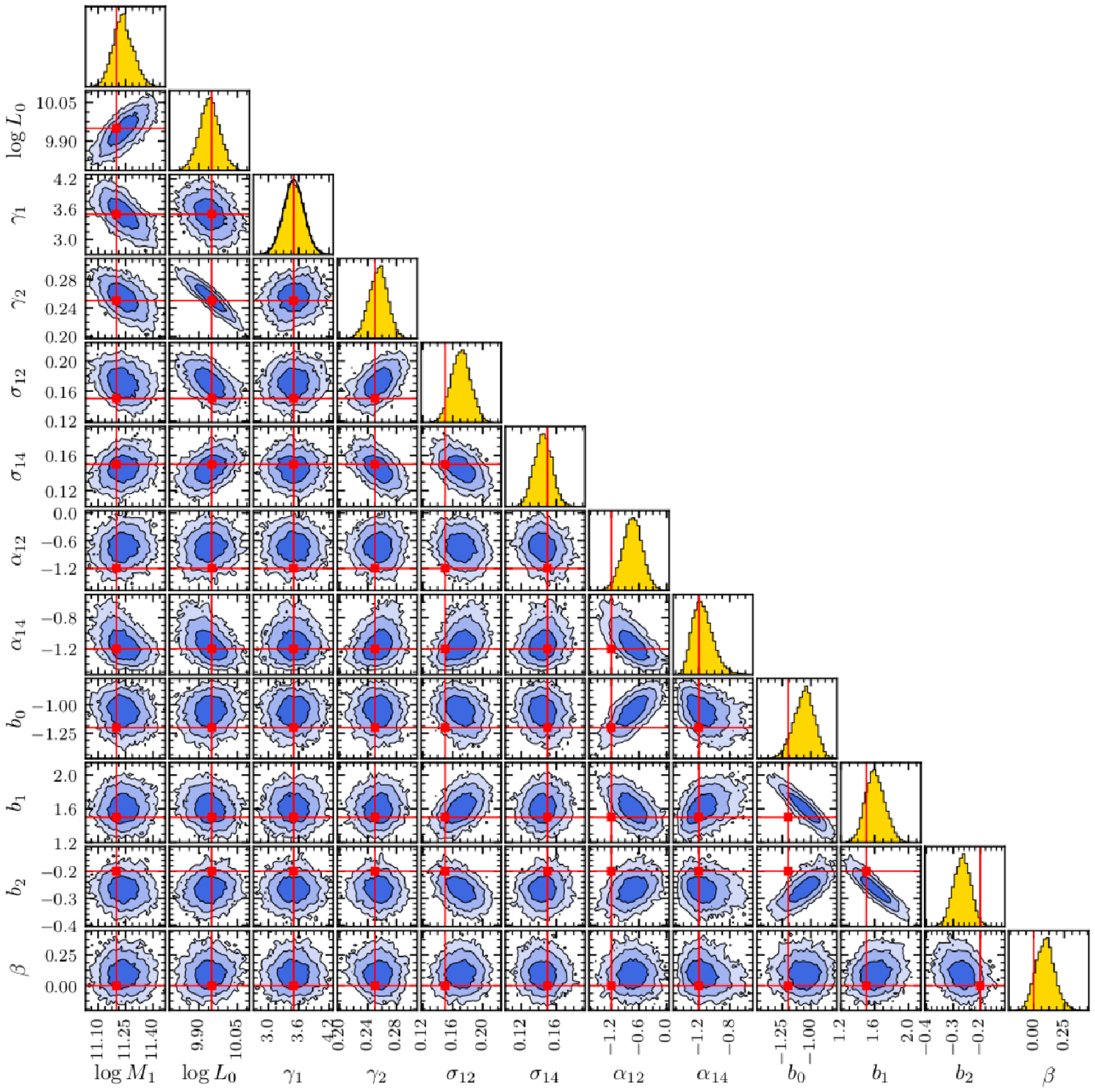}
\caption{Marginalized posteriors obtained by \Basilisk for the Tier-2 mock (assuming the best-fit radial profile for the satellites, with $\gamma=1$ and $\calR=1.34$). Results are shown for the 11 parameters that characterize the CLF, and for the anisotropy parameter, $\beta$. To avoid having panels that are too small, we do not show the nuisance parameters $\eta_0$, $\eta_1$, and $\eta_2$ that quantify the number density of interlopers. The diagonal shows marginalized 1D posteriors and off diagonal panels the 2D posteriors. In the latter case, contours demarcate the $68\%$, $95\%$ and $99\%$ containment of the posterior, while the red lines plus dot indicate the true input values used to create the mock data set. All parameters adopted a uniform prior, using the ranges indicated in Table~\ref{tab:parameters}, except for $\gamma_1$, which specifies the slope of the $\Lc$-$\Mh$ relation at the low-mass end, and which is poorly constrained. The Gaussian prior that was adopted instead is indicated with a thick, solid curve in the panel showing the 1D posterior for $\gamma_1$.}
\label{fig:posterior}
\end{figure*}
\begin{figure*}
\centering
\includegraphics[width=\textwidth]{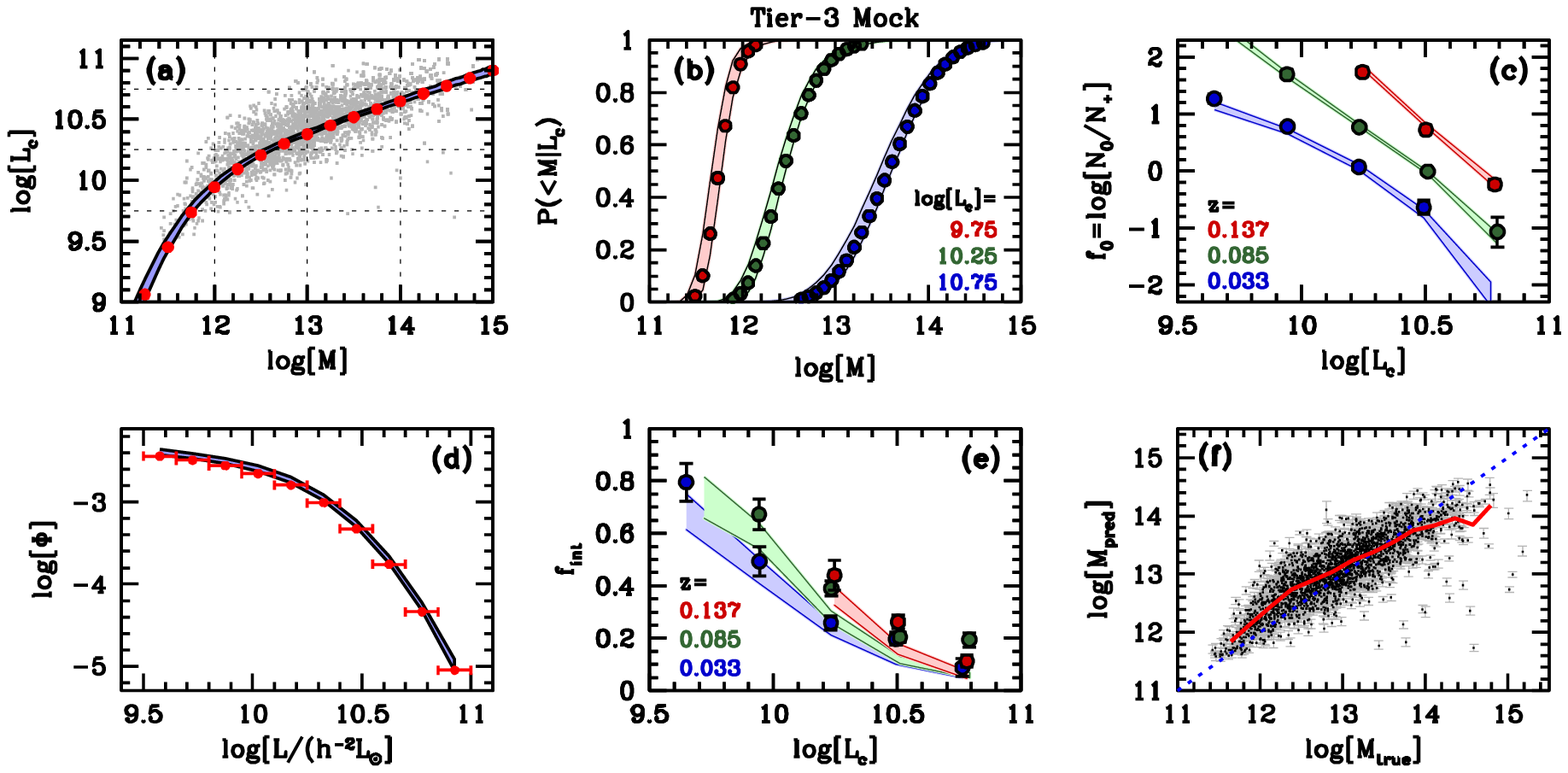}
\caption{Same as Fig.~\ref{fig:mock1res}, but for the Tier-3 Mock. Note that in this mock satellite galaxies follow the phase-space distribution of resolved subhaloes in the SMDPL simulation.}
\label{fig:mock3res}
\end{figure*}

The middle panel of Fig.~\ref{fig:contour} plots the constraints on $\calR$ and $\gamma$, marginalized over all other model parameters and obtained using the same method as for the Tier-1 mock. As is evident, the parameters are now less well constrained, and reveal a slight, systematic bias. In fact, the best-fit parameters are $\gamma = 1.0$ and $\calR = 1.34$. Note, though, that the true input-model, which has $\calR = \gamma = 1.0$ (indicated by the solid, black dot), falls well within the 95 percent confidence interval. The main reason for the reduced sensitivity is the fact that, as part of the mitigation of fibre-collisions, we exclude all secondaries separated from their primaries by less than $55''$.  Especially for fainter primaries at higher redshifts, this corresponds to an appreciable fraction of the halo's virial radius, and it should not come as a surprise that this significantly diminishes the data's potential to yield precise constraints on the phase-space distribution of satellite galaxies. However, this should not necessarily result in a significant, systematic bias, which instead arises from the combined effect of interlopers and impurity.  Our model does not account for the excess of interlopers at small $|\dV|$. Since these follow a radial profile that is more extended than that of the true satellites, this causes a systematic bias in $\calR$. And the same applies for impurities: the radial distribution of secondaries around primaries that in reality are satellites is less centrally concentrated than that around true centrals. Fortunately, as we demonstrate below and in more detail in \S\ref{sec:profile}, the resulting, systematic overestimate of $\calR$ does not have a significant impact on the inference regarding the galaxy-dark matter connection, which is our prime objective.

	
Fig.~\ref{fig:mock2res} shows the posterior results from the MCMC analysis of our fiducial Tier-2 subsample (see Table~\ref{tab:results} for the medians and 95 percent confidence intervals for all model parameters). Results for other subsamples are extremely similar. Here we have adopted the \CA-model with the best-fit $n_{\rm sat}(r|M)$ (i.e.,$\gamma = 1.0$ and $\calR = 1.33$). As for the Tier-1 mock, we infer the posterior constraints from a MCMC consisting of 5 million elements. A comparison with Fig.~\ref{fig:mock1res} shows that the posterior constraints are remarkably similar, despite the many additional complications that come with a Tier-2 mock and the fact that the value of $\calR$ adopted differs from that used to create the mock. In particular, the method yields constraints on $P(L_\rmc|\Mh)$ that are in good agreement with the input, both in terms of the median (panel a), and in terms of the full distributions (panel b).  As for the Tier-1 mock, the posterior predictions for $f_0$ (panel c) and the luminosity function (panel d) are in excellent agreement with the data. However, the predictions for the interloper fractions (panel e) are slightly too low. This is due to the fact that the model assumes a uniform velocity distribution for interlopers, which does not account for the central `peak' near $\Delta V = 0$ (Fig.~\ref{fig:fint}). Most importantly, though, other than a systematic underestimate of the number density of interlopers, which is merely a nuisance parameter, our oversimplified treatment of interlopers does not result in a systematic error in the inferred galaxy-halo connection. Finally, as is evident from panel (f), the predicted halo masses, which act as latent variables in our hierarchical Bayesian framework, are again in good agreement, in a statistical sense, with the true masses with $\langle \log(M_{\rm pred}/M_{\rm true}) \rangle = 0.15$ and a halo-to-halo scatter of $0.35$. Note that there are a small number of clear outliers, for which $M_{\rm pred} \ll M_{\rm true}$. These systems, which are absent in the Tier-1 mock, correspond to impurities in which the primary is a satellite in the outskirts of a massive halo. At $\sim 5$ percent the impurity fraction is sufficiently low that this does not notably impact our overall inference.

To get some insight as to how the various parameters are correlated, Fig.~\ref{fig:posterior} plots one- and two-dimensional marginalized posterior distributions, but only for the 11 parameters that characterize the CLF and the anisotropy parameter $\beta$. In order to avoid having the panels be too small, we do not show the results for the three (nuisance) parameters that characterize the effective bias of the interlopers ($\eta_0$, $\eta_1$, and $\eta_2$)\footnote{Neither of these parameters reveals significant covariance with any of the parameters shown in Fig.~\ref{fig:posterior}.}. The yellow histograms along the diagonal show the marginalized 1D posteriors, while the off-diagonal panels show marginalized 2D distributions, with the contours demarcating the $68\%$, $95\%$ and $99\%$ containment of the posterior. Red lines indicate the true input values used to create the mock data set. In general, the posteriors are in better agreement with the input values for the parameters that characterize the central part of the CLF ($M_1$, $L_0$, $\gamma_1$, $\gamma_2$, $\sigma_{12}$, $\sigma_{14}$), than for those that characterize the satellite component ($\alpha_{12}$, $\alpha_{14}$ $b_0$, $b_1$, $b_2$). This is true for all subsamples we examined, and also holds for the Tier-3 mock discussed below: whereas \Basilisk always yields constraints on $\Phi_\rmc(L|M)$ in excellent agreement with the input, often it will yield constraints on one or more parameters characterizing $\Phi_\rms(L|M)$ that are inconsistent with the input values. This owes to the impurities and the oversimplified treatment of interlopers (i.e., uniform velocity distribution) that is also responsible for the small but systematic error in $n_{\rm gal}(r|M)$.  
We suspect that the inference regarding the parameters of the satellite component of the CLF may significantly improve if the observed luminosities of the secondaries are included as data in our likelihood evaluation. Since the primary goal of analyzing satellite kinematics is to constrain the halo masses of central galaxies, we leave such an extention for future work.


Note how all 1D posterior distributions depicted in Fig.~\ref{fig:posterior} closely resemble Gaussians, and how most parameters are only weakly covariant with one another. Some notable exceptions are the sets $\{\log M_1, \log L_0, \gamma_1, \gamma_2\}$ and $\{b_0, b_1, b_2\}$, which reveal non-negligible covariance among each other \citep[see also][]{Cacciato.etal.13}. These mainly owe to the limiting dynamic range in halo mass covered by the (mock) data; especially the lack of data sampling the low mass end ($M \lta 10^{12} \Msunh$). Mainly for this reason we decided to adopt a restricting, Gaussian prior for $\gamma_1$, which is poorly constrained by the data. Indeed, as is evident from Fig.~\ref{fig:posterior}, the posterior distribution of $\gamma_1$ is in perfect agreement with its prior, indicated by the solid, black curve.

To summarize, we conclude that neither impurities, nor the complicated phase-space distribution of interlopers, nor fibre-collisions, nor realistic scatter in the concentration-mass relation (which is ignored in our modeling) prohibit \Basilisk from inferring accurate and precise constraints regarding the galaxy-halo connection of {\it central} galaxies. Although the aforementioned complications can cause a slight systematic bias in the inferred radial profile of satellite galaxies, and in the parameter $\alpha_{12}$, these do not significantly impact the inference regarding $\Phi_\rmc(L|M)$. As a cautionary note, though, we find that not including a treatment for fibre-collisions, i.e., not applying the correction factor, $f_{\rm corr}$, given by equation~[\ref{fcorr}], and not removing secondaries with $R_\rmp < R_\rmc(z_\rmc)$, typically results in significant biases that are very
similar to the biases one incurs when not accounting for the number of secondaries (see Appendix~\ref{App:noNsat}).

\subsection{Tier 3: using subhaloes to model the phase-space distribution of satellites}
\label{sec:tier3}

At the third and final stage of our validation, we move to mock data sets in which we no longer make assumptions regarding the phase-space distributions of satellite galaxies. We construct Tier-3 mocks exactly as in the case of Tier-2 mocks, but rather than assuming a radial profile, $n_{\rm sat}(r|M,z)$, or assuming that satellite galaxies obey the spherical Jeans equations, we now assume that satellite galaxies reside in subhaloes, and their phase-space distribution therefore follows that of resolved subhaloes in the SMDPL $N$-body simulation.

After drawing a satellite number $N_{\rm sat}$ for each halo, we assign those $N_{\rm sat}$ satellites the phase-space coordinates of the $N_{\rm sat}$ subhaloes with the highest $M_{\rm peak}$. It is possible that $N_{\rm sat}$ exceeds the number of resolved subhaloes in a specific halo. In that case, we randomly take phase-space positions of subhaloes hosted by other haloes of a similar mass. We then proceed to generate a mock SDSS-like catalogue and analyse it in the same way as the Tier-2 mock described in the previous section. In particular, we select a random subsample  containing $4,521$ secondaries around $2,431$ primaries, while an additional $23,360$ primaries have zero secondaries. We have verified that the results for other random subsamples of similar size are again extremely similar.

As is evident from Figs.~\ref{fig:impurity} and~\ref{fig:fint} the impurity and interloper contamination in the Tier-3 mock are very similar to those in the Tier-2 mock. The overall impurity is $5.2\%$, while the interloper fraction decreases from $\sim 80$\% for primaries with $L_\rmc = 10^{9.5} \Lsunh$ to $\sim 10$\% for the brightest primaries.

The right-hand panel of Fig.~\ref{fig:contour} shows that the best-fit radial profile for the satellites is very different from that of the Tier-1 and Tier-2 mocks. The data clearly prefers a large, constant density core with $\gamma=0$ and $\calR = 2.37$ (close to the limits on our adopted prior ranges). For comparison, the true, radial profile of satellite galaxies in the Tier-3 mock is best-fit with a gNFW profile with $\gamma=0$ and $\calR=2.57$ (indicated with a solid, black dot). This is not only in good agreement with the best-fit value inferred from \Basilisk, but also with the well-known fact that the radial distribution of dark matter subhaloes in dark-matter only simulations, such as the SMDPL simulation used here, is strongly anti-biased with respect to the dark matter particles \citep[e.g.,][]{Springel.etal.08}\footnote{This is, at least partially, an artifact due to artificial numerical disruption \citep[][]{vdBosch.etal.18a, vdBosch.etal.18b}.}. We analyze the Tier-3 mock, again using a MCMC with 5 million elements, keeping $\gamma$ and $\calR$ fixed at these best-fit parameters. The results are shown in Fig.~\ref{fig:mock3res}, while  Table~\ref{tab:results} lists the medians and 95 percent confidence intervals for all model parameters. The posterior constraints for the galaxy-halo connection of central galaxies, as characterized by $\Phi_\rmc(L|M)$, are once again in excellent agreement with the input model. In particular, the posteriors accurately reflect the full PDF of the relation between central luminosity and halo mass. Similar to what was found for the Tier-2 mock, the inferred fraction of interlopers is somewhat too low, while the predicted halo masses are in good agreement with the true masses, with $\langle \log(M_{\rm pred}/M_{\rm true}) \rangle = 0.10 \pm 0.37$.
Results for other subsamples are very similar. Hence, we conclude that \Basilisk is extremely reliable and robust in constraining the galaxy-halo connection of centrals, despite the oversimplified assumptions that satellite galaxies reside in spherical NFW haloes, obey the Jeans equation, and have Gaussian LOSVDs.

\section{Constraining the Phase-Space Distribution of Satellite Galaxies}
\label{sec:constrain}

In this section we discuss the accuracy and precision with which \Basilisk can constrain the orbital anisotropy of satellite galaxies, and we investigate how central velocity bias and errors in the  radial profile of satellite galaxies impact \Basilisk's inference regarding the galaxy-halo connection.

\subsection{The orbital anisotropy of satellite galaxies}
\label{sec:anisotropy}

\begin{figure}
\centering
\includegraphics[width=0.46\textwidth]{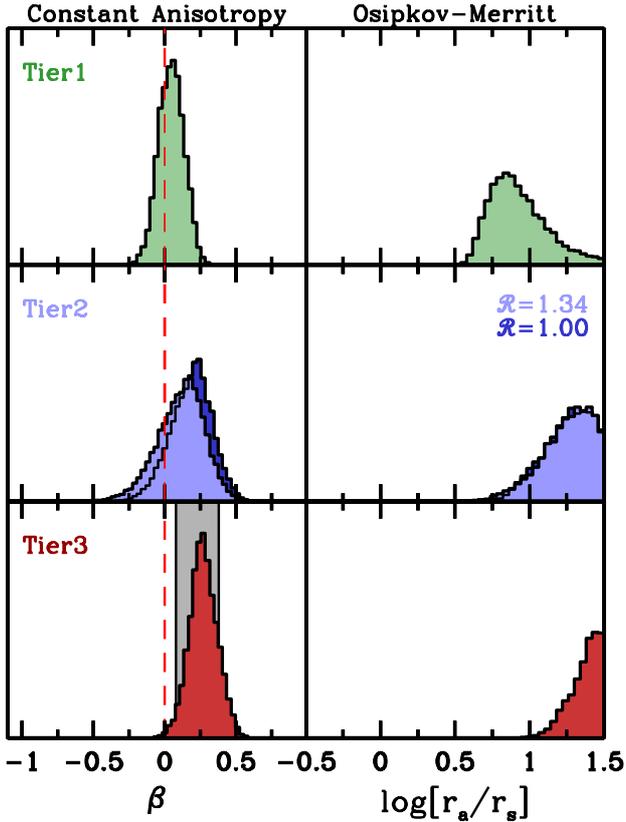}
\caption{Posterior distributions (normalized) for the anisotropy parameters inferred from the three tier mocks (different rows) for both the \CA-model (left) and the \OM-model (right). The Tier-1 and Tier-2 mocks were constructed with isotropic orbital distributions, corresponding to $\beta=0$ (indicated by the vertical, dashed line in the left-hand panels). For the Tier-2 mock, we show the posterior distributions for both our fiducial analysis, which adopts the best-fit value for $\calR$ (light-blue color), and for an analysis that adopts the true input value  of $\calR=1.0$ (dark-blue color). The grey, vertical bar in the lower-left panel indicates the typical range of anisotropy parameters for subhaloes in the SMDPL simulation (see Appendix~\ref{App:anisotropy}). In the Tier-3 mock, satellite galaxies are placed on subhaloes, and the grey bar therefore indicates the true, underlying anisotropy of the satellite galaxies in this mock.}
\label{fig:anisotropy}
\end{figure}

Fig.~\ref{fig:anisotropy} shows the posterior distributions inferred from the mocks from all three of the tiers. Left and right-hand panels show the results obtained from MCMCs in which we use the constant anisotropy (\CA) model and the Osipkov-Merritt (\OM) model, respectively. The upper panels correspond to Tier-1 and reveal posteriors in good agreement with the isotropy assumed, i.e., $\beta = 0$, which corresponds to a large $r_\rma/r_\rms$ in the case of an \OM-model. As indicated in Table~\ref{tab:results}, \Basilisk yields $\beta = 0.05^{+0.17}_{-0.18}$ and $\log[r_\rma/r_\rms] = 0.92^{+0.48}_{-0.28}$. The uncertainties indicate the 95\% confidence intervals centered on the medians. Note that the posterior distribution for  $\log[r_\rma/r_\rms]$ is restricted by our assumed prior, which is uniform over the range $[-1.0,1.5]$. The middle row of panels show the results for the Tier-2 mock, for two values of $\calR$; dark-blue histograms correspond to $\calR=1.0$, which is the true value of the mock, while the light-blue histograms correspond to the best-fit value of $\calR=1.34$. Both distributions are consistent with each other, and with the isotropic model used to construct the mock data (cf., Table~\ref{tab:results}). This indicates that the systematic error in the inferred radial profile of the satellite galaxies, which owes to interlopers and impurities in the sample, does not have a significant impact on the inferred anisotropy. Finally, in the case of the Tier-3 mock (lower panels), the posterior distribution of $\beta$ is only marginally consistent with isotropy; rather the data seems to prefer a mildly, radially anisotropic distribution with $\beta = 0.27_{-0.20}^{+0.19}$ (95\% CL). Recall that in the case of the Tier-3 mock we did not impose any orbital anisotropy, which instead derives from that of subhaloes in the SMDPL simulation. Interestingly, the \OM-model, which is isotropic in the center and becomes radially anisotropic for $r > r_\rma$, prefers a {\it large} value for the anisotropy radius; $\log[r_\rma/r_\rms] = 1.38^{+0.11}_{-0.26}$. 

For the Tier-1 and Tier-2 mocks, the best-fit \CA- and \OM-models have comparable values for $\chi^2_{\rm tot}$, with $|\Delta\chi^2_{\rm tot}| < 2$.  Hence, the data does not significantly prefer one anisotropy model over the other, which is to be expected given that the underlying model is isotropic and the \OM-model can be made isotropic by setting the anisotropy radius sufficiently large. In the case of the Tier-3 model, the \CA-model yields a significantly better fit than the \OM-model, with $\Delta\chi^2_{\rm tot} = \chi^2_{\rm tot,\CA} - \chi^2_{\rm tot,\OM} = 8.6$. As we demonstrate in Appendix~\ref{App:anisotropy}, the orbital anisotropy of dark matter subhaloes in the SMDPL simulation has a strong mass dependence \citep[see also][]{Cuesta.etal.08}. Whereas the average anisotropy is well described by $\beta \sim 0.2$, in excellent agreement with the constraints shown in the lower-left panel of Fig.~\ref{fig:anisotropy}, the anisotropy parameter has a strong mass and radius dependence. In massive hosts, with $M_{\rm vir} \gta 10^{13} \Msunh$, the anisotropy parameter $\beta$ increases with increasing radius, in good agreement with the findings by \citet{Diemand.Moore.Stadel.04}. However, in less massive haloes the anisotropy parameter is found to {\it decrease} with increasing radius \citep[see also][who studied the orbital anisotropy of Milky-Way sized host haloes]{Sawala.etal.17}. Since a negative radial gradient in $\beta$ is inconsistent with an Osipkov-Merrit model, for which $\rmd\beta/\rmd r > 0$, this explains why the best-fit \OM-model for the Tier-3 mock is significantly worse than for the \CA-model. 
\begin{figure*}
\centering
\includegraphics[width=0.85\textwidth]{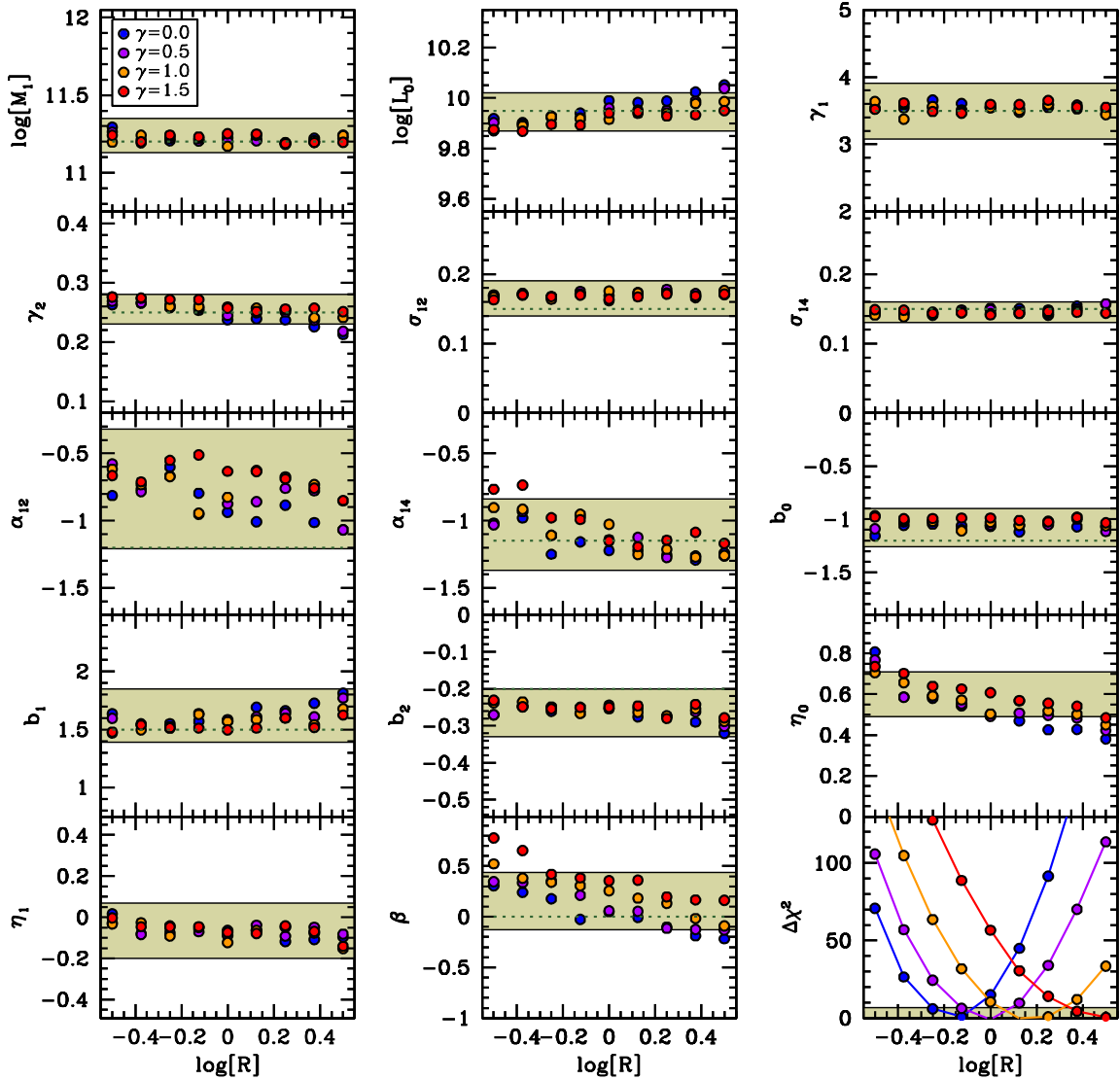}
\caption{The impact of changes in $\calR$ and $\gamma$, characterizing $n_{\rm sat}(r|M)$, on the inference of \Basilisk. Each panel plots the best-fit values of a different CLF parameter as a function of $\calR$, with different colors corresponding to different values of $\gamma$ (as indicated in the upper-left panel). The shaded regions indicate the 95 percent confidence intervals obtained in the case of our fiducial MCMC that adopts the best-fit parameters for $\calR$ and $\gamma$ (filled pentagon in the middle panel of Fig.~\ref{fig:contour}), while the horizontal, dotted lines mark the input values used to construct the mock data. The lower-right panel plots $\Delta\chi^2$, defined as the difference in the $\chi^2$ of the best-fit model compared to the overall best-fit (corresponding to the fiducial model), with the shaded region indicating $\Delta\chi^2 < 6.17$, which corresponds to the 95.4 percent confidence interval for a $\chi^2$-distribution with two degrees of freedom. Note that changes in $\calR$ and $\gamma$ have only a minor impact on the best-fit CLF parameters.}
\label{fig:dependence}
\end{figure*}

\subsection{The radial profiles of satellite galaxies}
\label{sec:profile}

As shown in \S\ref{sec:tier2}, the combination of fibre collision incompleteness, sample impurities and interlopers can cause a systematic error in the inferred radial profile of the satellite galaxies, as characterized by $\calR$ and $\gamma$. However, in the case of the Tier-2 mock,  we also found that the MCMC that adopts these biased, best-fit parameters yields constraints on the CLF parameters and orbital anisotropy that are not significantly biased, and thus that this does not have a significant impact on our inference regarding the galaxy-halo connection. 
To further gauge the impact of incorrect radial profiles, we use the downhill-simplex method to compute the best-fit models for the same Tier-2 subsample as used in \S\ref{sec:tier2}, but for different values of $\calR$ and $\gamma$. The results are shown in Fig.~\ref{fig:dependence}, where different panels plot the various best-fit parameters as function of $\calR$, for four different values of $\gamma$, as indicated.  As is evident, changing $\calR$ and/or $\gamma$ causes changes in the best-fit parameters that are small compared to the 95 percent confidence intervals of the fiducial model, shown as shaded regions.  Hence, even when the errors in $\calR$ and/or $\gamma$ are large, this has very little impact on the inference regarding the galaxy-halo connection. The impact is especially small for those parameters that characterize the halo occupation statistics of centrals. The parameters that reveal the largest dependence on $(\calR,\gamma)$ are the slope of the satellite luminosity function, characterized by $\alpha_{12}$ and $\alpha_{14}$, the normalization of the effective bias of the interlopers, $\eta_0$, and the anisotropy parameter $\beta$. In each case, though, the dependencies remain weak compared to the posterior uncertainties. Note, however, that the $\chi^2$ value of the best-fit model (lower-right panel) depends very strongly on $(\calR,\gamma)$, which is why we were able to obtain tight, albeit slightly biased, constraints on these parameters (cf. Fig.~\ref{fig:contour}). 
\begin{figure*}
\centering
\includegraphics[width=0.85\textwidth]{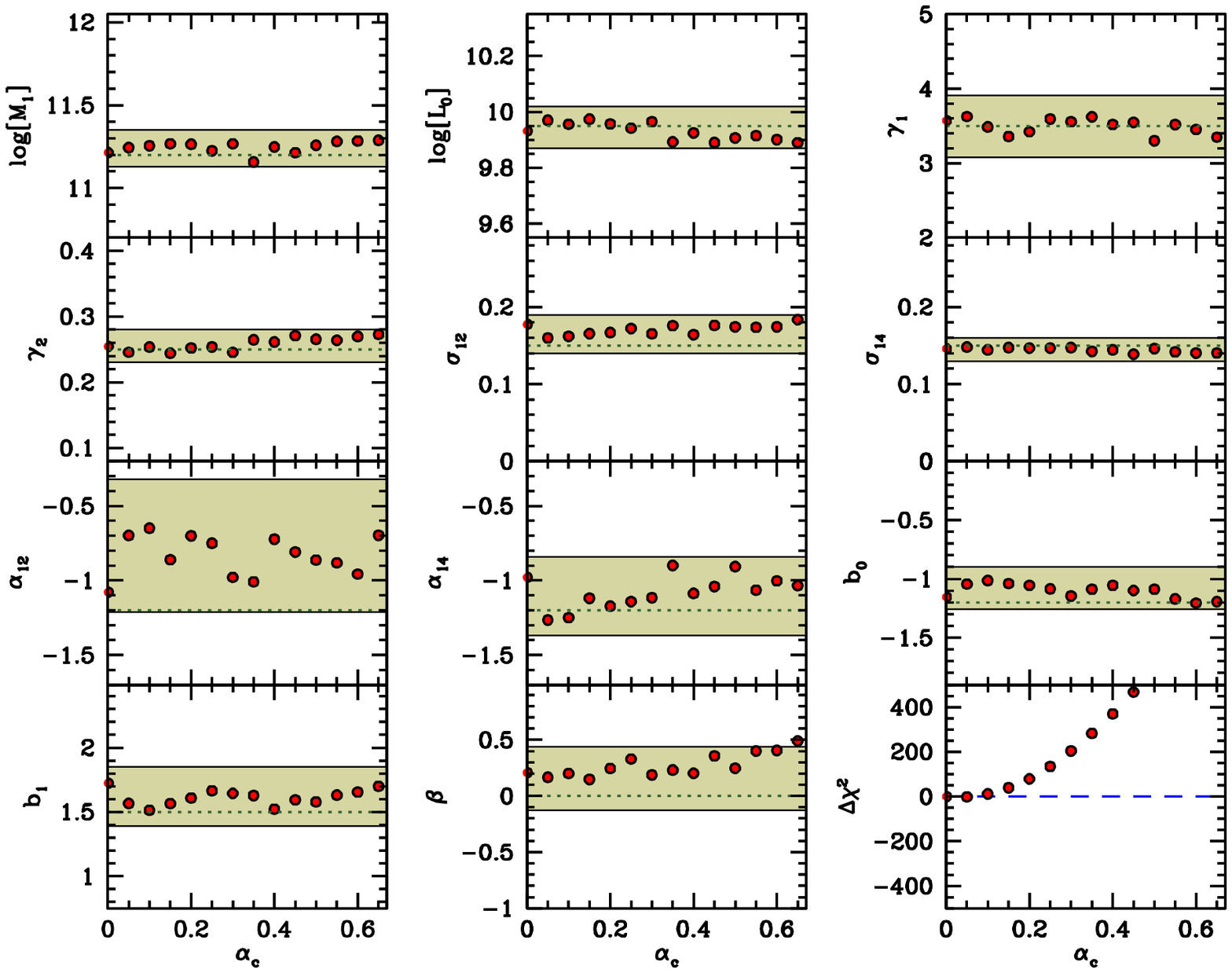}
\caption{Same as Fig.~\ref{fig:dependence}, but now as function of the central velocity bias, $\alpha_\rmc$.}
\label{fig:cenbias}
\end{figure*}

It may seem surprising that errors in $\calR$ and/or $\gamma$ have so little impact; after all, the radial profile of the satellite galaxies enters the Jeans equation that is used to compute the radial velocity disperion, $\sigma_r(r|M,z)$ (equations~[\ref{sig2beta}] and~[\ref{sig2OM}]). The main reason is that the line-of-sight velocity dispersion, which is used to compute the probability $P(\dV|\Rp,\Mh,z)$, is less sensitive to changes in $n_{\rm sat}(r|M,z)$ than the radial velocity dispersion. To first order, these changes are degenerate with (relatively modest) changes in orbital anisotropy and interloper contribution. Interestingly, though, there seems to be virtually no degeneracy with the parameters that characterize $\Phi_\rmc(L|M)$. Such degeneracy is inhibited by the additional constraints that we use in our inference (i.e., the galaxy luminosity function and the fraction of primaries with zero secondaries). Hence, we conclude that although fibre collisions, impurity, and an oversimplified treatment of interlopers can cause a slight, systematic error in \Basilisk's inference of the radial distribution of satellite galaxies, this has no discernible impact on the inference regarding the halo occupation statistics of central galaxies. And although it may impact the constraints on the velocity anisotropy, unless the systematic errors on $\gamma$ and/or $\calR$ are very large, which is unlikely to be the case, the magnitude of this effect is small compared to the typical statistical uncertainty.

\subsection{Central Velocity Bias}
\label{sec:cenbias}

Throughout we have assumed that central galaxies reside at rest at the center of the host halo. However, as pointed out in a number of studies \citep[]{vdBosch.etal.05b, Reid.etal.14, Guo.etal.15a, Guo.etal.15b, Guo.etal.16, Ye.etal.17}, this is not necessarily the case. Several studies have investigated this so-called central velocity bias in hydrodynamical simulations of galaxy formation \citep[e.g.,][]{Berlind.etal.03, Yoshikawa.etal.03}. More recently, \citet{Ye.etal.17} used hydrodynamical simulations from the Illustris-1 suite \citep[]{Vogelsberger.etal.14} to examine the motion of central galaxies with respect to the bulk velocity of their host haloes. They defined the central velocity bias
\begin{equation}
\alpha_\rmc \equiv \sqrt{\left\langle \frac{|\bv_\rmc - \bv_\rmh |^2}{\sigma^2_\rmh} \right\rangle}\,,
\end{equation}
where $\bv_\rmc$ is the velocity of the central galaxy, $\bv_\rmh$ is the centre-of-mass velocity of the host halo, $\sigma_\rmh$ is the 3D velocity dispersion of the dark matter particles of the host halo, and the angle brackets indicate an average over an ensemble of central galaxies. If central galaxies are at rest with respect to their host haloes, then $\alpha_\rmc = 0$ (`no bias'). \citet{Ye.etal.17} find that centrals in the Illustris-1 simulation have an average central velocity bias $\alpha_\rmc \sim 0.2$. In fact, $\alpha_\rmc$ depends strongly on halo age and the stellar mass-to-host halo mass ratio, $M_\rmc/M_\rmh$, with younger haloes and smaller $M_\rmc/M_\rmh$ corresponding to larger velocity bias, reaching values for $\alpha_\rmc$ as large as $0.6$.

If we assume that the average velocity (but not necessarily the rms velocity) of subhaloes coincides with that of the bulk of the host halo, then a non-zero central velocity bias will boost the root-mean-square velocity difference $\sqrt{\langle \dV^2 \rangle}$ by roughly a factor $\sqrt{1.0 + \alpha^2_\rmc}$, which can systematically bias the inference of \Basilisk. For $\alpha_\rmc=0.2$ the boost is only $\sim 1.02$, but since it is systematic and since, to good approximation, $\Mh \propto \langle \dV^2 \rangle^{3/2}$, the impact might be non-negligible. In order to test this, we proceed as follows. For each central galaxy, $i$, in a mock sample we first draw a random line-of-sight velocity, $\delta V_{\rmc, i}$, from a Gaussian with a velocity dispersion $\sigma_\rmh = V_{\rm vir}/\sqrt{2}$, with $V_{\rm vir}$ the halo's virial velocity. This approximation of the halo's 1D velocity dispersion is accurate to a few percent for NFW haloes with concentration parameters $c \sim 10$ \citep[]{Lokas.Mamon.01}. Next, for each central-satellite pair, we change the central-satellite velocity difference $\Delta V_{ij} \rightarrow \Delta V_{ij} + \alpha_\rmc \, \delta V_{\rmc,i}$, and we use the downhill-simplex method to find the best-fit parameters given that modified set of data. We repeat this exercise for different values of $\alpha_\rmc$, each time using the same $\delta V_{\rmc, i}$ in order to suppress the impact of realization noise. 

The resulting best-fit parameters as a function of $\alpha_\rmc$, for the same Tier-2 subsample as used in \S\ref{sec:tier2}, are shown as solid dots in Fig.~\ref{fig:cenbias}. The shaded regions indicate the 95 percent confidence intervals obtained in the case of zero velocity bias, while the horizontal, dotted lines mark the input values used to construct the mock data. The lower-right panel plots $\Delta \chi^2 \equiv \chi^2_{\rm tot}(\alpha_\rmc) - \chi^2_{\rm tot}(\alpha_\rmc=0)$, with $\chi^2_{\rm tot}$ the $\chi^2$-value of the best-fit model. For $\alpha_\rmc \lta 0.4$, central velocity bias has a negligible impact on the inference, with best-fit parameters that are almost indistinguishable from the case with no velocity bias. Even for $\alpha_\rmc = 0.6$, corresponding to the largest values found in the Illustris-1 simulation suite, the impact of the central velocity bias is extremely modest, in that the differences in the best-fit parameters remain small compared to the 95 percent confidence intervals.  The main impact of a non-zero velocity bias is a decrease in the goodness-of-fit, which is evident from the rapid increase of $\Delta \chi^2$ with increasing $\alpha_\rmc$. Together with the fact that $\alpha_\rmc$ does not display a significant degeneracy with any of the other model parameters, this therefore suggest that it should even be possible to include $\alpha_\rmc$ as a free parameter in our model, without significantly impacting \Basilisk's ability to constrain the galaxy-halo connection.

\section{Discussion and Conclusions}
\label{sec:conc}

As part of our ongoing efforts to mature satellite kinematics into an accurate, competitive probe of the galaxy-halo connection, complementary to galaxy clustering and galaxy-galaxy-lensing, this paper presents a new Bayesian hierarchical method for analyzing the kinematics of satellite galaxies. \Basilisk uses the spherically symmetric Jeans equations to model the kinematics of large ensembles of satellite galaxies associated with central galaxies that span a wide range in halo mass and luminosity (or stellar mass). The halo masses of the individual centrals act as latent variables in a hierarchical Bayesian framework that uses data from galaxy redshift surveys to constrain the detailed galaxy-halo connection, characterized by the conditional luminosity function. 

Unlike traditional methods for analysing satellite kinematics, \Basilisk does not resort to stacking the kinematics of satellite galaxies in bins of central luminosity, and does not make use of any summary statistic, such as satellite velocity dispersion. Rather, \Basilisk leaves the data in its raw form, which has the advantage that all data is used optimally. In addition, whereas traditional methods typically require volume-limited samples, \Basilisk can be applied to flux limited samples, thereby greatly enhancing the quantity and dynamic range of the data. And finally, \Basilisk is the only available method that simultaneously solves for halo mass and orbital anisotropy of the satellite galaxies, while properly accounting for `mass-mixing'.

Starting from a sample of primary and secondary galaxies selected from a galaxy redshift survey, representing centrals and satellites, respectively, \Basilisk uses the projected phase-space coordinates of the secondaries with respect to their primaries to constrain the galaxy-halo connection under the following assumptions:
\renewcommand{\labelenumi}{\bf \Roman{enumi}}
\begin{enumerate}
\item Dark matter halos are characterized by spherical NFW density profiles that follow the concentration-mass relation of \citet{Maccio.etal.08} with zero scatter.
\item Satellite galaxies are a virialized, steady-state tracer population of their underlying host halo potential. 
\item The LOSVD, $P(\dV|\Rp,M,z)$, is a Gaussian with zero mean and a dispersion, $\sigma_{\rm los}$, that follows from the spherical Jeans equations.
\item Interlopers have a uniform distribution in redshift space.
\end{enumerate}

Each of these assumptions is expected to be violated at some level. Dark matter haloes are triaxial, rather than spherical \citep[e.g.,][]{Jing.Suto.02, Allgood.etal.06}, and their concentration-mass relation relation has an appreciable amount of (log-normal) scatter \citep[e.g.][]{Bullock.etal.01, Maccio.etal.07}. In addition, due to the continued disruption and merging of existing satellites, and the accretion of new ones, the system of satellite galaxies is not expected to be perfectly virialized or to be in a steady-state, and their kinematics therefore do not necessarily obey the Jeans equations \citep[e.g.,][]{Ye.etal.17, Wang.etal.17, Wang.etal.18, Adhikari.etal.18}. Furthermore, there is no reason why the local LOSVD, $P(\dV|\Rp,M,z)$, be perfectly Gaussian. Finally, several studies have shown that interlopers do not have a uniform distribution in redshift space \citep[e.g.,][]{vdBosch.etal.04, Wojtak.etal.07}. Clearly, then, there are numerous reasons why one might expect \Basilisk to fail, and its performance therefore needs to be tested and validated in detail.

We have done so using a three-tiered validation process, in which we test the performance of \Basilisk on a series of mock data sets of increasing complexity and realism. The Tier-1 mocks are highly idealized, abiding by all four assumptions listed above, and are mainly used to test the main engine of \Basilisk and to gauge the constraining power given a particular amount of data. For the Tier-2 mocks we place mock galaxies inside dark matter haloes in a large, cosmological $N$-body simulation. The satellite galaxies are given phase-space coordinates within their host haloes that, by construction, still obey assumptions {\bf II} and {\bf III}, but the masses, concentrations, positions and velocities of the dark matter haloes are taken directly from the simulation. A mock redshift survey is constructed by placing a virtual observer at a random location within the simulation volume, after which primaries and secondaries are selected using the cylindrical isolation criteria described in \S\ref{sec:selection}. Consequently, the Tier-2 mocks have realistic distributions of interlopers, and suffer from impurities and incompleteness effects (in particular fibre-collisions) in the same way as data extracted from a realistic galaxy redshift survey. Finally, the Tier-3 mocks are similar to the Tier-2 mocks, except that now the phase-space distributions of the satellite galaxies are assumed to be identical to those of the subhaloes (i.e., satellite galaxies are placed on subhaloes within the $N$-body simulation). Hence, in the Tier-3 mocks the satellite galaxies only obey assumptions {\bf II} and {\bf III} in as far as subhaloes do. 

As shown in \S\ref{sec:validation}, \Basilisk is able to yield precise and accurate constraints on the galaxy-halo connection in the case of all three tier mocks. The combined effect of impurities and fibre-collisions causes a small, systematic bias in the inferred radial profile of satellite galaxies, but we have demonstrated that this does not significantly impact the inference regarding the galaxy-halo connection. As demonstrated in Appendix~\ref{App:noNsat}, it is crucial, though, to include the number of secondaries per primary as a constraint on the data; failing to do so means that the model is unable to constrain the scatter in the galaxy-halo connection, and results in posteriors that are systematically biased. In line with our previous studies \citep[][]{Lange.etal.19a, Lange.etal.19b}, we therefore conclude that the kinematics of satellite kinematics are a powerful probe of the galaxy-halo connection, complementary to and competitive with galaxy clustering and galaxy-galaxy lensing. 

In fact, satellite kinematics has several advantages with respect to these alternative methods. Unlike galaxy-galaxy lensing, which requires tangential shear measurements that rely on accurate photometry, accurate characterization of the point-spread function (`seeing'), and a non-trivial method to extract reliable shape measurements, satellite kinematics can be measured from any redshift survey without the need for any additional data. And unlike galaxy clustering, which only probes halo mass indirectly through the mass dependence of the halo bias, satellite kinematics, similar to galaxy-galaxy lensing, directly probes the gravitational potential of the dark matter haloes, thus giving a more direct handle on the galaxy-halo connection. 

Although the discussion presented here has focused exclusively on constraining the conditional luminosity function, \Basilisk is easily modified so that it can constrain the conditional stellar mass function. This is most easily done by first defining a redshift-dependent, stellar-mass complete sample as in \citet{vdBosch.etal.08a}, and subsequently replacing the minimum luminosity, $L_{\rm min}(z)$, used in the computation of the expectation value for the number of satellites (equation~[\ref{lambdasat}]) with the corresponding minimum stellar mass, $M_{\ast, {\rm min}}(z)$. In addition, one has to account for the fact that stellar masses are not directly measurable, but instead are inferred from (multi-wavelength or spectroscopic) data in a model-dependent fashion. This implies that the data is effectively `convolved' with an unknown probability function, $P(M_\ast^{\rm obs}|M_\ast^{\rm true})$, relating the true and `observed' stellar masses, $M_\ast^{\rm true}$ and $M_\ast^{\rm obs}$, respectively. The Bayesian hierarchical framework that underlies \Basilisk is ideally suited to account for such a complication \citep[see][]{Sonnenfeld.Leauthaud.18}.

We end this paper by discussing potential future advances and applications. There are several opportunities for further development of the \Basilisk framework. First of all, \Basilisk in its current form assumes that the halo occupation statistics depend only on halo mass. If, instead, the occupation statistics also depend on other halo properties that impact satellite kinematics, such as halo concentration, the inference of \Basilisk may be significantly impacted. A particular concern is that haloes of a given mass that are more concentrated have fewer subhaloes on average \citep[e.g.,][]{vdBosch.etal.05a, Zentner.etal.05, Giocoli.etal.10, Jiang.vdBosch.17, Fielder.etal.19}\footnote{These studies reveal a strong anti-correlation between halo formation redshift and subhalo occupation; the dependence on halo concentration follows from its strong correlation with formation time \citep[][]{Wechsler.etal.02}.}. This implies a correlation between the number of satellite galaxies and host halo concentration, at fixed host halo mass.Since the kinematics of satellites depend on the concentration of the host halo, this correlation, which is currently not accounted for, could potentially impact \Basilisk's inference. We intend to examine this issue in the near-future, and upgrade \Basilisk accordingly, if needed. Another possible extension of \Basilisk is to include satellite luminosities as constraints on the model. In principle this should tighten the constraints on the satellite component of the CLF, and may well be important for improving the constraints on $\alpha_{12}$ and $\alpha_{14}$. On the other hand, this also causes complications, as it probably requires a careful treatment of luminosity-segregation, i.e., the fact that satellites of different luminosities have different radial profiles \citep[e.g.,][]{Rood.Turnrose.68, Biviano.etal.02, vdBosch.etal.08b, Balogh.etal.14}. Since this introduces additional degrees of freedom, i.e., $n_{\rm sat}(r|M) \rightarrow n_{\rm sat}(r|M,L_\rms)$, it remains to be seen to what extent including the satellite luminosities actually aids in constraining the galaxy-halo connection. And finally, based on the strong mass and radius dependence of the orbital anisotropy of dark matter subhaloes (cf. Fig.~\ref{fig:subhalo_anisotropy}), it may be worthwhile to consider more sophisticated anisotropy models for the satellite galaxies, going beyond the constant anisotropy (\CA) and Osipkov-Merritt (\OM) models considered here.

As for applications, we will use \Basilisk to analyze existing and forthcoming galaxy redshift surveys, including the SDSS Main Galaxy Sample \citep[][]{York.etal.00}, the Baryon Oscillation Spectroscopic Survey \cite[BOSS,][]{Dawson.etal.13}, the Galaxy and Mass Assembly Survey \citep[GAMA,][]{Driver.etal.11}, and the Dark Energy Spectroscopic Instrument (DESI). In particular, since \Basilisk can yield accurate constraints for relatively small subsamples, compared to the full extent of these surveys, such an analysis will yield a much richer, multi-dimensional characterization of the galaxy-halo connection. We will also compare the inferred relation between galaxies and their dark matter haloes to constraints inferred from a combined analysis of galaxy clustering and galaxy-galaxy lensing. The latter consistently reveals tension with the cosmological parameters inferred from the Planck cosmic microwave background data \citep[see e.g.,][]{Cacciato.etal.13, Mandelbaum.etal.13, Leauthaud.etal.17, DES_18a}, and it will be interesting to see whether such tension persists when satellite kinematic data are included in the analysis. Another interesting opportunity is to compare the halo masses inferred for samples of primaries from satellite kinematics, using \Basilisk, and galaxy-galaxy lensing. This can test the law of gravity on the scale of galaxy- and group-sized haloes by constraining the gravitational slip \citep[see e.g.,][]{Daniel.etal.08, Pizzuti.etal.19}.  In short, we envision a bright future for the hitherto under-utilized method of satellite kinematics as a probe of the galaxy-halo connection.


\section*{Acknowledgments}

We are grateful to the referee for constructive comments that helped to improve the presentation. FvdB and JUL are supported by the US National Science Foundation through grant AST 1516962. ARZ is funded by the Pittsburgh Particle physics Astrophysics and Cosmology Center (Pitt PACC) at the University of Pittsburgh and by the NSF through grant NSF AST 1517563. This research was supported in part by the National Science Foundation under Grant No. NSF PHY-1125915 and NSF PHY-1748958. FvdB received additional support from the Klaus Tschira foundation, and from the National Aeronautics and Space Administration through Grant No. 17-ATP17-0028 issued as part of the Astrophysics Theory Program. 


\bibliographystyle{mnras}
\bibliography{references_vdb}


\onecolumn
\appendix

\section{Computing likelihoods}
\label{App:comp}

In the main text we have given the general expressions that are relevant for computing the likelihood $\calL_{\rm SK}(\bD|\btheta)$, where $\btheta$ is the model vector, and $\bD$ is the data vector given by
\begin{equation}
\bD = \sum_{i=1}^{N_\rmc} \left( \{\dVij, \Rij | j=1,...,\Nsi \} | \Lci, \zci, \Nsi\right)\,.
\end{equation}
Here we give the corresponding expressions for the specific model adopted throughout.  As shown in \S\ref{sec:method} the log-likelihood can be written in compact form as $\ln\calL_{\rm SK}(\bD|\btheta) = \sum_{i=1}^{N_\rmc} \left(\ln G_i - \ln F_i\right)$ (Equation~[\ref{lnLD}]). Here $G_i$ and $F_i$ are integrals over halo mass, which we compute by integrating $\ln M$ from $M=10^{10} \Msunh$ to $10^{15}\Msunh$ using Gaussian quadrature. This allows us to write
\begin{equation}
F_{i} = \sum\limits_k  w_k \exp[{\hat{F}_{ik}}]\,, \,\,\,\,\,\,\,\,\,\,\,\,\,\,\,\,\,\,\,\,\,\,\,\,
G_{i} = \sum\limits_k  w_k \exp[\hat{F}_{ik}+Q_{ik}]\,,
\end{equation}
with $\hat{F}_{ik} = \ln F_{ik}$ (cf. equations~[\ref{Gi}]-[\ref{Fi}]). Using the expressions for $P(\Lci | \Mk, \zci)$ and  $P(\Nsi | \Mk, \Lci, \zci)$, we have that
\begin{equation}
\hat{F}_{ik} = \ln \Mk + \ln\left[n(\Mk,z_i)\right] - \ln\Gamma(N_{\rms,i} + 1)- \ln[\sigma_\rmc(\Mk)] -
\left(\frac{\log \Lci - \log \bar{L}_\rmc(\Mk)}{\sqrt{2}\sigma_\rmc(\Mk)} \right)^2
 + N_{\rms,i} \ln\lambda_{ik} - \lambda_{ik}\,. 
\end{equation}
Here $\lambda_{ik}$ is the expectation value for the total number of secondaries (satellites plus interlopers) around a primary of luminosity $L_{\rmc,i}$ residing in a halo of mass $\Mk$ at redshift $z_{\rmc,i}$, and is given by
\begin{equation}
\lambda_{ik} = \lambda_{\rm tot}(\Mk, \Lci, \zci) =  \lambda_{\rm sat}(\Mk, \Lci, \zci) + \lambda_{\rm int}(\Lci, \zci)\,,
\end{equation}
with $\lambda_{\rm sat}(M,\Lc,\zc)$ and $\lambda_{\rm int}(\Lc,\zc)$ given by equations~(\ref{lambdasat}) and (\ref{interlopermodel}), respectively. For our particular model for the CLF we have that 
\begin{equation}
\int_{L_{\rm min}(z)}^{\infty} \Phi_\rms(L|M) \, \rmd L = \frac{\phi^{\ast}_\rms(M)}{2} \,\Gamma\left( \frac{\alpha_\rms+1}{2},\left[\frac{L_{\rm min}(z)}{L^*_\rms(M)}\right]^2 \right)\,,
\end{equation}
with $\Gamma(a,x)$ the incomplete Gamma function, while the generalized NFW profile implies an aperture fraction (equation~[\ref{fap}])
\begin{equation}
f_{\rm ap}(\Mh, \Lc, \zc) =\frac{1}{\mu_\gamma(c/\calR)} \, \int\limits_0^{c/\calR} \frac{x^{2-\gamma}}{(1+x)^{3-\gamma}} \, \Big(\zeta[x \calR r_\rms,\Rmax] -  \zeta[x \calR r_\rms, \Rmin]\Big)\, \rmd x\,.
\end{equation}
Here $\Rmax = \Rs(\Lc)$ and $\Rmin = R_{\rm cut}(\zc)$, while 
\begin{equation}\label{mugamma}
\mu_\gamma(x) = \int_0^x \frac{y^{2-\gamma}\,\rmd y}{(1+y)^{3-\gamma}}\,,
\end{equation}
and $\zeta(r,R)$ is given by equation~(\ref{zetafunc}).

The expression for $Q_{ik}$ is given by
\begin{equation}
Q_{ik} = \sum\limits_{j=1}^{\Nsi} \ln\left[f_{\rm int} P_{\rm int}(\dVij, \Rij|\Lci) + (1-f_{\rm int}) P_{\rm sat}(\dVij, \Rij|\Mk, \Lci, \zci) \right]\,.
\end{equation}
Here $f_{\rm int} = f_{\rm int}(\Mk, \Lci, \zci)$ is the interloper fraction [equation~(\ref{fint})], $P_{\rm int}(\dV,\Rp|L,z)$ is given by equation~(\ref{Pint}), and $P_{\rm sat}(\dV,\Rp|M,L,z)$ is written as the product of two probabilities, $P(\Rp|M,L,z)$ [equation~(\ref{PrP})], and $P(\dV|\Rp,M,z)$ [equation~(\ref{PdVgauss})]. These depend on the normalized, projected number density, $\bar{\Sigma}(\Rp|M,z)$ [equation~(\ref{ProjNsat})], which for the gNFW profile (equation~[\ref{nsatprof}]) is given by
\begin{equation}
\bar{\Sigma}(\Rp|M,z) = \frac{\calQ(\gamma,\calR)}{2 \pi \, \calR^2 \, r^2_\rms \, \mu_\gamma(c/\calR)} \,,
\end{equation}
where
\begin{equation}
\calQ(\gamma,\calR) \equiv \int\limits_{\Rp/\calR r_\rms}^{c/\calR} \frac{y^{1-\gamma} \, \rmd y}{(1+y)^{3-\gamma} \, \sqrt{y^2 - (\Rp/\calR r_\rms)^2}}\,,
\end{equation}
and on the projected line-of-sight velocity dispersion, $\sigma^2_{\rm los}(\Rp|M,z)$. In the case of a constant anisotropy (\CA-model), the latter is given by equation~(\ref{sig2beta}), which for a gNFW tracer population in a NFW halo becomes
\begin{equation}
\sigma^2_{\rm los}(\Rp|M,z) = \Vvir^2 \frac{c}{\mu_{1}(c)} \, \frac{\calR^{2\beta-2}}{\calQ(\gamma,\calR)} \, \int\limits_{\Rp/r_\rms}^{c} \left[1 - \beta \frac{s^2_\rmp}{s^2}\right] \frac{s^{1-2\beta} \, \rmd s}{\sqrt{s^2 - s^2_\rmp}} \int\limits_{s/\calR}^{\infty} \frac{x^{2\beta - 2 - \gamma}}{(1+x)^{3-\gamma}} \, \mu_{1}(x\calR) \, \rmd x\,.
\end{equation}
Here $s_\rmp \equiv \Rp/r_\rms$ is the projected radius in units of the scale radius of the dark matter host halo, $\Vvir^2 = G M/\rvir$, and both $\cvir$ and $\rvir$ are functions of $M$ and $z$. For the former we adopt the fitting function of \citet{Maccio.etal.08}, ignoring scatter, and the latter follows directly from our definition of halo mass. In the case of an Osipkov-Merritt (\OM) model, the expression for the line-of-sight velocity dispersion becomes
\begin{equation}
\sigma^2_{\rm los}(\Rp|M,z) = \Vvir^2 \frac{c}{\mu_{1}(c)} \, \frac{1}{\calQ(\gamma,\calR)} \, \int\limits_{\Rp/r_\rms}^{c} \frac{s^2 + s^2_\rma - s^2_\rmp}{(s^2 + s^2_\rma)^2} \, \frac{s \, \rmd s}{\sqrt{s^2 - s^2_\rmp}} \int\limits_{s/\calR}^{\infty} \frac{x^2 + (s_\rma/\calR)^2}{x^{2+\gamma} \, (1+x)^{3-\gamma}} \, \mu_{1}(x\calR) \, \rmd x \,.
\end{equation}
Here $s_\rma \equiv r_\rma/r_\rms$ is the anisotropy radius in units of the scale radius of the dark matter host halo.
\begin{figure*}
\centering
\includegraphics[width=\textwidth]{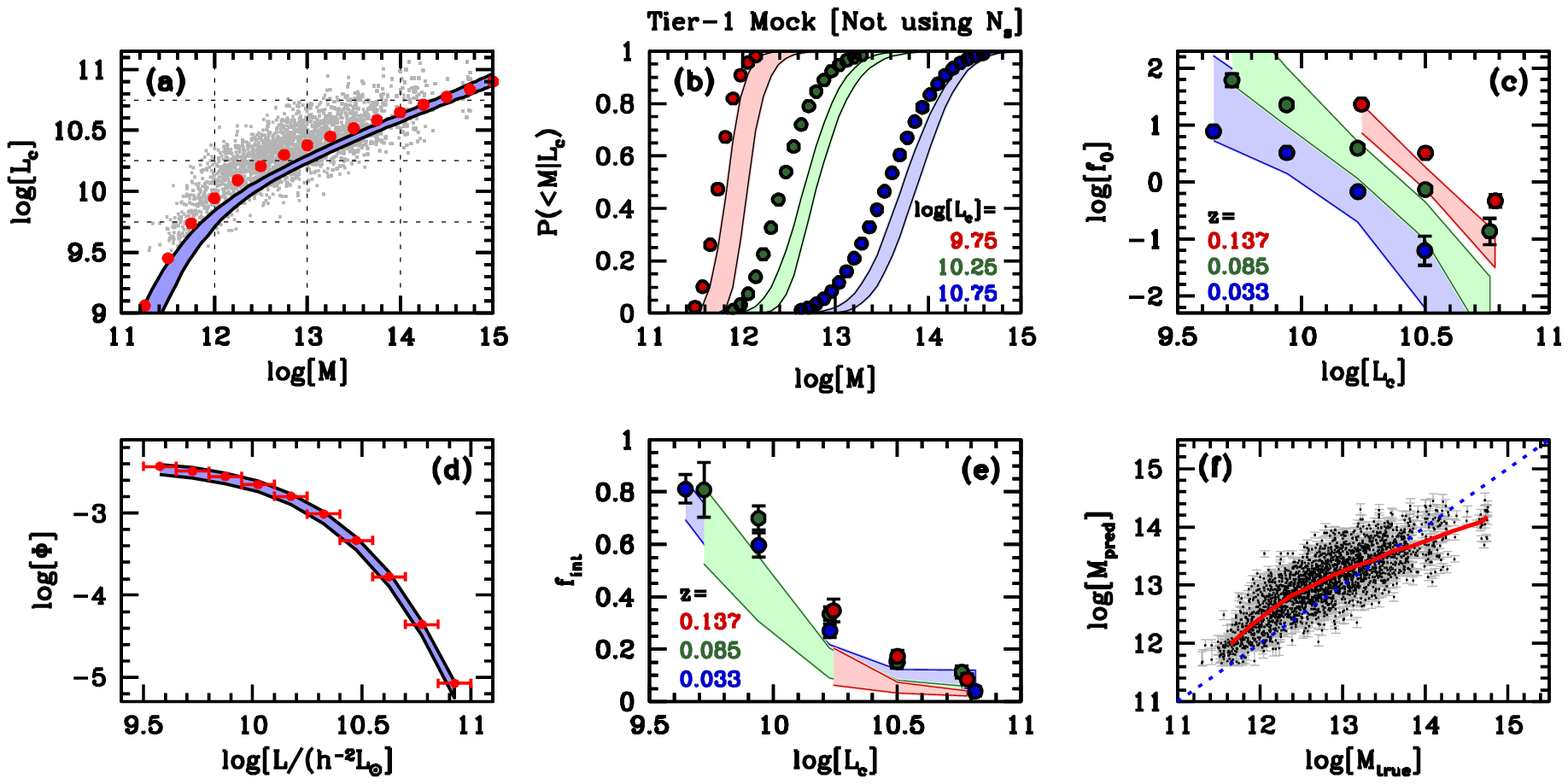}
\caption{Same as Figure~\ref{fig:mock1res}, but here we do not use $N_{\rms,i}$, the number of secondaries associated with primary $i$, as constraints. Due to reasons explained in the text, this results in a predicted galaxy-halo connection that is systematically offset from the true values.}
\label{fig:mock1test}
\end{figure*}

Along similar lines, when computing the log-likelihood $\ln \calL_0$ (equation~[\ref{calL0}]), we compute the probabilities $P_{ij}$ that a primary with luminosity $L_i$ at redshift $z_j$ has zero secondaries (equation~[\ref{Pij}]) using 
\begin{equation}
P_{ij} = \frac{\sum_k  w_k \exp[{\hat{E}_{ijk} - \lambda_{ijk}}]}{\sum_k  w_k \exp[{\hat{E}_{ijk}}]}\,.
\end{equation}
Here
\begin{equation}
\hat{E}_{ijk} = \ln \Mk + \ln\left[n(\Mk,z_j)\right] - \ln[\sigma_\rmc(\Mk)] -
\left(\frac{\log L_i - \log \bar{L}_\rmc(\Mk)}{\sqrt{2}\sigma_\rmc(\Mk)} \right)^2\,,
\end{equation}
and $\lambda_{ijk}$ is the expectation value for the total number of secondaries (satellites plus interlopers) around a primary of luminosity $L_{\rmc,i}$ residing in a halo of mass $\Mk$ at redshift $z_{\rmc,j}$.

By keeping the Gaussian quadrature points $(\Mk, \wk)$ fixed, we can pre-compute the first two and three terms of $\hat{E}_{ijk}$ and $\hat{F}_{ik}$, respectively, as they are independent of the model, $\btheta$. Furthermore, if we keep the satellite distribution, $n_{\rm sat}(r|M)$, and anisotropy, $\beta(r|M)$, fixed, $P(\dVij, \Rij | \Mk, \zci)$  and $f_{\rm apt}(\Mk, \Lci, \zci)$ also become independent of $\btheta$ and can thus be pre-computed. In that case, computing $\ln\calL_{\rm SK}(\bD|\btheta)$ and $\ln\calL_{0}(\bD_0|\btheta)$ becomes a simple summation of terms that only require the computation of $\bar{L}_{\rmc}(\Mk)$, $\sigma_\rmc(\Mk)$, and $\langle N_\rms | \Mk \rangle$. This is sufficiently fast to allow for an accurate inference of the posterior $P(\btheta|\bD)$ in a reasonable amount of time, and using only standard resources. In particular, a single evaluation of $\ln\calL_{\rm tot}(\bD|\btheta)$ for a data set $\bD$ containing $5,000$ satellites only requires $\sim 10$ milli-seconds on a single-core processor. 

\section{The importance of accounting for the number of secondaries}
\label{App:noNsat}

As described in \S\ref{sec:method}, the data vector used in our analysis is $\bD = \sum_{i=1}^{N_\rmc} \bD_i$, with
\begin{equation}\label{dataprimB}
\bD_i = \left( \{\dVij, \Rij | j=1,...,\Nseci \} | L_{\rmc,i}, N_{\rms,i}, z_{\rmc,i}\right)\,.
\end{equation}
Here $\Nseci$ is the number of secondaries associated with primary $i$, which in the model is used to inform the halo mass in the conditional probability $P(\Mh | L_{\rmc,i}, z_{\rmc,i}, N_{\rms,i})$ given by equation~(\ref{ProbMass}). Using Bayes theorem, this implies that the likelihood for each primary is multiplied with the probability $P(N_{\rms,i}|\Mh,z_{\rmc,i},L_{\rmc,i})$, described in detail in \S\ref{sec:PNsat}.

In principle, we could also opt to not include $N_\rms$ as part of the data vector. This implies that equation~(\ref{ProbMass}) simplifies to
\begin{equation}\label{ProbMassnoN}
P(\Mh | L_\rmc, z_\rmc) = \frac{P(L_\rmc|\Mh, z_\rmc) \, P(\Mh, z_\rmc)}{\int \rmd \Mh \, P(L_\rmc|\Mh, z_\rmc) \, P(\Mh,z_\rmc)}\,.
\end{equation}
which removes $P(N_{\rms,i}|\Mh,z_{\rmc,i},L_{\rmc,i})$ from the likelihood. Effectively this means that the number of secondaries per primary is no longer used as a constraint in the inference. In addition, ignoring $N_\rms$ also implies that we have no constraints on $f_0$ (i.e., the fraction of primaries with zero secondaries). Although this speeds up the likelihood evaluation by roughly a factor of three, this is not a viable option, as it results in a large, systematic bias in the inferred galaxy-halo connection. This is demonstrated in Fig.~\ref{fig:mock1test} where we show the results of an analysis of our Tier-1 mock (cf. Fig.~\ref{fig:mock1res}), in which we have ignored $N_{\rms,i}$ as observational constraints. Clearly, the predicted masses at given central luminosity are now systematically and significantly too high, while the predicted luminosities of centrals at given halo mass are too low.  The systematic bias arises from the non-negligible scatter in the galaxy-halo connection, and from the fact that more massive haloes contribute, on average, more satellites. As described in \S\ref{sec:constraining}, unless one accounts for this, the massive haloes receive more weight in the analysis, causing an overestimate in the predicted halo masses. In the standard analysis of satellite kinematics, described in \S\ref{sec:standard}, this problem can be circumvented using host-weighting (ideally in combination with satellite weighting), while in \Basilisk it is avoided by using the number of secondaries as constraints in the hierarchical Bayesian inference.

\section{The orbital anisotropy of dark matter subhaloes in the SMDPL simulation}
\label{App:anisotropy}

\begin{figure}
\centering
\includegraphics[width=3.33in]{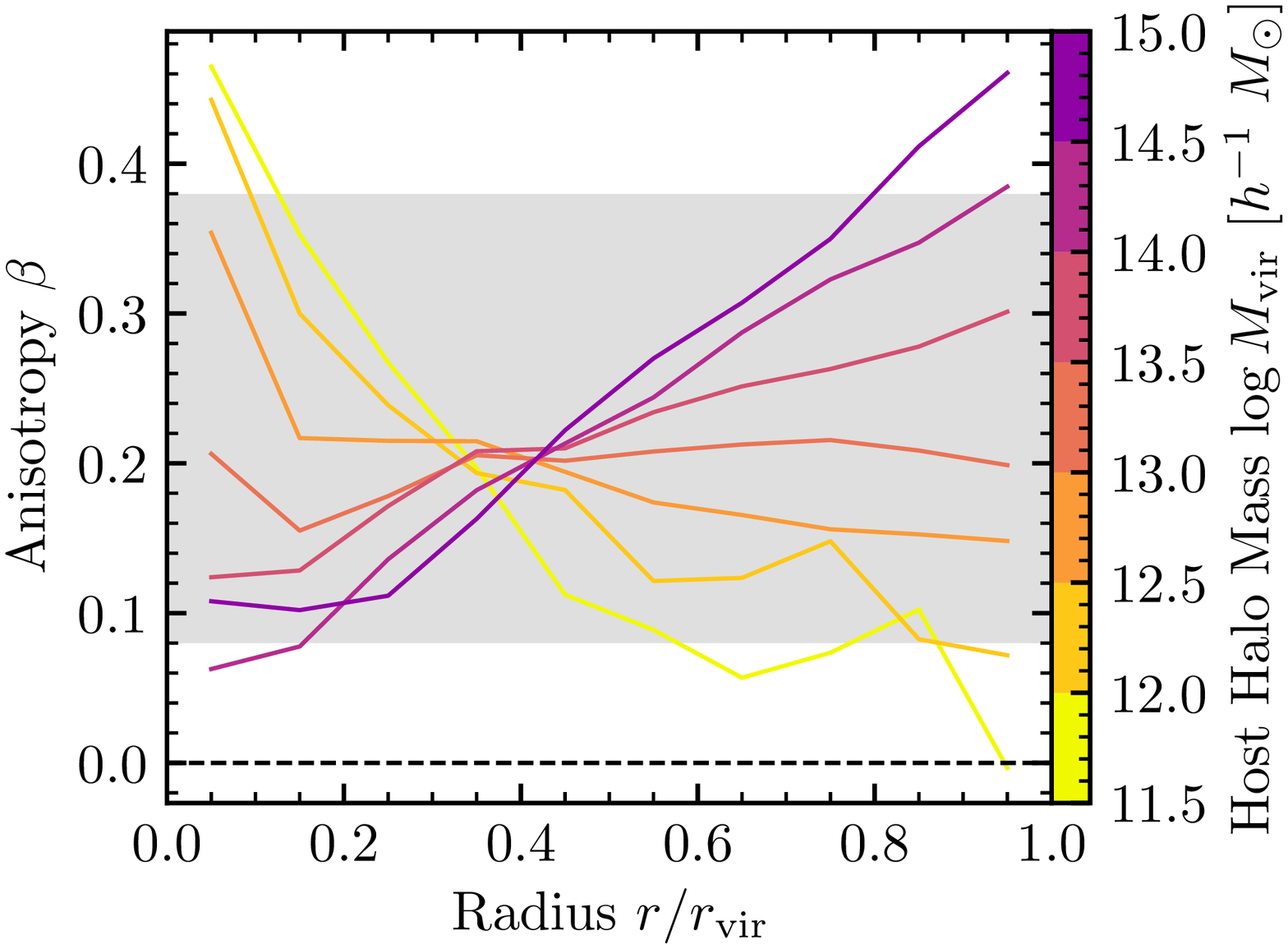}
\caption{The orbital anisotropy $\beta$ as a function of the 3D radial distance $r$ for resolved subhaloes with $M_{\rm peak} > 3 \times 10^{10} \Msunh$ in the SMDPL simulation. Different lines correspond to different host halo masses, as indicated by the color bar. All radii are scaled by the virial radius of the host halo. Overall, the orbits of subhaloes are radially anisotropic, but with a radial dependence that depends strongly on host halo mass. The gray-shaded region, which is reproduced as a vertical band in Fig.~\ref{fig:anisotropy}, corresponds to $\beta = [0.08,0.38]$ and roughly indicates the range of anisotropy parameters found in the majority of host haloes in the mass range $12 \lta \log[\Mvir/(\Msunh)] \lta 14.5$.}
\label{fig:subhalo_anisotropy}
\end{figure}

Here we study the orbital anisotropy of dark matter subhaloes in the SMDPL simulation  \citep {Klypin.etal.16}. We select all subhaloes with peak halo masses $M_{\rm peak} > 3 \times 10^{10} \Msunh$, and calculate their coordinates $\boldsymbol{r}$ and velocities $\boldsymbol{v}$ with respect to the halo core identified by {\sc ROCKSTAR}. We then determine the distance to the core $r = |\boldsymbol{r}|$, as well as the radial velocity $v_\rmr = \boldsymbol{v} \cdot \boldsymbol{r} / r$ and tangential velocity $v_\rmt = (\boldsymbol{v}^2 - v_\rmr^2)^{1/2}$. We use these to compute the local anisotropy parameter $\beta$ (see equation~[\ref{betadef}]) as a function of radius. The results are shown in Fig.~\ref{fig:subhalo_anisotropy}, with different curves corresponding to different bins in host halo mass, as indicated. We confirm previous findings on the mass dependence of the subhalo anisotropy parameter \citep[][]{Cuesta.etal.08}. For massive haloes with $M_{\rm vir} \gta 10^{13} \Msunh$, the velocity anisotropy increases with increasing radius \citep[][]{Diemand.Moore.Stadel.04}, while the trend reverses for less massive haloes \citep[][]{Sawala.etal.17}. Interestingly, the anisotropy at $r \sim 0.4 \, r_{\rm vir}$ is roughly constant for all halo masses studied here. We checked that assuming different lower mass limits for $M_{\rm peak}$ does not qualitatively affect these findings. Overall, if we were to assume a halo mass and radius-independent anisotropy parameter, as done in our modelling, $\beta \sim 0.2$ would best describe the anisotropy parameter of subhaloes and therefore also the anisotropy of satellites of the Tier 3 mocks. This is in excellent agreement with the constraints on $\beta$ as inferred from the Tier-3 mock data (cf. lower-left panel of Fig.~\ref{fig:anisotropy}).  Note that the results of Fig.~\ref{fig:subhalo_anisotropy} are consistent with an Osipkov-Merrit model, which transits from isotropic at small radii to radially anisotropic at $r > r_\rma$, only for the most massive haloes ($M_{\rm vir} \gta 10^{14} \Msunh$). For less massive host haloes, the radial trend of $\beta$ is not adequately described by an OM model. This explains why, in the case of the Tier-3 mock, the \CA-model provides a significantly better fit to the data than the \OM-model (see discussion in \ref{sec:anisotropy}). 

\label{lastpage}

\end{document}